\documentclass[desactivate]{aa}  
\usepackage{natbib}
\usepackage{graphicx}
\usepackage{txfonts}
\usepackage{ulem}
\usepackage{threeparttable}
\usepackage{mathrsfs}
\usepackage{amsmath}
\usepackage{subfig}
\usepackage{float}
\usepackage{comment}
\usepackage{mathtools}
\usepackage{hyperref}
\hypersetup{
    colorlinks=true,
    linkcolor=blue,
    citecolor=blue,
    urlcolor=green,
    }
    \usepackage{placeins}
\usepackage{afterpage}

\begin{document}

\title{Ageing and dynamics of the tailed radio galaxies in Abell 2142}

   \author{L. Bruno 
          \inst{1,2},
          T. Venturi
          \inst{2,3,4},
            D. Dallacasa
          \inst{1,2},
            M. Brienza
         \inst{3,1},
            A. Ignesti
          \inst{5},
            G. Brunetti
         \inst{2},
         C. J. Riseley
           \inst{1,2},
        M. Rossetti
         \inst{6},
        F. Gastaldello
         \inst{6},
         A. Botteon
          \inst{2},   
        L. Rudnick
         \inst{7},
          R. J. van Weeren
         \inst{8},
        A. Shulevski
         \inst{9},
         D. V. Lal
         \inst{10}
          }

   \institute{
    Dipartimento di Fisica e Astronomia (DIFA), Universit\`a di Bologna (UNIBO), via Gobetti 93/2, 40129 Bologna, Italy
    \and
    Istituto Nazionale di Astrofisica (INAF) - Istituto di Radioastronomia (IRA), via Gobetti 101, 40129, Bologna, Italy
             \and
    Istituto Nazionale di Astrofisica (INAF) - Osservatorio di Astrofisica e Scienza dello Spazio (OAS) di Bologna, Via P. Gobetti 93/3, 40129, Bologna, Italy
             \and
    Center for Radio Astronomy Techniques and Technologies, Rhodes University, Grahamstown 6140, South Africa
             \and
    Istituto Nazionale di Astrofisica (INAF) - Osservatorio Astronomico di Padova (OAPD), Vicolo dell'Osservatorio 5, I-35122 Padova, Italy
    \and
    Istituto Nazionale di Astrofisica (INAF) - Istituto di Astrofisica Spaziale e Fisica cosmica (IASF) di Milano, Via A. Corti 12, I-20133, Milano, Italy
             \and
             Minnesota Institute for Astrophysics, University of Minnesota, 116 Church St SE, Minneapolis, MN 55455, USA
                \and
             Leiden Observatory, Leiden University, PO Box 9513, 2300 RA Leiden, The Netherlands
             \and
             ASTRON, Netherlands Institute for Radio Astronomy, Oude Hoogeveensedijk 4, 7991 PD, Dwingeloo, The Netherlands
              \and
             Tata Institute of Fundamental Research, Post Box 3, Ganeshkhind P.O., Pune 411007, India
             \\
\email{luca.bruno4@unibo.it}
}


 
  \abstract
   {Tailed radio galaxies are shaped by ram pressure owing to the high-velocity motion of their host through the intracluster medium (ICM). Recent works have reported on the increasing complexity of the phenomenology of tailed galaxies, with departures from theoretical ageing models and evidence of re-energising mechanisms, which are yet unclear.  }
   {The nearby ($z=0.0894$) galaxy cluster Abell 2142 hosts two tailed galaxies, namely T1 and T2, which exhibit peculiar morphological features. We aim to investigate the properties of T1 and T2 and constrain their spectral evolution, dynamics, and interactions with the ICM.  }
   {We combined LOw Frequency Array (LOFAR), upgraded Giant Metrewave Radio Telescope (uGMRT), Very Large Array (VLA), and MeerKAT data (from 30 MHz to 6.5 GHz) to carry out a detailed spectral analysis of T1 and T2. We analysed surface brightness profiles,
 measured integrated and spatially-resolved spectral indices, and performed a comparison with single injection ageing models. \textit{Chandra} X-ray data were used to search for discontinuities in the ICM properties in the direction of the targets.    } 
   {The spectral properties of T1 at low frequencies are predicted by ageing models, and provide constraints on the 3D dynamics of the host by assuming a constant velocity. However, sharp transitions along sub-regions of the tail, local surface brightness enhancements, and a spectral shape at high frequencies that is not predicted by models suggest a more complex scenario, possibly involving hydrodynamical instabilities and particle mixing. T2 exhibits unusual morphological and surface brightness features, and its spectral behaviour is not predicted by standard models. Two AGN outburst events during the infall of T2 towards the cluster centre could explain its properties.}
   {}

   
\titlerunning{Tailed galaxies in A2142}
\authorrunning{Bruno et al. 2024}
   \maketitle
%

\section{Introduction}

A variety of discrete and diffuse radio sources can be found in galaxy clusters, such as radio galaxies, radio halos, mini-halos, and radio relics (see \citealt{brunetti&jones14,vanweeren19} for reviews). These targets offer the chance to probe the evolution and dynamics of galaxy clusters and their members, the cosmic magnetism, and the energy transfer processes on various scales. 

Radio galaxies in clusters are typically of Fanaroff-Riley I-type (FRI; \citealt{fanaroff}) exhibiting two jets lacking hotspots at their tips, which are launched in opposite directions from the core at relativistic velocities, and become sub-relativistic at a distance of a few kpc. In galaxy clusters, FRIs can be reshaped due to the high-velocity (${\rm v}\sim 1000 \; {\rm km \; s^{-1}}$) motion of the host galaxy throughout the rarefied ($n_{\rm ICM}\sim 10^{-2}-10^{-4} \; {\rm cm^{-3}}$) thermal intracluster medium (ICM). Specifically, ram pressure ($P_{\rm ram}\propto {\rm v}^2n_{\rm ICM}$;  \citealt{gunn&gott72}) can deflect the radio jets by angles $\gtrsim 90^{\rm o}$ and generate narrow-angle tail (NAT) galaxies \citep[e.g.][]{miley72,pfrommer&jones11,ternidegregory17,pal&kumari23}, which exhibit a bright core (the head) and roughly parallel jets that rapidly diffuse as tails of length $\sim 100-500$ kpc. Ram pressure is also responsible for moderate deflections by angles $\ll 90^{\rm o}$ observed in some targets, possibly associated with merging subclusters, forming in this case a wide-angle tail (WAT) galaxy \citep[e.g.][]{owen&rudnick76,owen78,pinkney93,missaglia19,odea&baum23}. When the jets are not resolved (due to the intrinsic bending, projection effects, or resolution), NAT and WAT galaxies are generally referred to as head-tail (HT) galaxies. Owing to their origin, tailed radio galaxies are useful tracers of high-$z$ and/or low-mass groups and clusters \citep[e.g.][]{blanton00,blanton03,giacintucci&venturi09}, which are hardly detected by current X-ray telescopes.

In typical tailed galaxies, the radio brightness decreases from the core along the tail due to particle ageing. Therefore, the spectral index steepens with the distance from the core \citep[e.g.][]{cuciti18,botteon21,rudnick21}. However, departures from these trends have been observed in a number of targets \citep[e.g.][]{jones&mcadam94,degasperin17,sebastian17,vanweeren17b,lal20,muller21,rudnick21,edler22,riseley22b,lusetti24}, which suggest the presence of particle re-acceleration processes and/or amplification of magnetic fields. Furthermore, tailed galaxies, especially those extending throughout large fractions of the radii of clusters \citep[e.g.][]{owen14,wilber18,srivastava20,ignesti22b}, likely release relativistic electrons and magnetic fields into the ICM \citep[e.g.][]{degasperin17,vanweeren17b,vazza&botteon24}. Therefore, tailed galaxies are interesting targets to probe the interplay between thermal and non-thermal components in galaxy clusters, and the complex re-acceleration mechanisms that are yet poorly constrained \citep[e.g.][]{brunetti&jones14}.

Abell 2142 (A2142) is a nearby and massive galaxy cluster in an intermediate dynamical state between relaxed and merging systems \citep{rossetti13}. It is characterised by a complex dynamics \citep{liu18} and hosts peculiar discrete and diffuse radio sources \citep{venturi17,bruno23b,riseley24}. In the present work, we focus on two morphologically interesting HT galaxies in A2142, labelled as T1 and T2 in \cite{venturi17}. We aim to analyse their morphological and spectral properties over a wide range of wavelengths (30 MHz - 6.5 GHz), search for possible interactions between the tails and the ICM, and constrain their dynamics.

Throughout this paper we adopted a standard $\Lambda$CDM cosmology with $H_0=70\;\mathrm{km\; s^{-1}\; Mpc^{-1}}$, $\Omega_{\rm M}=0.3$ and, $\Omega_{\rm \Lambda}=0.7$. At the cluster redshift of $z=0.0894$, $1''=1.669$ kpc (or $1'\sim 100$ kpc). We adopted the convention on the spectral index $\alpha$ as defined from the flux density $S(\nu) \propto \nu^{-\alpha}$. The paper is organised as follows. In Sect. \ref{sect: The galaxy cluster Abell 2142}, we describe the galaxy cluster A2142. In Sect. \ref{sect: Observations and data reduction}, we present the radio and X-ray data and their processing. In Sect. \ref{sect: Results}, we report on the results of our analysis. In Sect. \ref{sect: Discussion} we discuss scenarios to explain the properties of T1 and T2. In Sect. \ref{sect: Summary and conclusions}, we summarise our work.

\section{The galaxy cluster Abell 2142}  
\label{sect: The galaxy cluster Abell 2142}

\begin{figure*}
	\centering
	\includegraphics[width=0.75\textwidth]{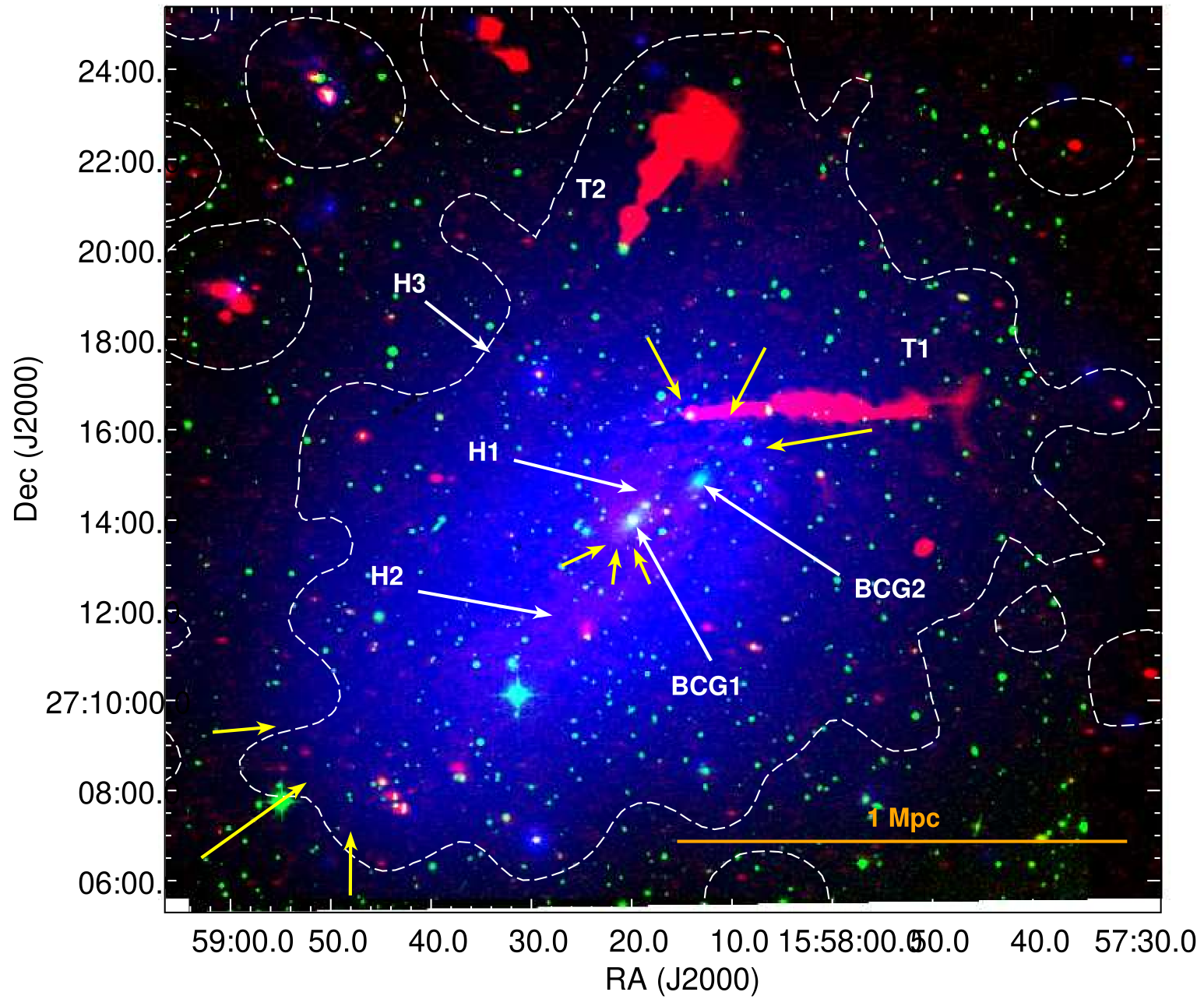}
	\smallskip
	
	\caption{Composite RGB image of A2142: radio (LOFAR at 143 MHz, $9''\times 6''$) in red, optical (DSS-2, r-filter) in green, and X-rays (XMM-Newton, 0.7-1.2 keV) in blue. Labels indicate the tailed radio galaxies (T1, T2), the primary and secondary brightest cluster galaxies (BCG1, BCG2), and the two components (H1, H2) of the radio halo. The dashed white contours represent the $3\sigma$ level of the $73''\times 66''$ LOFAR image at 143 MHz reported in \cite{bruno23b}, showing the third component (H3) of the radio halo. The yellow arrows indicate the location of the NW, central, and SE cold fronts reported by \cite{rossetti13}.}
	\label{ROX}
\end{figure*}

A2142 (RA$_{\rm J2000}=15^{\rm h}58^{\rm m}20^{\rm s}$, Dec$_{\rm J2000} =  27^{\rm o}14'00''$) is a nearby ($z=0.0894$) galaxy cluster of mass{\footnote{$M_{500}$ is the mass within a radius $R_{500}$, which encloses $500\rho_{\rm c}(z)$, where $\rho_{\rm c}(z)$ is the critical density of the Universe at a given redshift.}} $M_{500}=(8.8\pm0.2)\times 10^{14} \;  M_\odot$ within a radius $R_{500}=14.07\pm 0.70 \; {\rm arcmin}$ \citep[$1408.5\pm 70.4$ kpc at the cluster redshift;][]{planckcollaboration16,tchernin16}. The galaxy cluster A2142 is located at the centre of the A2142 supercluster, after which it is named \citep{einasto15,gramann15}. In Fig. \ref{ROX} we provide a multi-wavelength (optical, X-ray, radio) view of A2142, and label the various sources and features that we discuss below.  

A2142 hosts about 900 member galaxies gathered in small groups and hierarchically organised in structures and substructures \citep{owers11,einasto18,liu18}. Furthermore, several groups are infalling towards the richest structure in the cluster centre \citep[e.g.][]{owers11,eckert17,liu18}, which hosts the brightest cluster galaxy `BCG1' ${\rm (RA_{J2000} =239.5834, \;  DEC_{J2000} = 27.2334}$, $z=0.09081$, $M_*=1.9\times 10^{11} \; M_{\odot}$). The secondary brightest cluster galaxy, BCG2 ${\rm (RA_{J2000} =239.5554, \;  DEC_{J2000} = 27.2481}$, $z=0.0965$, $M_*=1.5\times 10^{11} \; M_{\odot}$) is located at a projected distance of $\sim 180$ kpc from BCG1 and is thought to be the main member of a merging group. Minor mergers are likely responsible for the uncommon properties of the ICM, owing to their intermediate dynamical state in between that of relaxed cool cores and unrelaxed (major) mergers \citep[e.g.][]{cavagnolo09,rossetti13,tchernin16,wang&markevitch18,cuciti21b}.

Radio observations of A2142 with the Low Frequency Array (LOFAR) revealed diffuse emission from the ICM in the form of a hybrid radio halo (Fig. \ref{ROX}) consisting of three distinct components \citep{bruno23b}. The brightest component (the `core', `H1') in the cluster centre has a roundish morphology and a diameter of $\sim  200$ kpc, the second component (the `ridge', `H2') is elongated towards south-east for $\sim  400$ kpc, and the third component (`H3') extends for $\sim 2$ Mpc embedding both H1 and H2 \citep{venturi17,bruno23b,riseley24}. The origin of the hybrid halo is likely associated with turbulent re-acceleration triggered by mergers taking place on different spatial scales and/or timescales \citep[see][for a detailed discussion]{bruno23b}.

 Besides the hybrid halo, extended radio emission is associated with two prominent HT galaxies, T1 and T2, which are the focus of the present work. The host of T1 ${\rm (RA_{J2000} =239.5596, \;  DEC_{J2000} = 27.2721}$, $z=0.09540$, $M_*=0.7\times 10^{11} \; M_{\odot}$) is an elliptical galaxy likely being a member of the merging group associated with BCG2 \citep{einasto18}. Its optical spectrum\footnote{\url{https://skyserver.sdss.org/dr12/en/tools/chart/navi.aspx}} shows the presence of weak emission lines (e.g. O[III], H$\alpha$) that indicate AGN activity. The core is  active at radio wavelengths \citep[e.g.][]{colla72,govoni10,ternidegregory17} and bright in the 0.5-10 keV band ($L_{0.5-10}=2.1\times 10^{42} \; {\rm erg \; s^{-1}}$; \citealt{sun09}), further supporting the association with an AGN. The host of T2 ${\rm (RA_{J2000} =239.5870, \;  DEC_{J2000} = 27.3337}$,  $z=0.08953$, $M_*=2.1\times 10^{11} \; M_{\odot}$)  is an elliptical member galaxy of A2142. Its optical spectrum lacks emission lines that would confirm the presence of current AGN activity, while its radio and X-ray emission will be discussed throughout this paper.

\section{Observations and data reduction}
\label{sect: Observations and data reduction}

\begin{table*}
 \centering
 \caption[]{Details of the radio data analysed in this work. A detailed description of the data marked with `$^*$' and `$^{**}$' and their processing is reported in \cite{bruno23b} and \cite{riseley24}, respectively.}
 \label{datiRADIO}
\resizebox{\textwidth}{!}{
 \begin{tabular}{cccccc}
\hline
   	\noalign{\smallskip}
   	Instrument & Band name &  Frequency coverage &  Observation date & On-source time & Project code  \\
   	&  &  (MHz) & & (h) &  \\
   	\noalign{\smallskip}
  	\hline
   	\noalign{\smallskip}
   	 LOFAR$^*$  & LBA  & 30-78 & 08,17,23-Dec.-2021 & 16.0  &  LC17\_012  \\
   	 LOFAR$^*$ & HBA & 120-168 & 15-Sep.-2018; 25,31-Oct.-2020; 13-Nov.-2020 & 32.0  &  P239+27; LC14\_018 \\
    GMRT & - & 225-240 & 23-Mar.-2013 & 6.0  & 23\_017  \\
    GMRT$^*$ & - & 305-340 & 27-Mar.-2013 & 5.0 & 23\_017    \\
    uGMRT$^*$  & band-3 & 300-500 & 15-Mar.-2018 & 3.0  &  33\_052   \\
    GMRT  & - & 590-625 & 22-Jun.-2013 & 5.0  & 23\_017    \\
    uGMRT & band-5  & 1050-1450 & 20-Mar.-2018 & 4.0 & 33\_052   \\
    MeerKAT$^{**}$ & L-band  & 872-1712 & 10-Oct.-2021; 12-Nov.-2021  & 5.5 & SCI-20210212-CR-01   \\
   $\rm VLA^*_{\rm [C-array]}$  & L-band & 1000-2000 & 27-Apr.-2012 & 0.5 & 11B-156    \\
   $\rm VLA^*_{\rm [D-array]}$  & L-band & 1000-2000 & 9-Oct.-2011 & 1.5  & 11B-156  \\ 
   $\rm VLA_{\rm [A-array]}$  & L-band & 1000-2000 & 10-Sep.-2015 & 0.25 & 15A-016  \\
   $\rm VLA_{\rm [A-array]}$  & C-band & 4500-6500 & 10-Sep.-2015 & 0.25 & 15A-016  \\
   
   	\noalign{\smallskip}
   	\hline
 
 \end{tabular}
}
\end{table*}

    \begin{table}
      \centering
        \caption[]{Details of the \textit{Chandra} X-ray data analysed in this work. A detailed description of the data marked with `$^*$' and their processing is reported in \cite{bruno23b}. }
        \label{datiX}
  \begin{tabular}{cccc}
\hline
   	\noalign{\smallskip}
        ObsID &  CCDs  &  Observation date &   Clean time \\
        &  &  & (ks)  \\
        \noalign{\smallskip}
        \hline
        \noalign{\smallskip}
        5005$^*$ & S2,I0,I1,I2,I3  & 13-Apr.-2005 &   41.5  \\
        15186$^*$ & S1,S2,S3,I2,I3 & 19-Jan.-2014 &   82.7  \\
        16564$^*$ & S1,S2,S3,I2,I3 & 22-Jan.-2014 &   43.2  \\
        16565$^*$ & S1,S2,S3,I2,I3 & 24-Jan.-2014 &   19.5   \\
        17168 & S2,S3,S4,I2,I3 & 01-Dec.-2014 &   86.9   \\
        17169 & S2,S3,S4,I2,I3 & 04-Oct.-2014 &   14.8   \\
        17492 & S2,S3,S4,I2,I3 & 03-Dec.-2014 &   67.5   \\
        \noalign{\smallskip}
        \hline
        \end{tabular}

   \end{table}

In this section we present the radio and X-ray data to study the HT radio galaxies in A2142. We first provide a brief overview of the radio data recently analysed in \cite{bruno23b} and in \cite{riseley24} for the study of the cluster's hybrid radio halo. The additional radio and X-ray data used in the present work are described in Sects. \ref{sect: GMRT data} to \ref{sect: Chandra data}. The details of all radio and X-ray observations are reported in Table \ref{datiRADIO} and Table \ref{datiX}, respectively.

Some of the radio data exploited in this work have been recently presented in \cite{bruno23b} to characterise the hybrid radio halo. These include data from LOFAR, the Giant Metrewave Radio Telescope (GMRT), the upgraded GMRT (uGMRT), and the Very Large Array (VLA). Furthermore, MeerKAT data have been recently used by \cite{riseley24} for a follow-up analysis. Here we briefly summarise these observations, and refer to \cite{bruno23b} and \cite{riseley24} for details on the telescope-specific setup and calibration strategies.

LOFAR observed A2142 for 16 and 32 hours with the Low Band Antenna (LBA) and High Band Antenna (HBA) arrays operating at 30-70 MHz and 120-168 MHz, respectively. The GMRT observed A2142 for 5 hours at 305-340 MHz \citep[see also][]{venturi17}, and 3 hour observations were carried out with the uGMRT at 300-500 MHz (band-3). Mosaicked pointings (each having a field of view of $\sim 30'$) on A2142 were obtained with the Very Large Array (VLA) at 1-2 GHz (L-band) in C-array and D-array configurations, for a total of 2 hours. MeerKAT observed A2142 at 872-1712 MHz (L-band) for 5.5 hours as part of the `MeerKAT-meets-LOFAR mini-halo census' project \citep{riseley22,riseley23}.

\subsection{GMRT data}
\label{sect: GMRT data}
Archival GMRT observations of A2142, first presented by \cite{venturi17}, are available in the 225-240 MHz and 590-625 MHz bands, for 6 and 5 hour on-source, respectively. The total bandwidths of 16 and 32 MHz are split into 128 and 256 channels, respectively. The sources 3C286 and 3C48 were used as absolute flux density scale calibrators.

Following the same procedure described in \cite{bruno23b} for the 305-340 MHz band data, we reprocessed the 225-240 and 590-625 MHz band data by means of the Source Peeling and Atmospheric Modeling ({\tt SPAM}) automated pipeline \citep{intema09}, which corrects for ionospheric effects by deriving directional-dependent gains from bright sources across the field of view. We reached noise levels of $\sim 180 \; \mu {\rm Jy \; beam^{-1}}$ at $13''\times 10''$ and $\sim 40 \; \mu {\rm Jy \; beam^{-1}}$ at $6''\times 4''$ for the 234 and 608 MHz datasets, respectively, which are consistent with those reported in \cite{venturi17}.

\subsection{uGMRT band-5 data}
\label{sect: uGMRT band-5 data}

New observations of A2142 were performed with the uGMRT in the 1050-1450 MHz (band-5) frequency range in March 2018. The total 400 MHz bandwidth is split into 8192 channels of width $\sim 49$ kHz each. The cluster was covered in its full extent with five different pointings (each having a field of view of $\sim 27'$), for a total of 7 hours. For the study presented in this paper, only the three pointings covering the northern part of the cluster ($\sim 40'$ in total) are considered (namely ${\rm `A2142\_2', \; `A2142\_3', \; `A2142\_5'}$), for a total of 4 hours. The sources 3C286 and 1602+334 were used as amplitude and phase calibrators, respectively. 

Data reduction of band-5 observations with {\tt SPAM} has not been tested in depth. However, direction-dependent corrections are negligible for these gigahertz frequency and small field of view data, and thus we did not used {\tt SPAM}. To process these data, we split the total bandwidth in 6 sub-bands of $\sim 67$ MHz each and carried out a standard data reduction with the Common Astronomy Software Applications \citep[{\tt CASA} v. 6.5;][]{mcmullincasapaper07} by iteratively performing flagging of Radio Frequency Interference (RFI), and bandpass, amplitude, and phase calibrations for each sub-band and pointing. We then recombined the calibrated sub-bands of each pointing to perform rounds of phase and phase plus amplitude self-calibration. The self-calibrated pointings were imaged separately with {\tt WSClean} v. 2.10 \citep{offringa14,offringa17} with multi-frequency and multi-scale synthesis options. After correcting each of the three images by the corresponding primary beam attenuation at the central frequency of 1250 MHz and convolving them to the same resolution, they were combined by means of the tool {\tt lm.makemosaic} in {\tt CASA} to produce a single mosaic image. At a resolution of $2.5''$, the final noise level is in the range $\sim 25-40 \; {\rm \mu Jy \; beam^{-1}}$. 

\subsection{VLA L-band and C-band radio data}
\label{sect: dati JVLA}

As first presented by \cite{ternidegregory17}, the head-tail galaxy T1 was studied with the VLA in A configuration at 1-2 GHz (L-band) and 4.5-6.5 GHz (C-band), for 15 minutes on source in each band. In both observations, the sources 3C286 and 1602+3326 were used as absolute flux density and phase calibrators, respectively. Data were recorded in 16 spectral windows of 128 MHz each. 

The field of view of these VLA observations include the core of T2, which we aim to study. We reprocessed the L-band and C-band data in {\tt CASA} following standard calibration procedures (see Sect. \ref{sect: uGMRT band-5 data}), and performing an additional cycle of phase self-calibration. Our data processing improved the quality of the images with respect to those reported in \cite{ternidegregory17} in terms of noise (improvement by factors of $\sim1.6$ and $\sim 1.2$ in L-band and C-band, respectively). We reached a noise level of $\sim 27 \; {\rm \mu Jy \; beam^{-1}}$ in L-band at $\sim 1''$ resolution and $\sim 11 \; {\rm \mu Jy \; beam^{-1}}$ in C-band at $\sim 0.3''$ resolution.

\subsection{Radio imaging}
\label{sect: Radio imaging}

Imaging of all radio data was carried out with {\tt WSClean} v. 2.10 \citep{offringa14,offringa17} to account for wide-field, multi-frequency, and multi-scale synthesis. For both VLA and uGMRT, mosaicked observations were imaged separately and then properly combined following the procedure described in Sect. \ref{sect: uGMRT band-5 data} for uGMRT.

In the following, uncertainties on the reported radio flux densities $S$ are computed as: 
\begin{equation}
\Delta S= \sqrt{ \left( \sigma^2 \cdot N_{\rm beam} \right) + \left(  \xi_{\rm cal} \cdot S \right) ^2} \; \; \; ,
\label{eq: erroronflux}
\end{equation}
where $N_{\rm beam}$ is the number of independent beams within the considered region, and $\xi_{\rm cal}$ is the calibration error. We assumed standard calibration errors of $\xi_{\rm cal}=10\%$ for LOFAR \citep{shimwell22LOTSSDR2,degasperin23}, $\xi_{\rm cal}=7\%, \; 6\%, \; 6\%, \; 5\%, \; 5\%$ for GMRT at 234 MHz, 323 MHz, 407 MHz, 608 MHz, and 1250 MHz, respectively \citep{chandra04}, $\xi_{\rm cal}=5\%$ for VLA and MeerKAT in \textit{L}-band \citep{perley&butler13}, and $\xi_{\rm cal}=3\%$ for VLA in \textit{S}-band and \textit{C}-band \citep{perley&butler13}.

\subsection{\textit{Chandra} X-ray data}
\label{sect: Chandra data}

In \cite{bruno23b} we analysed deep \textit{Chandra} observations of A2142. These consist of 4 pointings of 187 ks in total that mainly cover the central regions of the clusters. In the present work, we considered 3 additional pointings covering the northern and north-eastern regions of A2142, in the direction of T2. These observations were carried out in 2014, in VFAINT mode, with both ACIS-I and ACIS-S CCDs, and were first presented in \cite{eckert17}. A summary of all \textit{Chandra} data considered in this work is reported in Table \ref{datiX}.

As for the other pointings, we reprocessed the additional data by means of {\tt CIAO} v. 4.13, with {\tt CALDB} v. 4.9.4. After extracting light curves in source-free regions, soft proton flares were filtered out with the {\tt lc\_clean} algorithm, leaving a clean time of $169.1$ ks. Overall, the 7 pointings provide a total clean time of $356$ ks. In this work, we will use these data to investigate the local conditions of the ICM towards T1 and T2 with a combination of resolution and signal-to-noise ratio (${\rm S/N}$) that depends on the depth of the pointings covering the same regions. We refer to \cite{bruno23b} for the description of background treatment for imaging and spectral analysis.


\section{Results}
\label{sect: Results}

\subsection{Radio morphology and sub-regions}
\label{sect:Radio morphology}

\begin{figure*}
	\centering
	\includegraphics[width=0.48\textwidth]{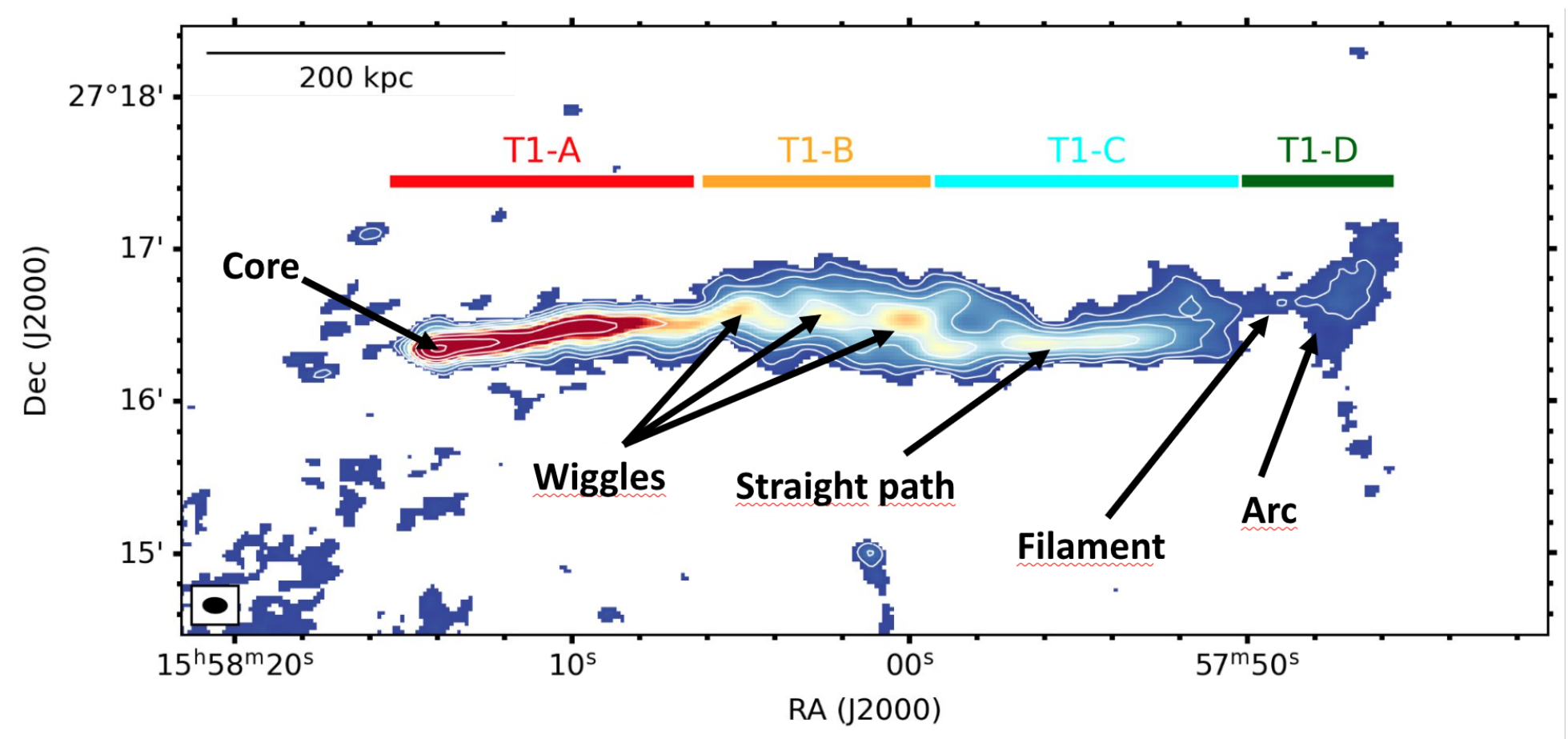}
\includegraphics[width=0.48\textwidth]{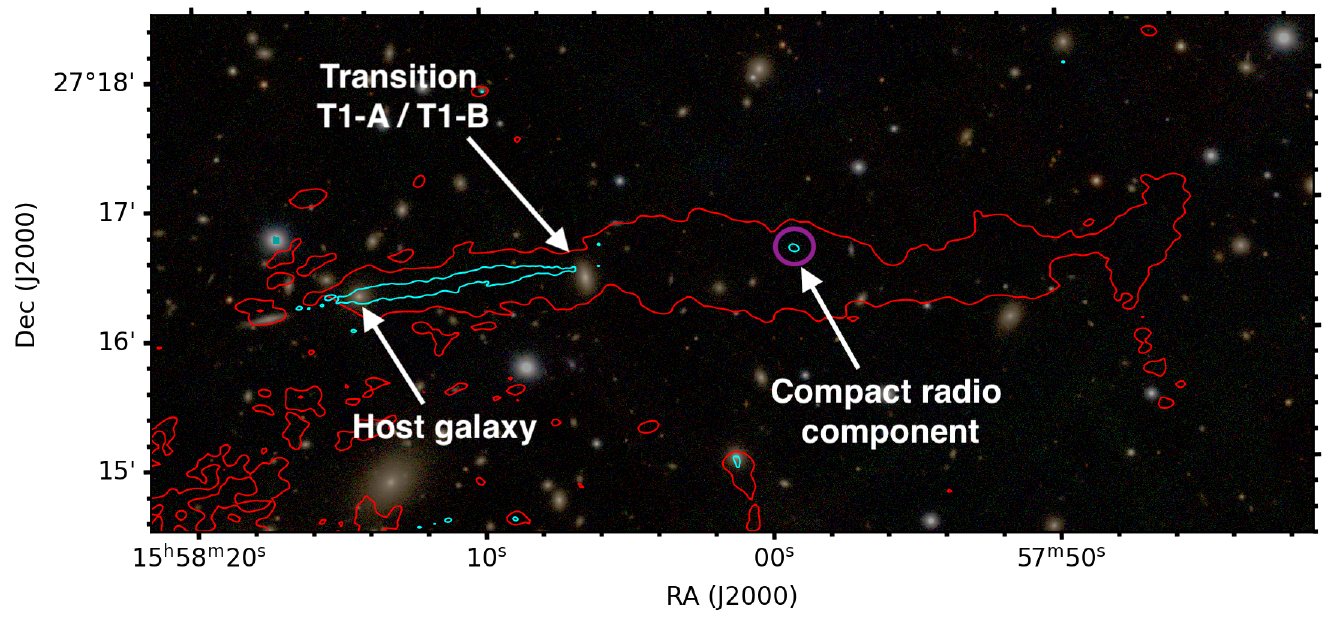}
	\includegraphics[width=0.4\textwidth]{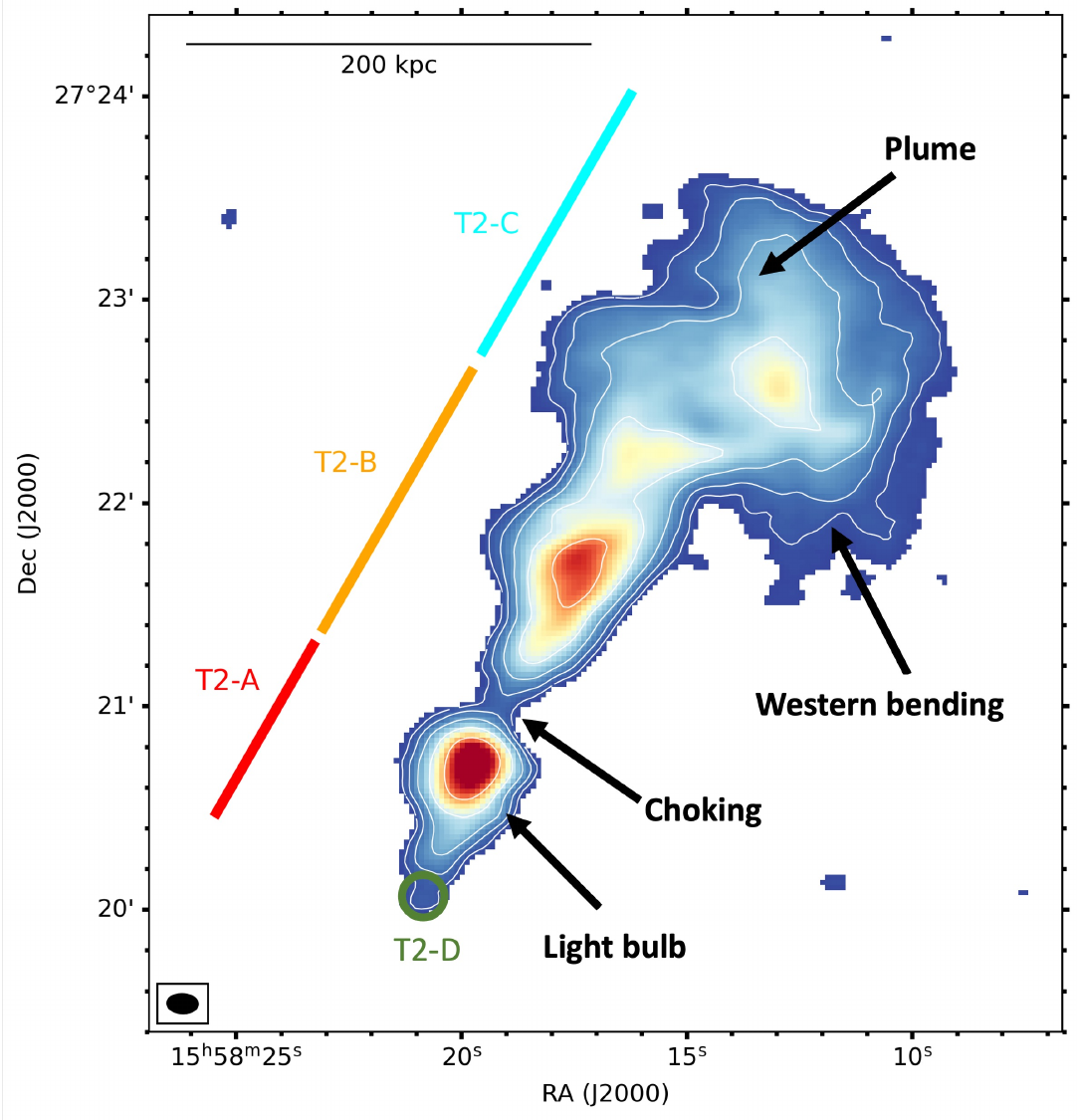}
\includegraphics[width=0.4\textwidth]{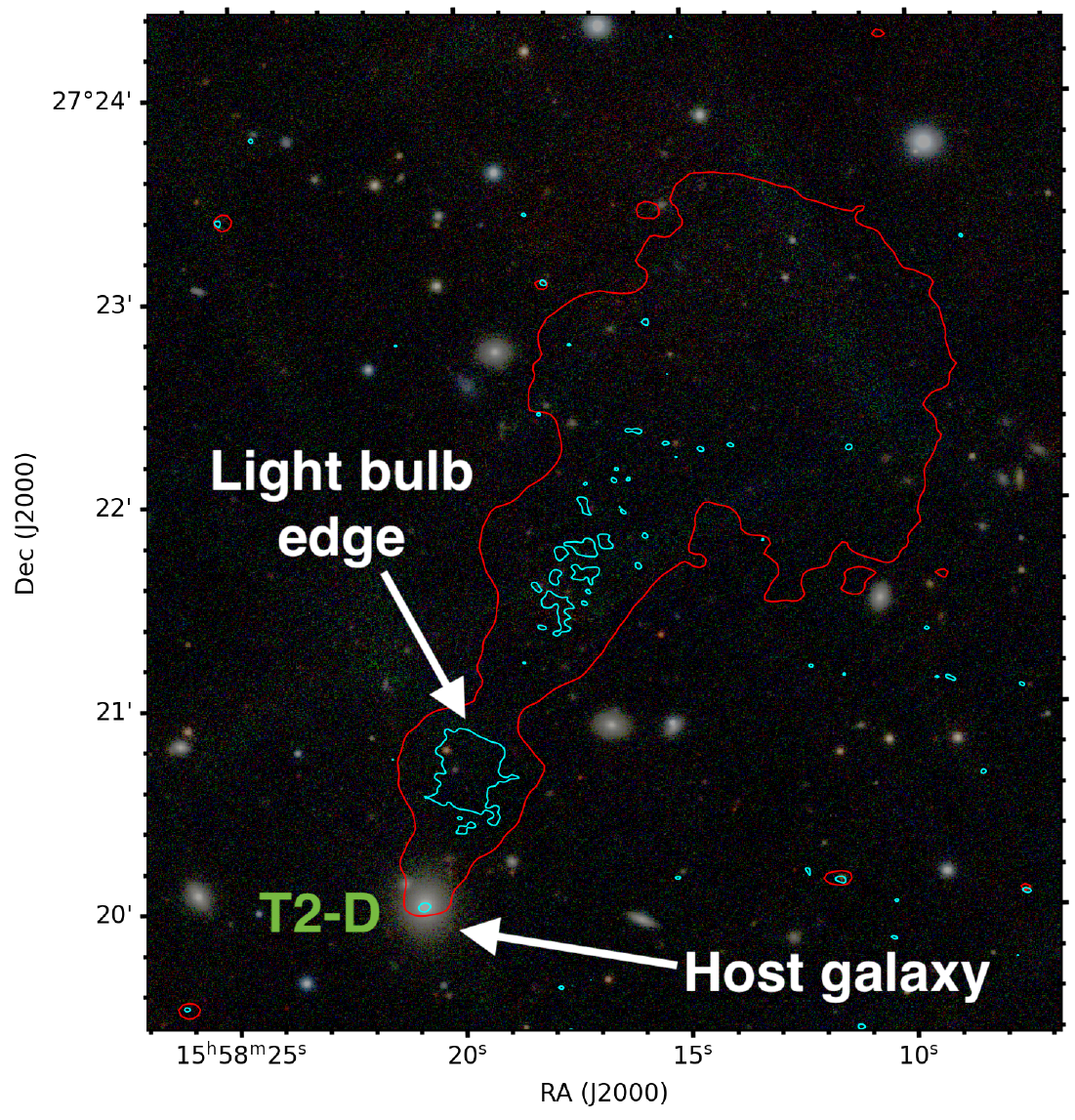}

	\smallskip
 \caption{Radio images and contours labelling the sub-regions and features discussed in the text. \textit{Left panels}: 143 MHz images (see Figs. \ref{fig: radio images full res}, \ref{fig: radio images full res 2}) of T1 (top) and T2 (bottom). \textit{Right panels}: Pan-STARSS optical (composite i, r, g filters) images zoomed towards T1 (top) and T2 (bottom). Radio contours at 143 MHz ($5\sigma$, in red) and 1250 MHz ($5\sigma$ for T1 and $3\sigma$ for T2, in cyan) are overlaid.}
	\label{fig: substructures}
\end{figure*} 

\begin{figure*}[!h]
	\centering
	\includegraphics[width=0.49\textwidth]{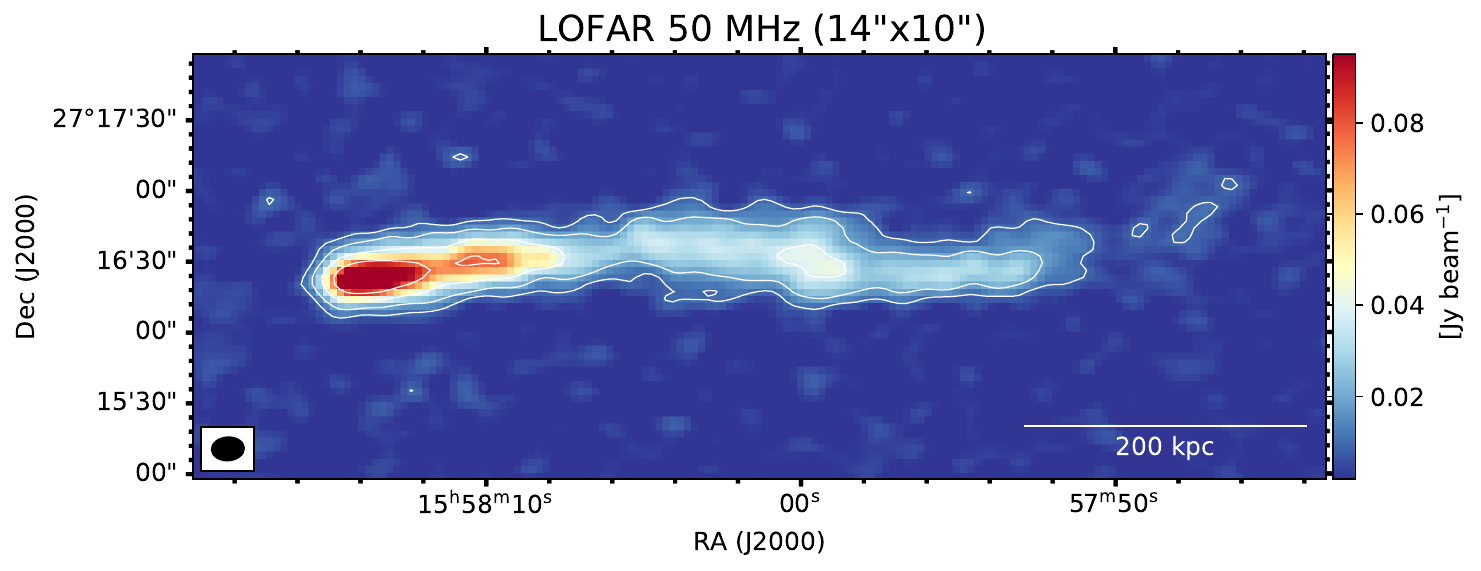}
	\includegraphics[width=0.49\textwidth]{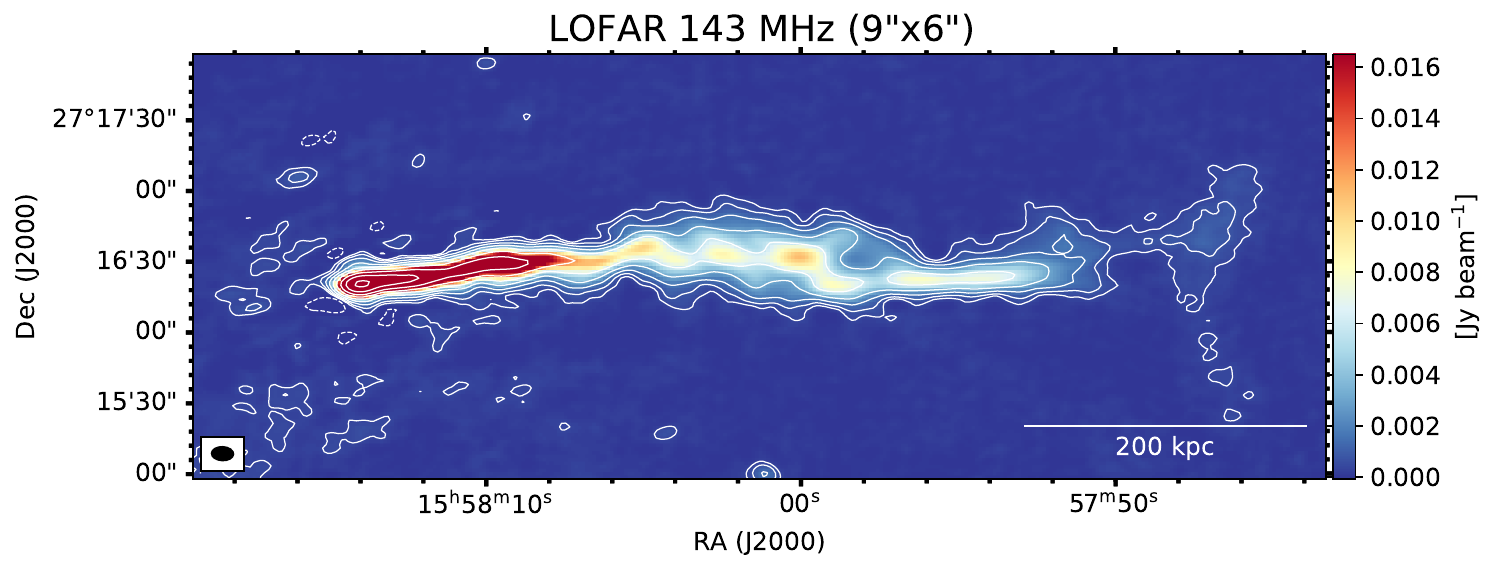}
	\includegraphics[width=0.49\textwidth]{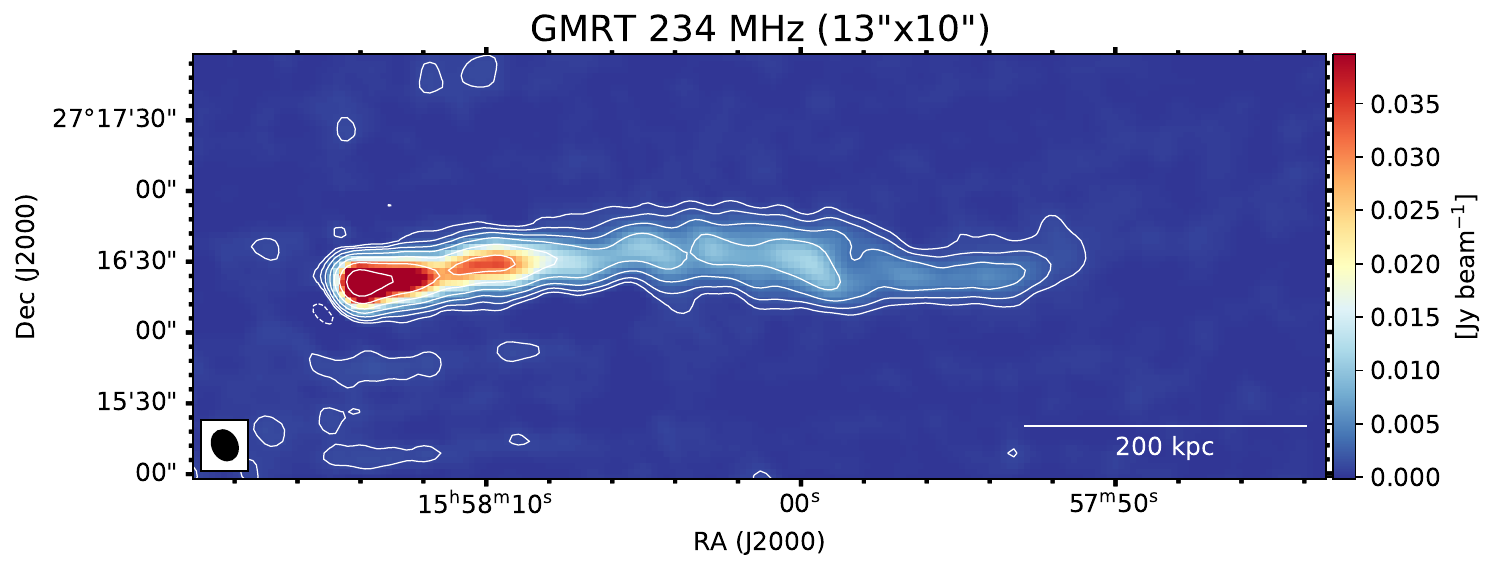}
	\includegraphics[width=0.49\textwidth]{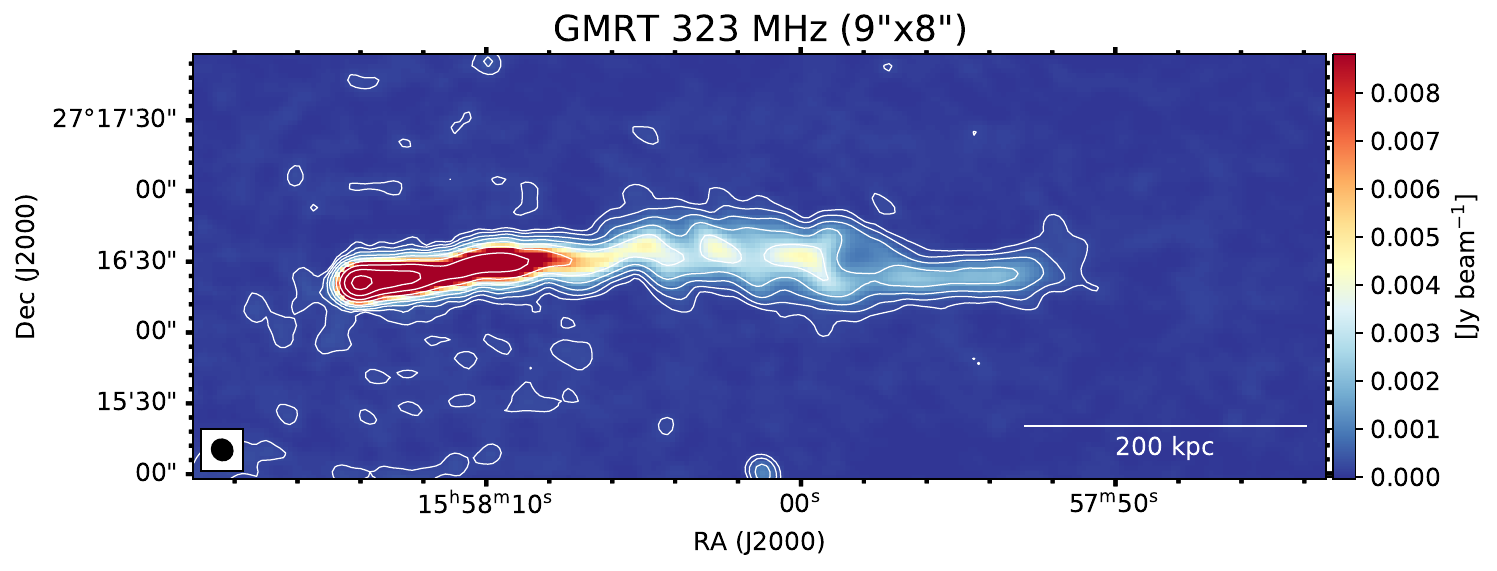}    \includegraphics[width=0.49\textwidth]{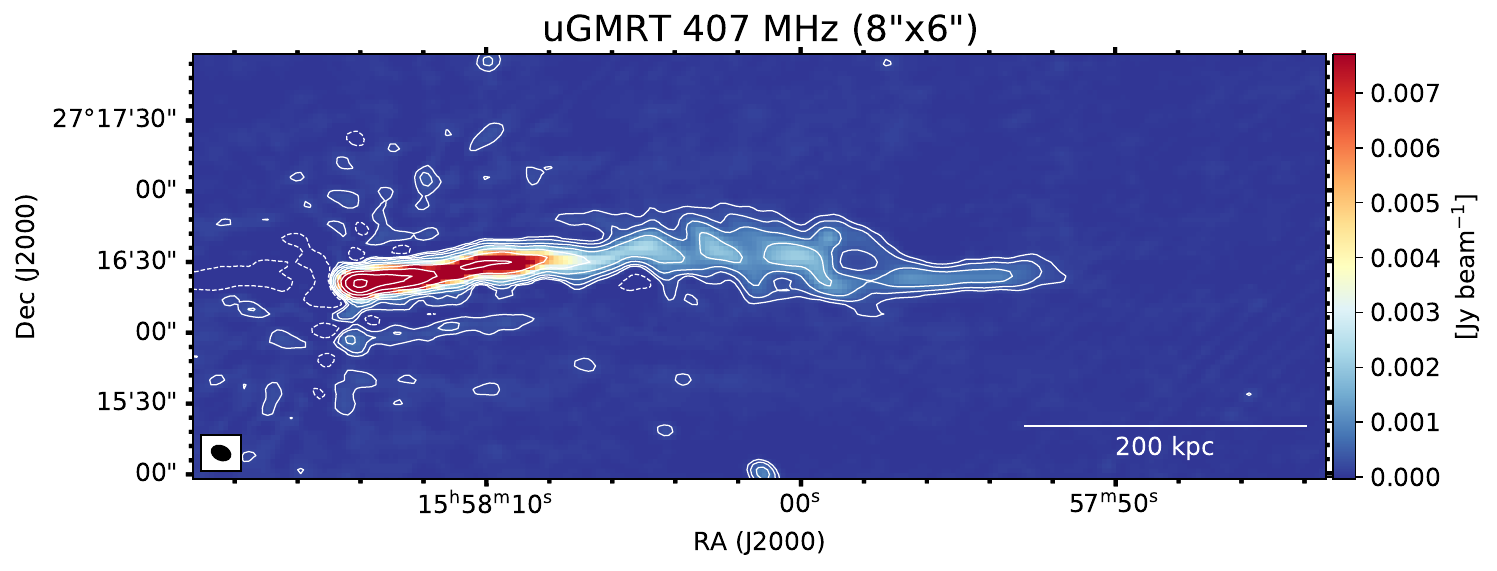}
	\includegraphics[width=0.49\textwidth]{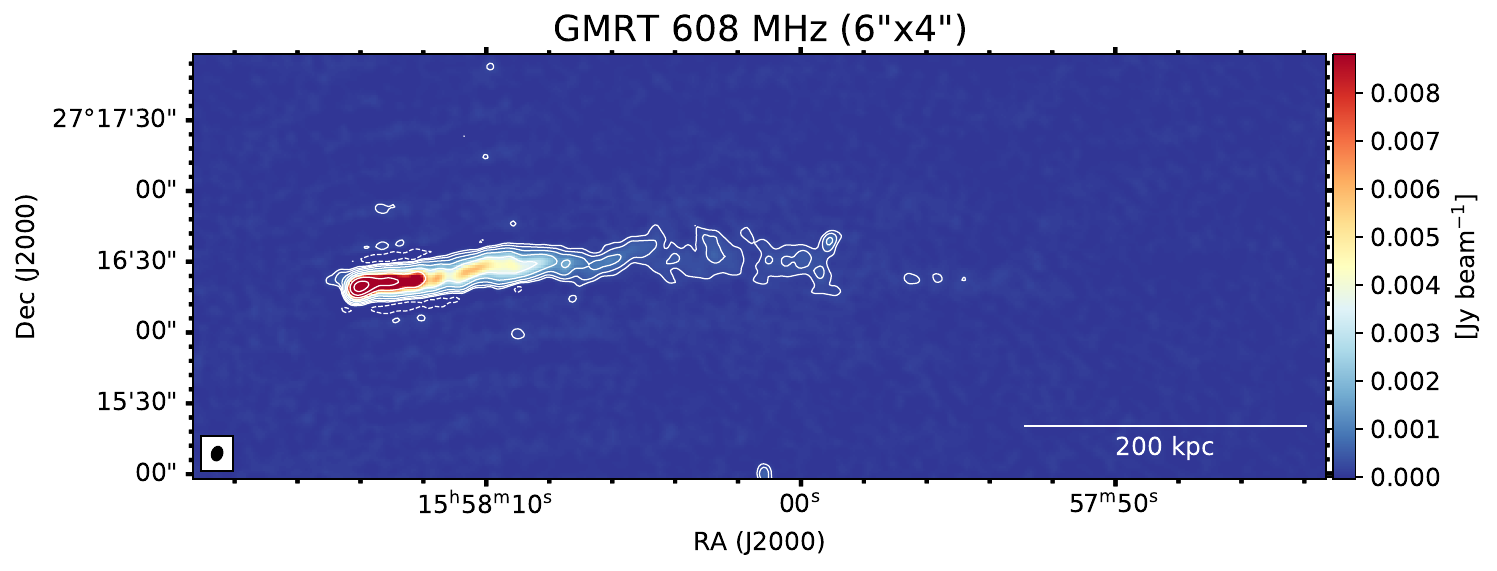}
 \includegraphics[width=0.49\textwidth]{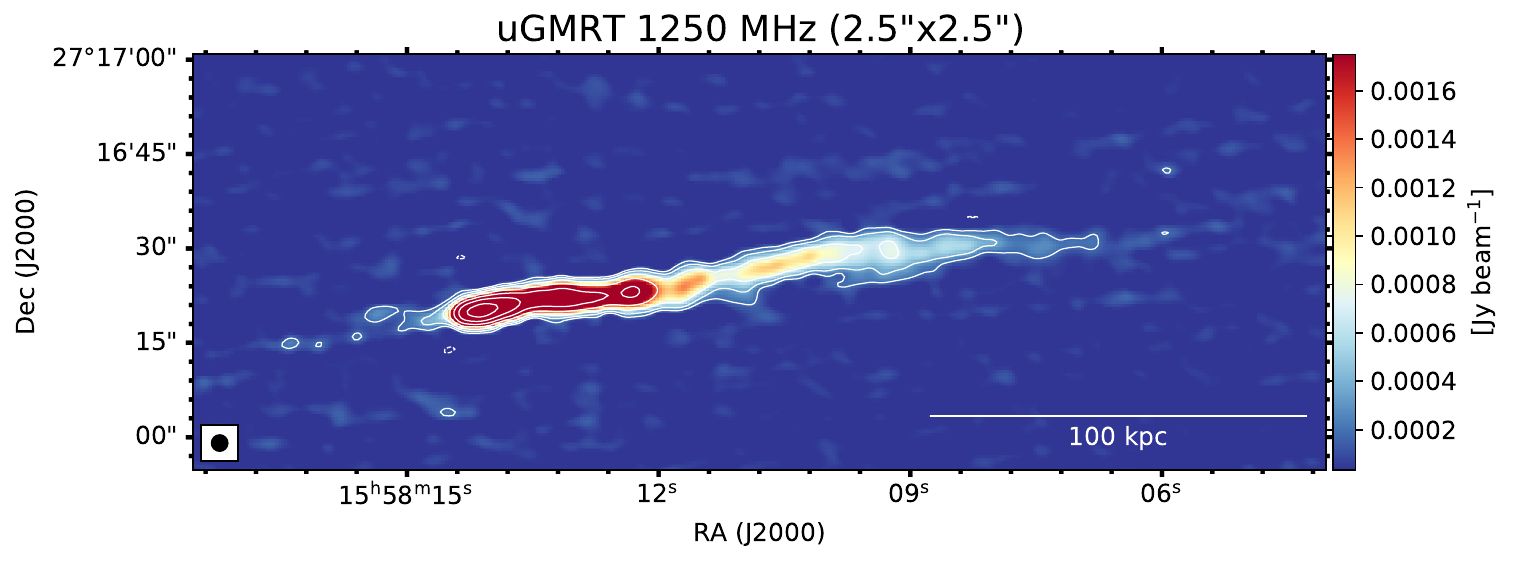}
  \includegraphics[width=0.49\textwidth]{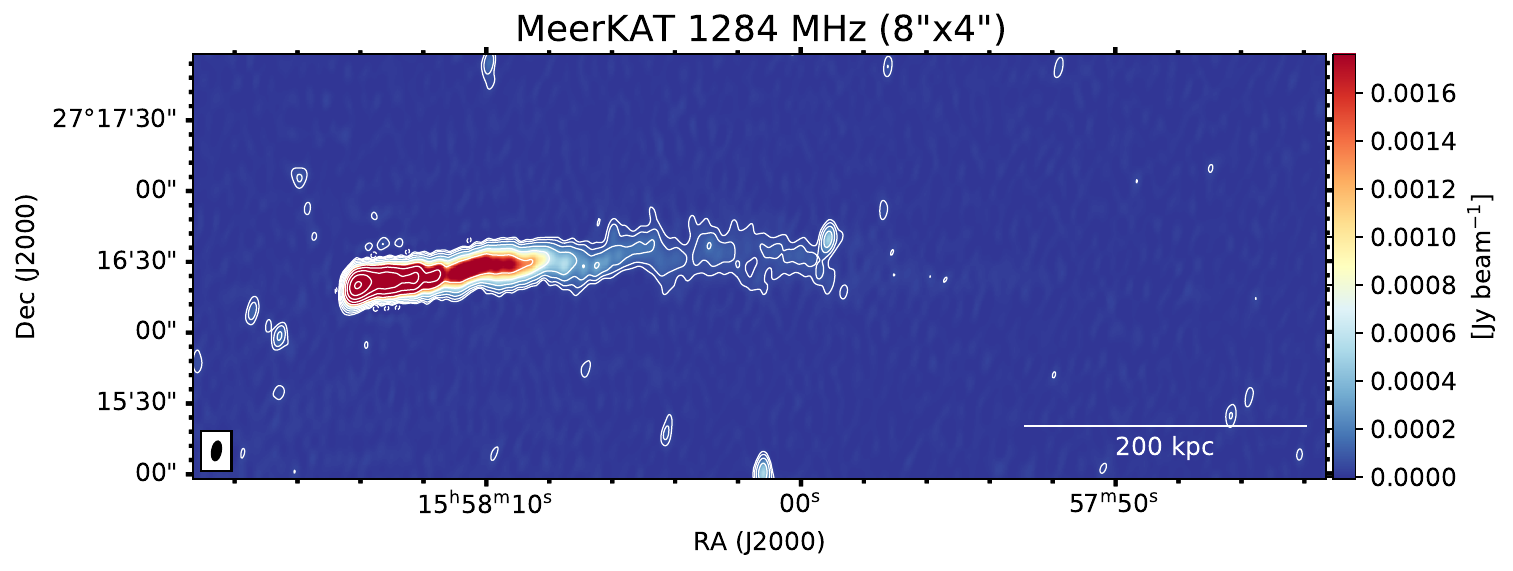}
 \includegraphics[width=0.49\textwidth]{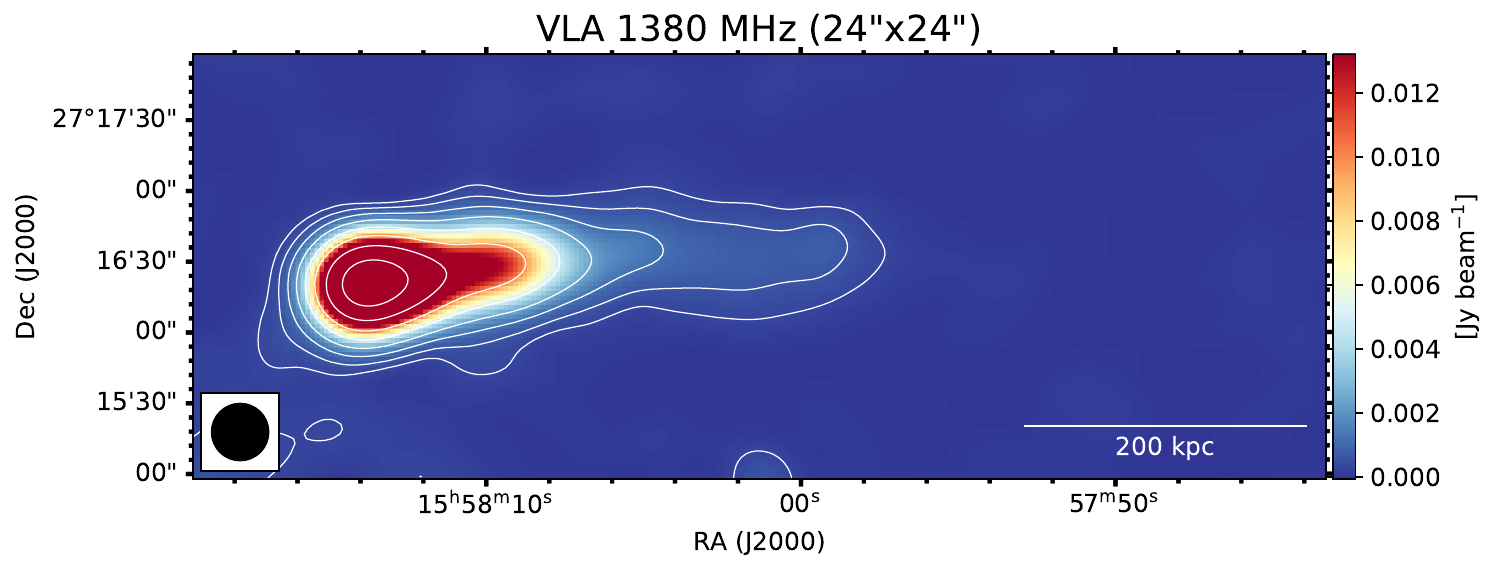}
 \includegraphics[width=0.49\textwidth]{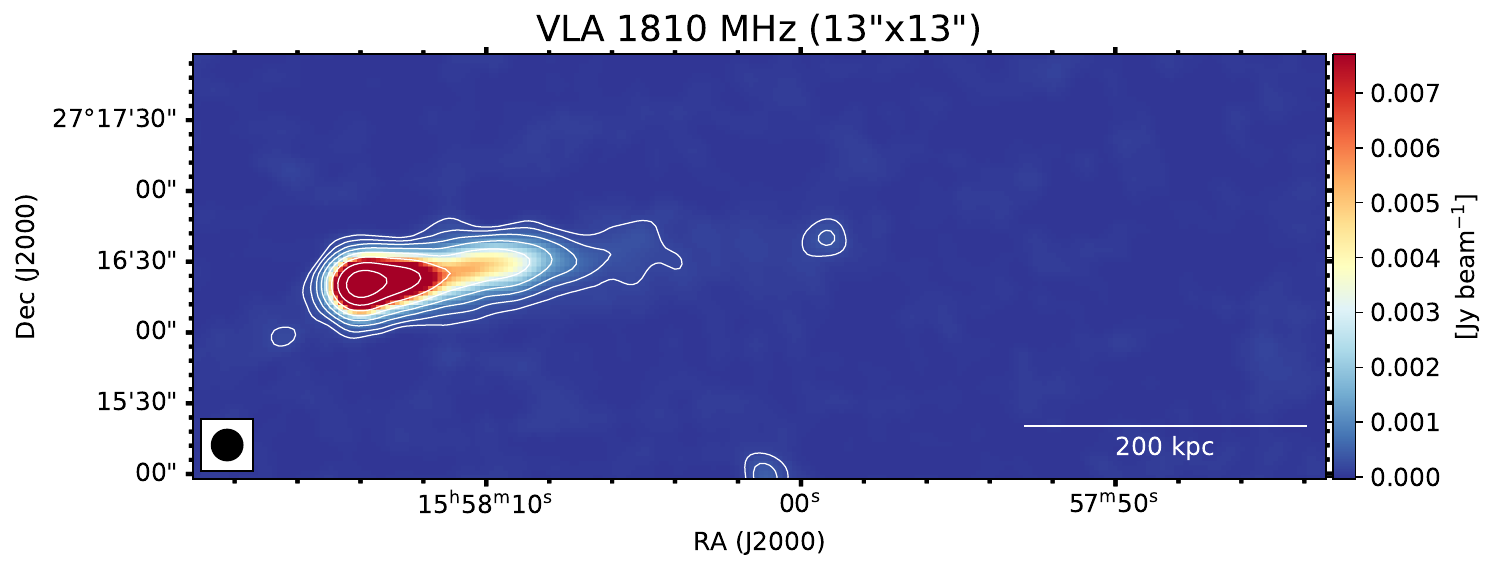}
	\smallskip
 \caption{Radio images of T1. Details on resolution and noise are reported in Table \ref{tab: image full res}. The overlaid contour levels are $[\pm5, \;10, \;20, ...]\times \sigma$. {\it From top left to bottom right}: LOFAR LBA at 50 MHz, LOFAR HBA at 143 MHz, GMRT at 234 MHz, GMRT at 323 MHz, uGMRT at 407 MHz, GMRT at 608 MHz, uGMRT at 1250 MHz (cropped to a smaller area for inspection purposes), MeerKAT at 1284 MHz, VLA at 1380 MHz, VLA at 1810 MHz.}
	\label{fig: radio images full res}
\end{figure*}

\begin{figure*}
	\centering
	\includegraphics[width=0.33\textwidth]{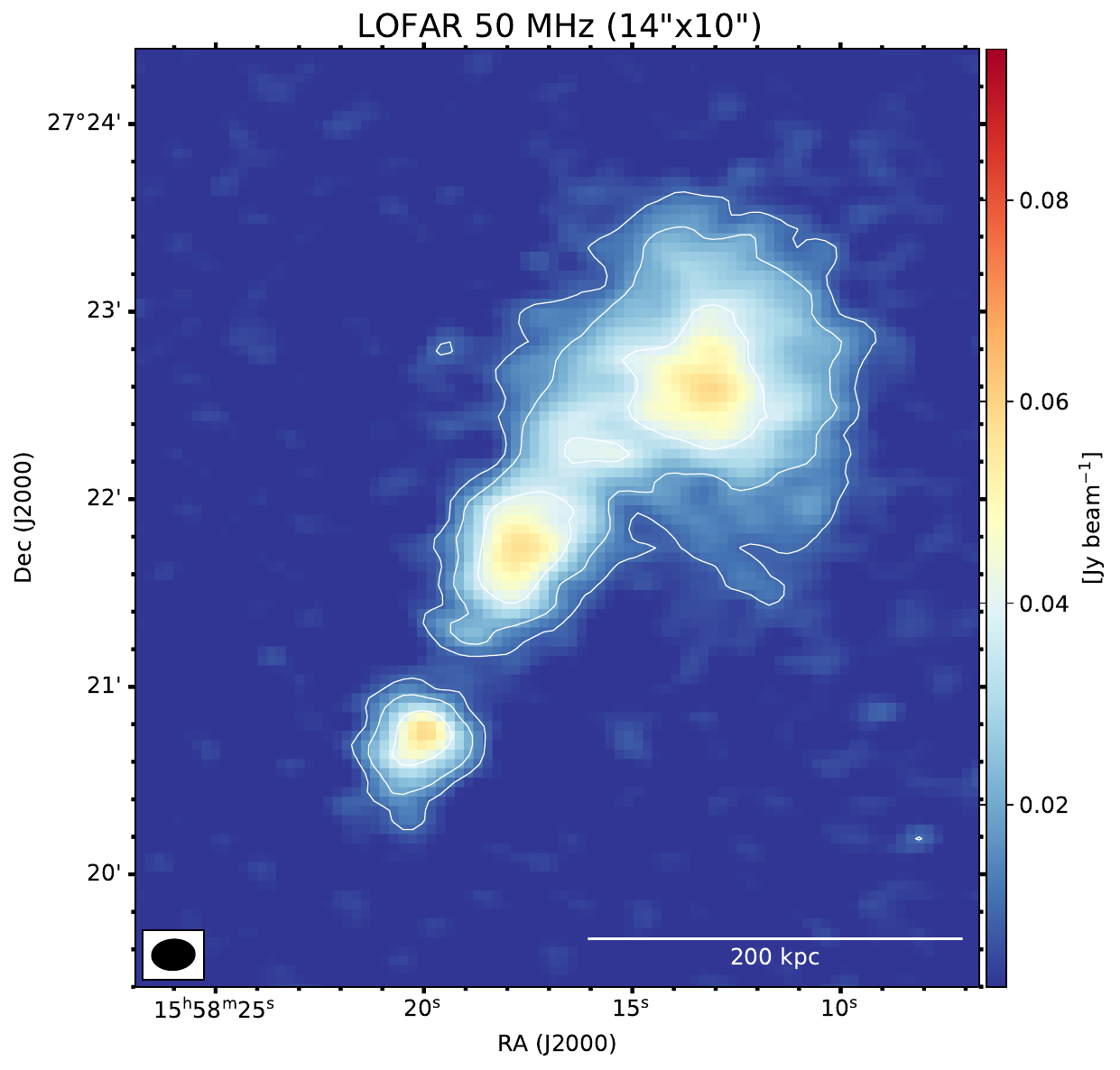}
	\includegraphics[width=0.33\textwidth]{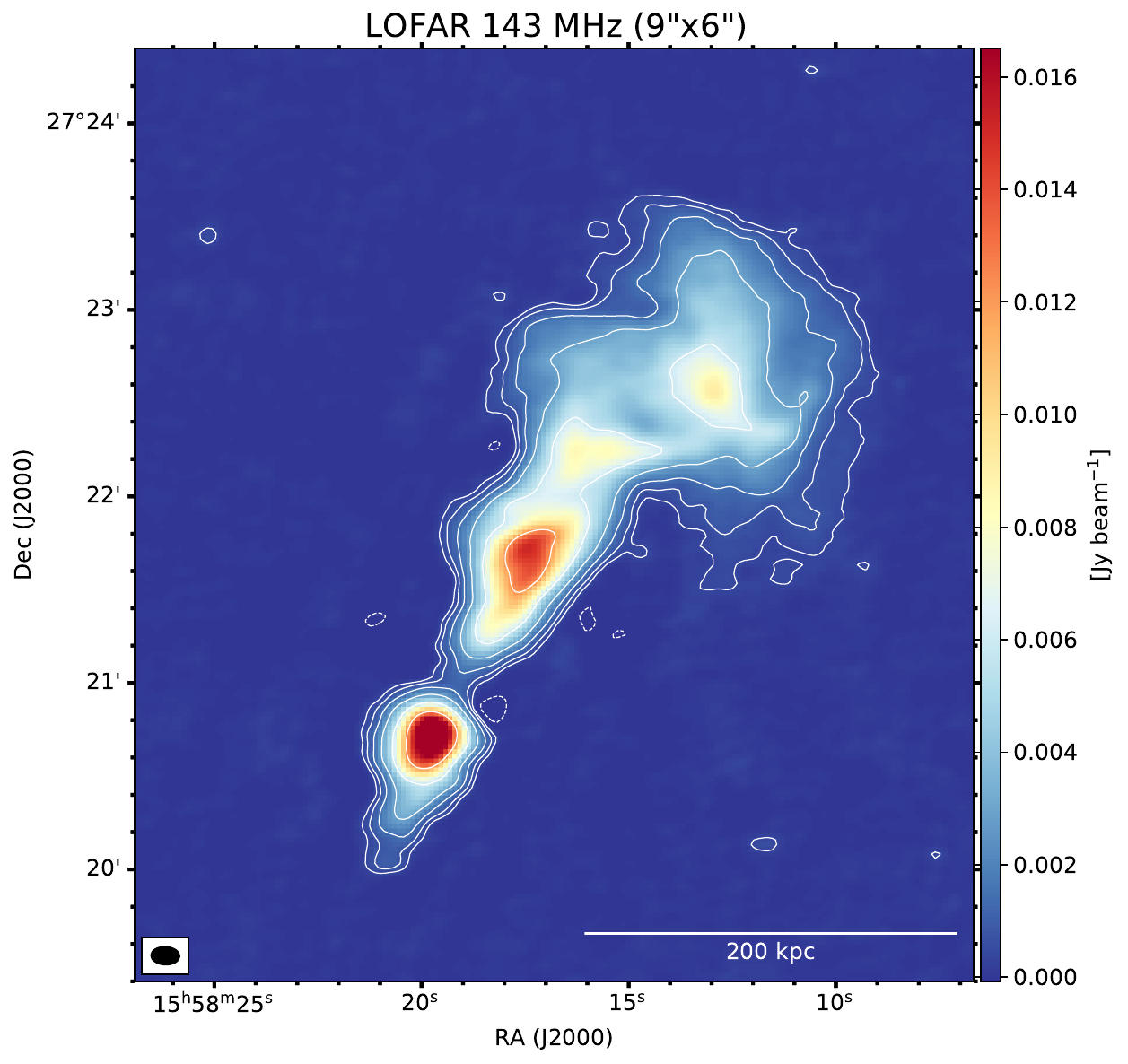}
	\includegraphics[width=0.33\textwidth]{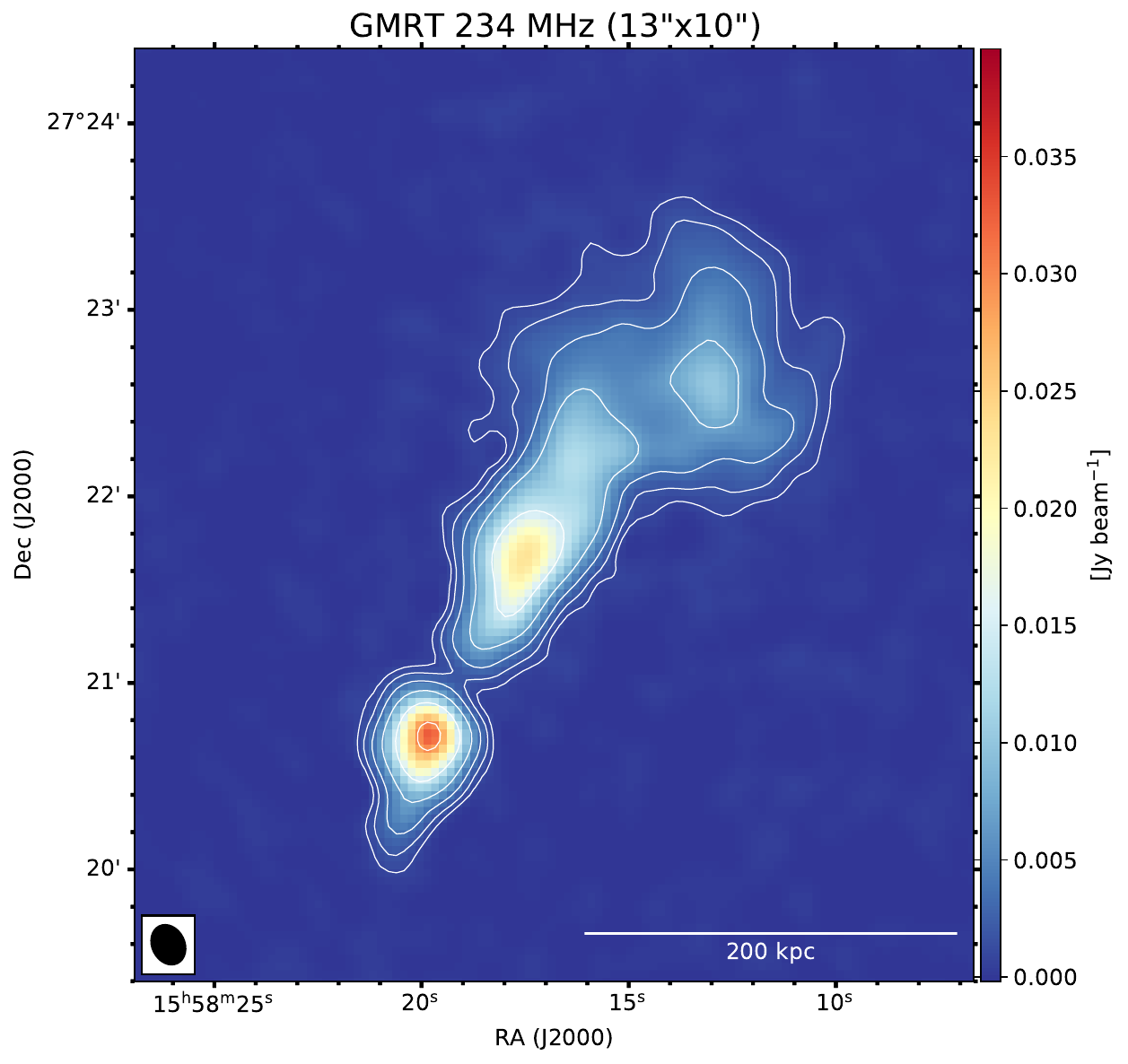}
	\includegraphics[width=0.33\textwidth]{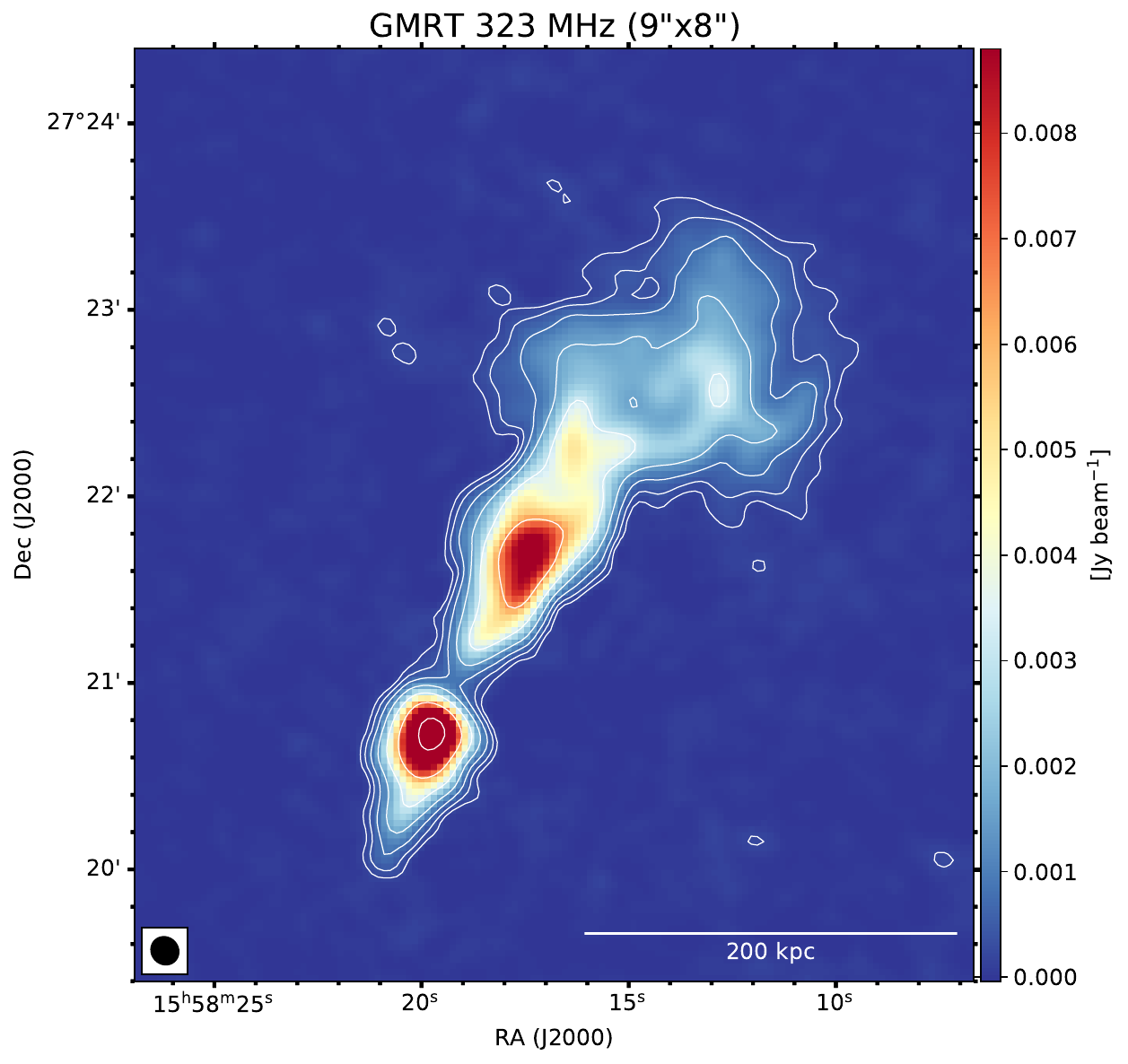}   
 \includegraphics[width=0.33\textwidth]{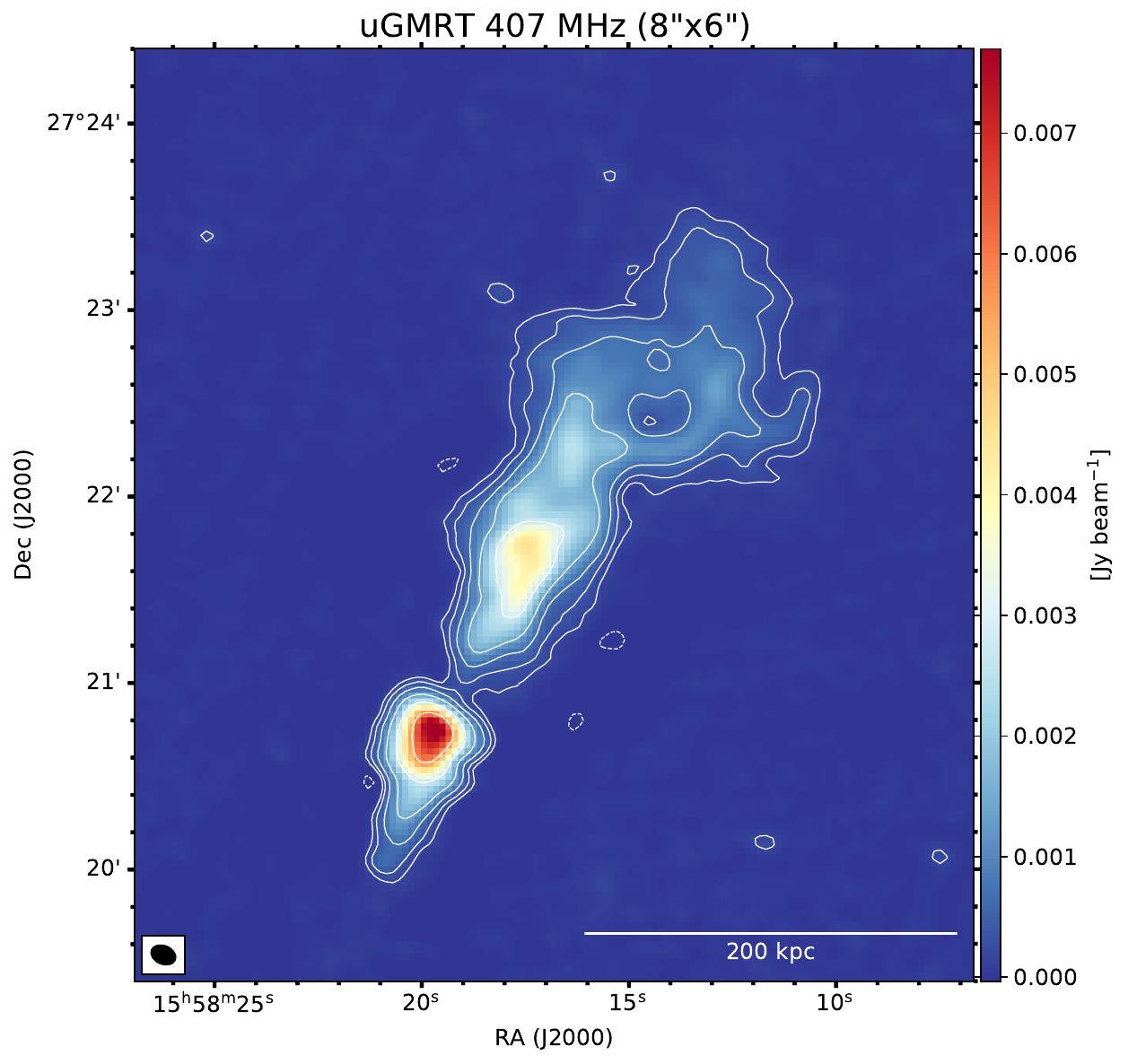}
	\includegraphics[width=0.33\textwidth]{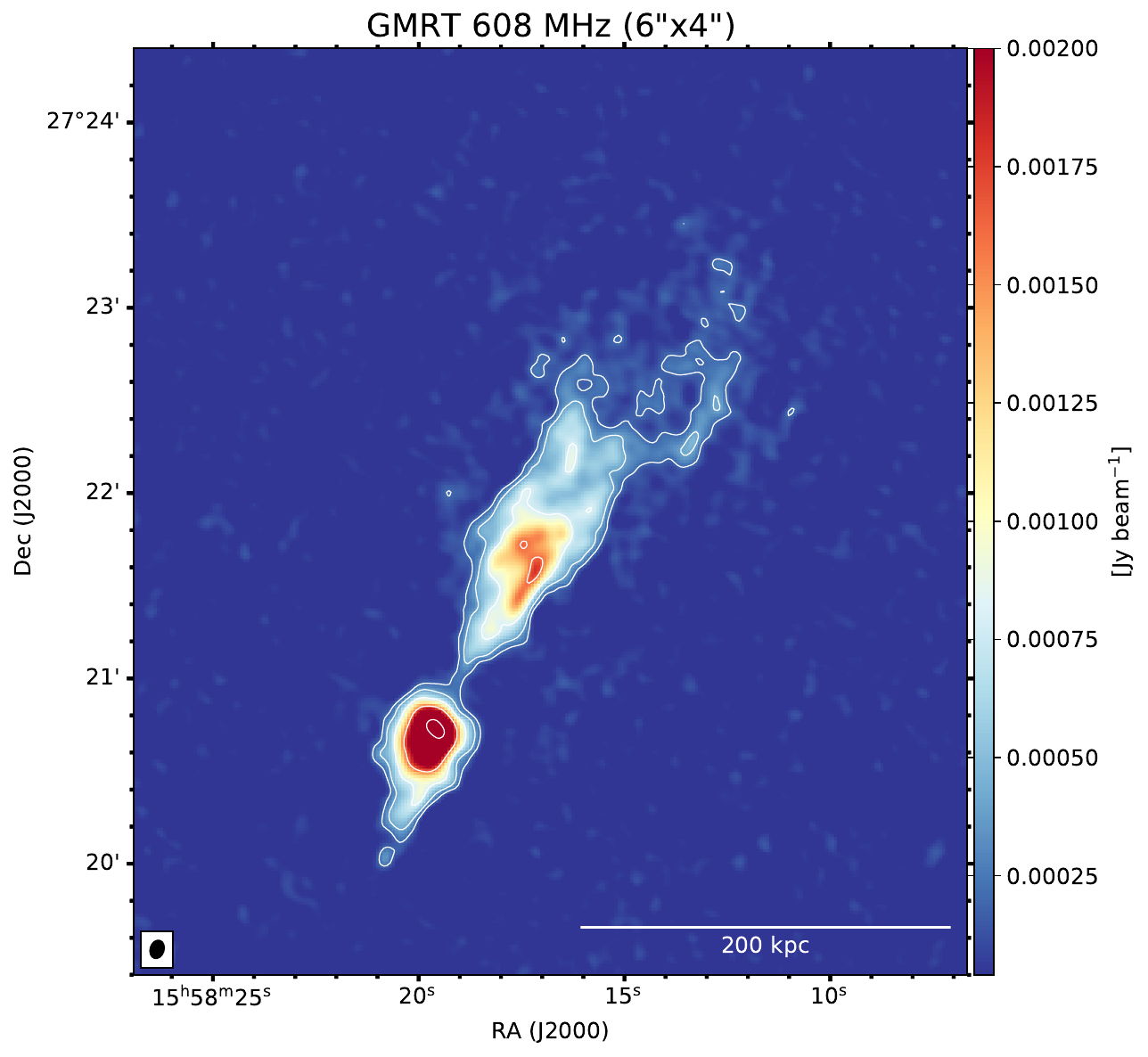}
 \includegraphics[width=0.33\textwidth]{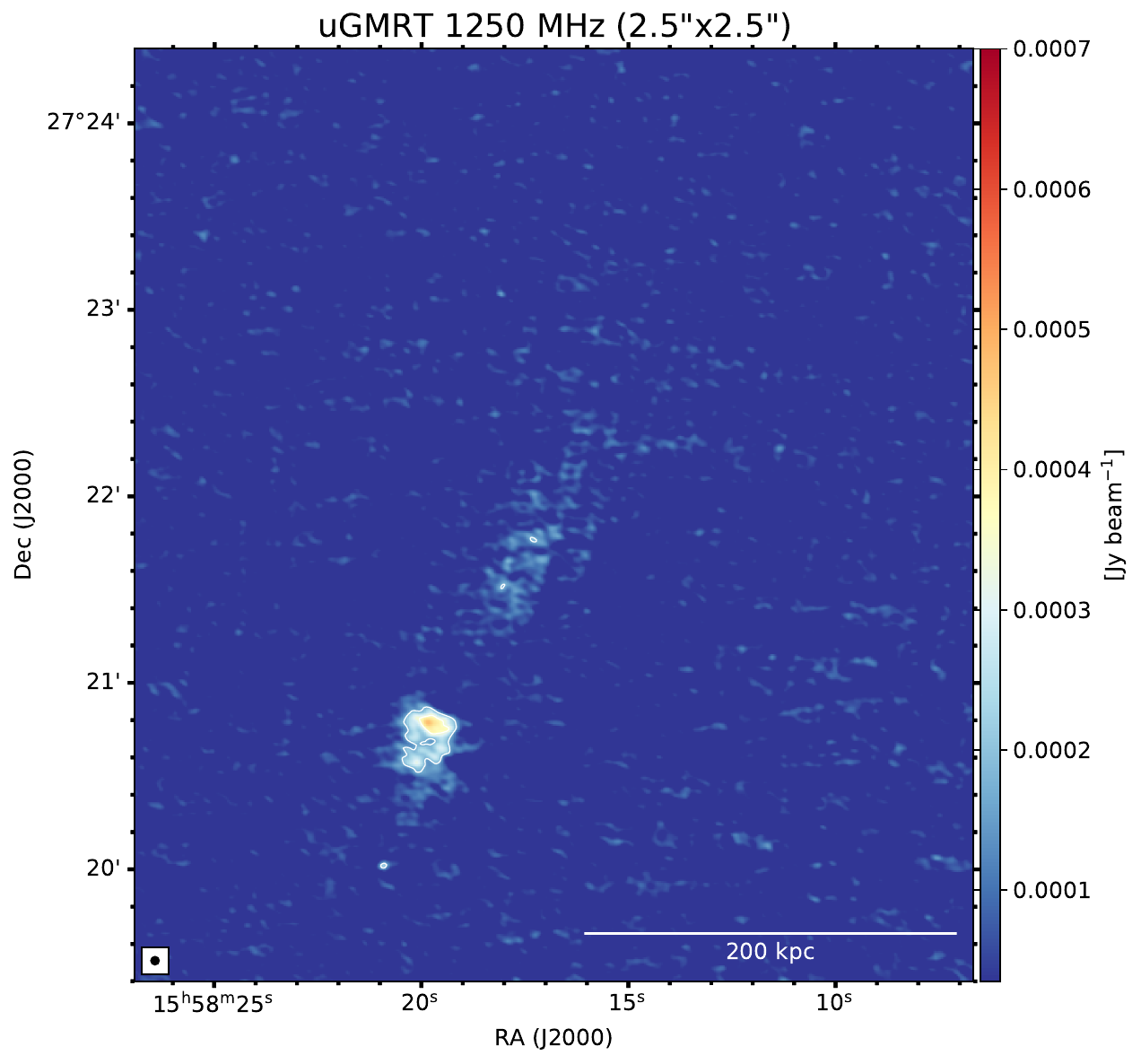}
  \includegraphics[width=0.33\textwidth]{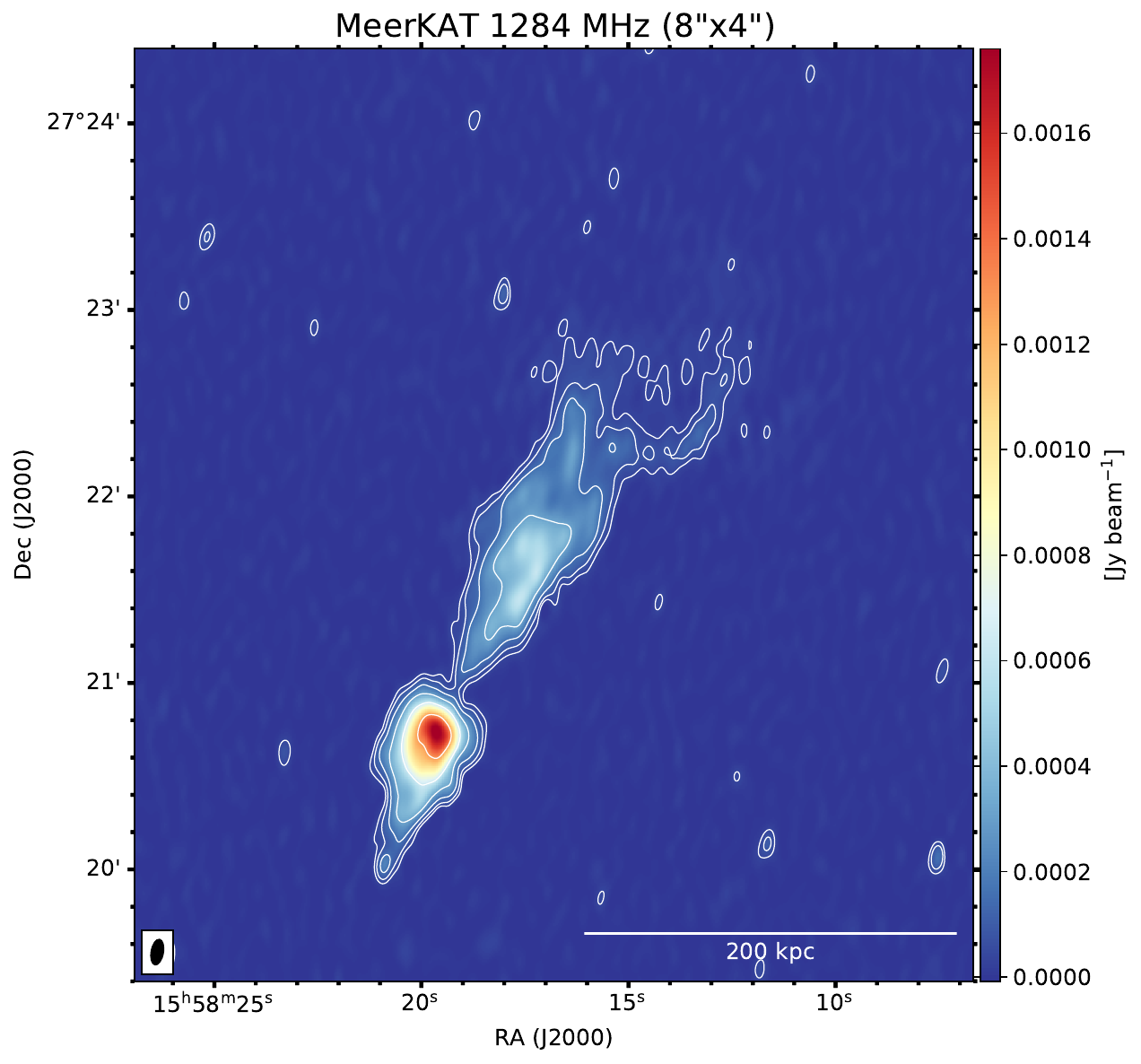}
 \includegraphics[width=0.33\textwidth]{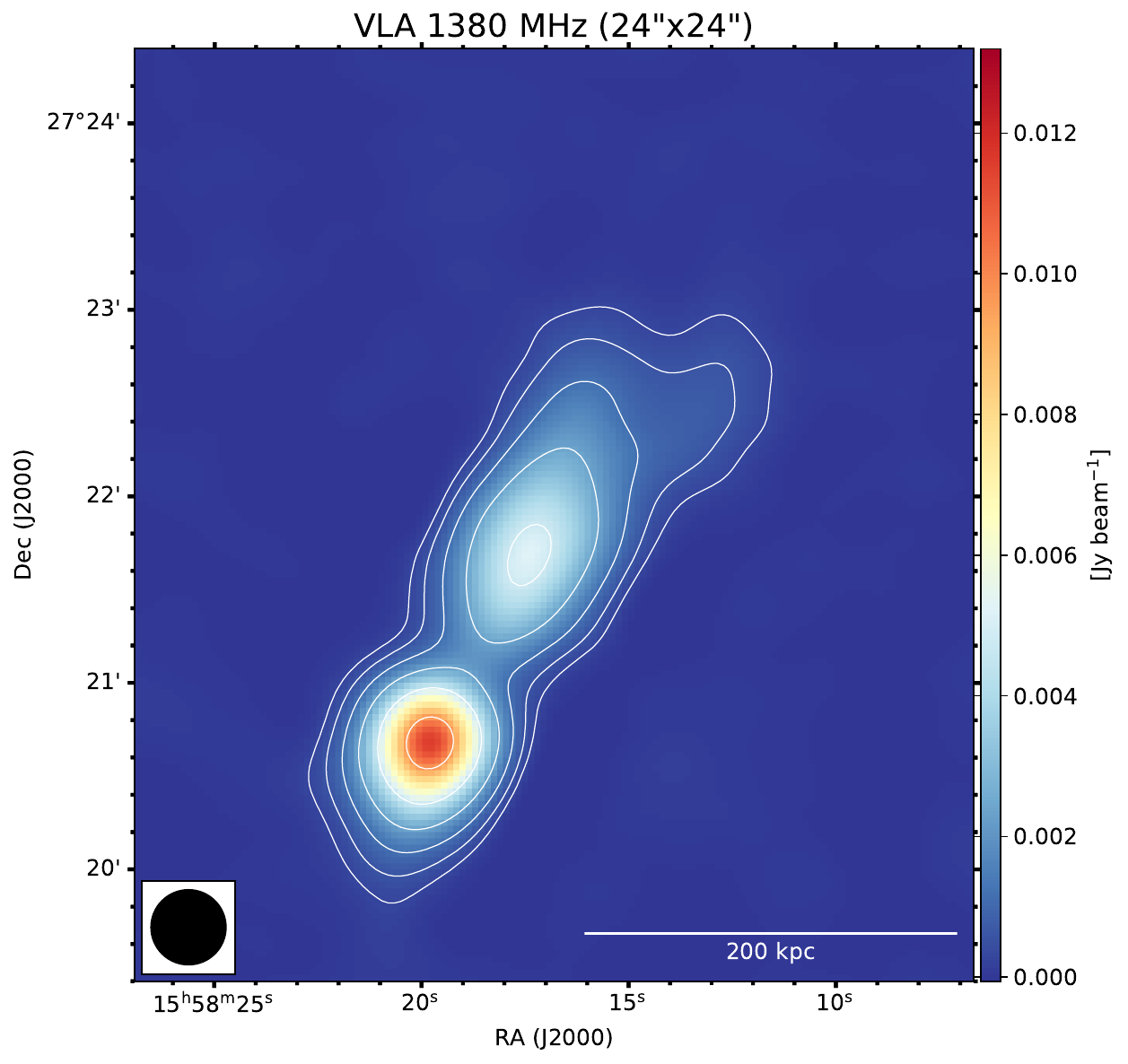}
 \includegraphics[width=0.33\textwidth]{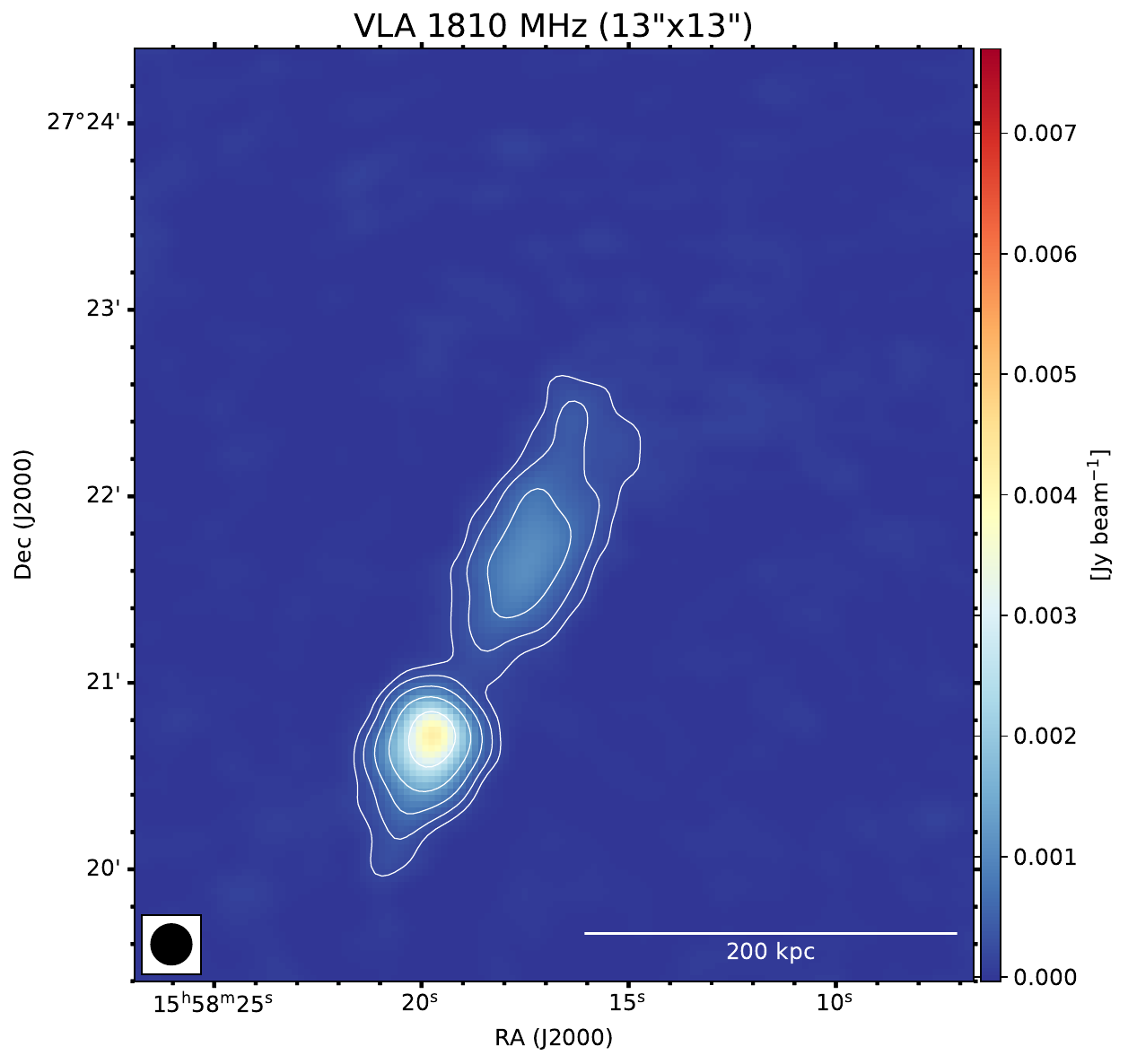}
	\smallskip
 \caption{Radio images of T2. Details on resolution and noise are reported in Table \ref{tab: image full res}. The overlaid contour levels are $[\pm5, \;10, \;20, ...]\times \sigma$. {\it From top left to bottom right}: LOFAR LBA at 50 MHz, LOFAR HBA at 143 MHz, GMRT at 234 MHz, GMRT at 323 MHz, uGMRT at 407 MHz, GMRT at 608 MHz, uGMRT at 1250 MHz, MeerKAT at 1284 MHz, VLA at 1380 MHz, VLA at 1810 MHz.  }
	\label{fig: radio images full res 2}
\end{figure*}

 \begin{table}
      \centering
   	\caption[]{Summary of the radio images shown in Fig. \ref{fig: radio images full res}. Cols. 1-5: Instrument, central frequency ($\nu$), restoring beam ($\theta$), beam position angle ($P.A.$) measured northeastwards, and noise ($\sigma$).}
   	\label{tab: image full res}
   	\begin{tabular}{ccccccc}
   	\hline
   	\noalign{\smallskip}
   	 Instrument & $\nu$ & $\theta$ & $P.A.$ & $\sigma$ \\
   	&  (MHz) & ($'' \; \times \; ''$) & (deg) & (${\rm \mu Jy \; beam^{-1}}$) \\
   	\noalign{\smallskip}
   	\hline
   	\noalign{\smallskip}
    LOFAR &	50  & $14\times10$ & 86 & 1900 \\
    LOFAR &	143 & $9\times6$ & 94 & 75 \\
    GMRT & 234 & $13\times10$ & 26 & 180 \\
    GMRT &   323 & $9\times8$ & 41 & 40 \\
    uGMRT & 407 & $8\times6$ & 67 & 35 \\
    GMRT & 608 & $6\times4$ & 163 & 40 \\
     uGMRT & 1250 & $2.5\times2.5$ & 0 & 35 \\
     MeerKAT & 1284 & $8\times4$ & 170 & 8 \\
     VLA & 1380 & $24 \times24$ &  0 & 60 \\
     VLA & 1810 & $13 \times13$ &  0 & 35 \\
   	\noalign{\smallskip}
   	\hline
   	\end{tabular}
   \end{table}  

Within the tailed radio galaxies T1 and T2 we define sub-regions that are discussed throughout this section and labelled in Fig. \ref{fig: substructures}, along with optical overlays from the Panoramic Survey Telescope \& Rapid Response System (Pan-STARSS; \citealt{flewelling20PANSTARRS}). Radio images of T1 and T2 are presented in Figs. \ref{fig: radio images full res}, \ref{fig: radio images full res 2} at different frequencies and resolutions (see details in Table \ref{tab: image full res}).

The host of T1 is located at a projected distance of $\sim 270$ kpc from the brightest cluster galaxy. Starting from the head, T1 extends from east towards west for a total projected length of $\sim 5.5'$ (at 143 MHz), corresponding to $\sim 550$ kpc at the cluster redshift. The sub-region T1-A, which corresponds to the initial $\sim 200$ kpc, is the brightest part of the source and is well imaged at all frequencies. In this region, the width of tail is $\sim 35$ kpc, thus suggesting that the whole structure (made by the two merged jets into the tail) is highly collimated. For the subsequent $\sim 150$ kpc (T1-B), the tail doubles its width up to $\sim 70$ kpc and shows clear brightness fluctuations, which we will refer to as wiggles due to their oscillating pattern. We note the presence of a galaxy (${\rm RA_{J2000} =239.5267, \;  DEC_{J2000} = 27.2742}, \; z=0.08684$) at the transition from T1-A and T1-B (Fig. \ref{fig: substructures}). At the end of sub-region T1-B, a bright compact component is visible at higher frequencies (its location is indicated by the purple circle in Fig. \ref{fig: substructures}), which is likely a compact radio source seen in projection. For the last $\sim 200$ kpc (T1-C), the tail follows a straight path. Interestingly, additional emission (T1-D) is well detected at 143 MHz (and partly visible at 50 and 323 MHz in lower resolution images). A thin, $\sim 50$ kpc long filament connects T1-C with an arc-shaped structure extending for $\sim 50 \; {\rm kpc \; and} \; 200 \; {\rm kpc}$ along east-west and north-south, respectively. If the arc were the termination of the tail, the total length of T1 would be $\sim 650$ kpc (in lower resolution images, more emission from T1-D is recovered, reaching $\sim 700$ kpc in total). We aim to shed light on this feature with a spectral analysis in Sect. \ref{sect: spectral properties 2}. 

The morphology of T2, which extends along the SE-NW axis for $\sim 400$ kpc, is more peculiar than classical tailed galaxies. The optical counterpart of T2 is located at a projected distance of $\sim 600$ kpc from the brightest cluster galaxy, and hosts a weak radio core (T2-D; see also Fig. \ref{fig: substructures}, bottom right panel) that is resolved from the rest of the source at 608 and 1250 MHz only (this is likely due to a favourable combination of higher resolution and lower sensitivity to extended components than other images). The core is the base of the first sub-region (T2-A), which has a light bulb shape of width $\sim 75$ kpc and length $\sim 100$ kpc. As highlighted by our high resolution ($2.5''$) 1250 MHz data, the light bulb ends with a sharp edge (see also Fig. \ref{fig: substructures}). The global morphology of T2-A may be presumably due to the backward bending of the jets by the ram pressure, but these are not resolved by any of our images.  A second sub-region (T2-B) is defined by a choking, that is the abrupt shrinking of the width of the tail and a drop in the radio surface brightness. T2-B extends for $\sim 150$ kpc, has a fairly constant width of $\sim 75$ kpc, and exhibits a single bright spot. The last sub-region (T2-C) is defined by the spread of the tail into a diffuse and filamentary plume of length $\sim 150$ kpc, maximum width $\sim 200$ kpc, and non-uniform brightness. The plume is detected in our images only at low frequencies ($\nu < 407$ MHz), thus suggesting a very steep spectral index for this region. Interestingly, the western part of the plume is bent towards SW \citep[see also LBA images at lower resolution in][]{bruno23b}.

\subsection{Radio surface brightness fluctuations}
\label{sect: Radio surface brightness fluctuations}

\begin{figure*}
	\centering
 \includegraphics[width=0.7\textwidth]{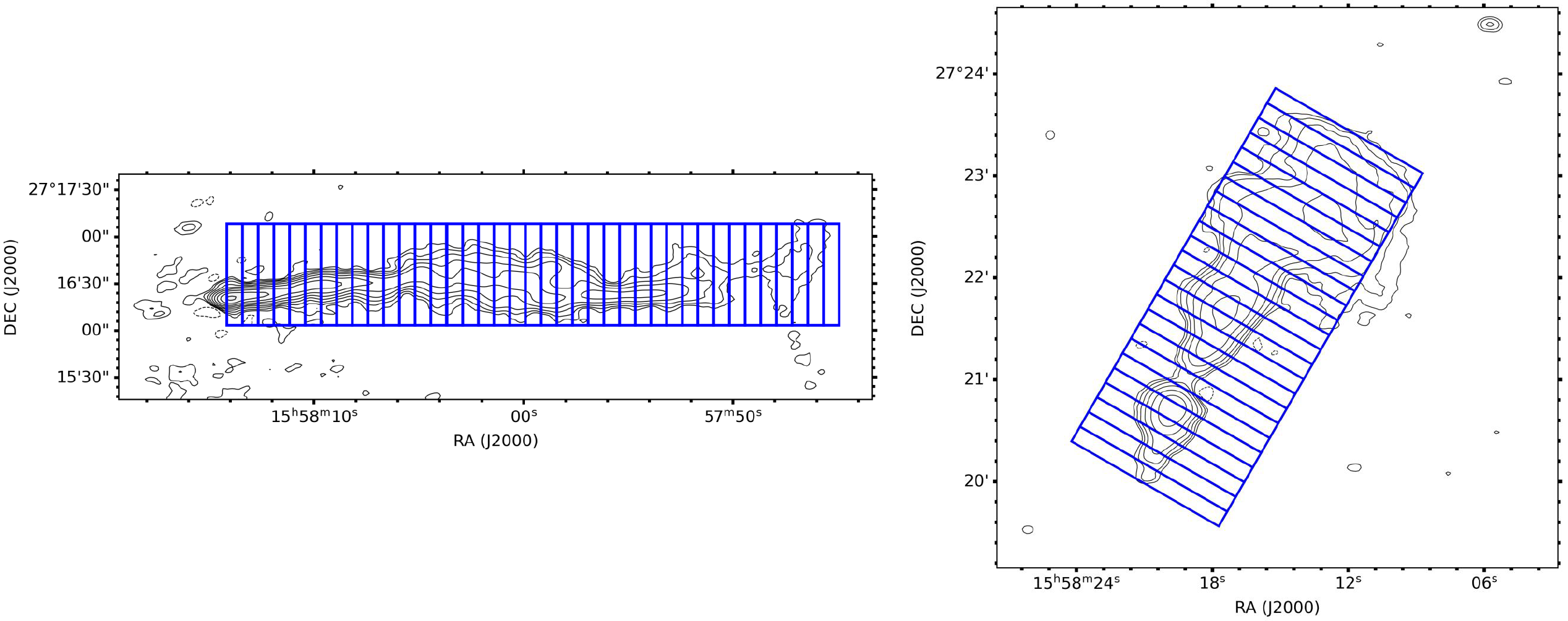} \\
	\includegraphics[width=0.49\textwidth]{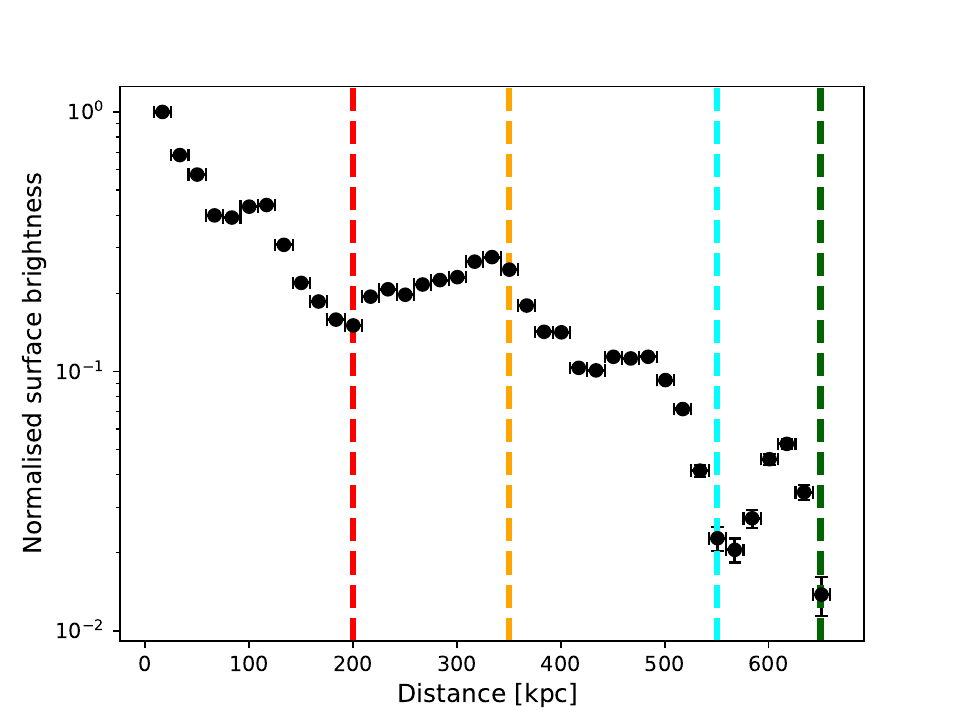}
 	\includegraphics[width=0.49\textwidth]{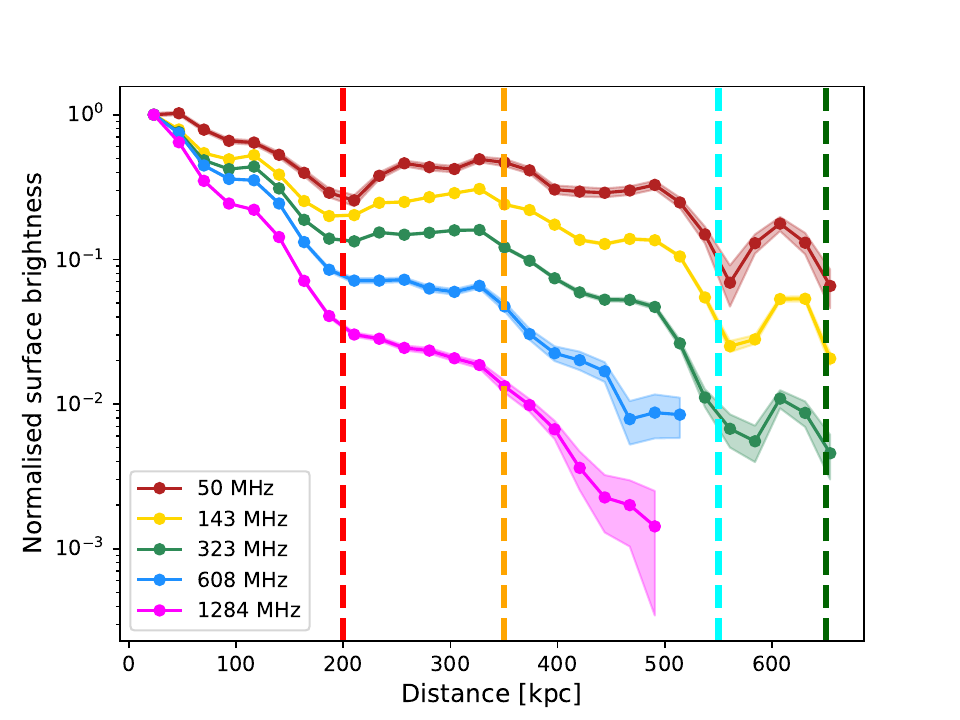}
	\includegraphics[width=0.49\textwidth]{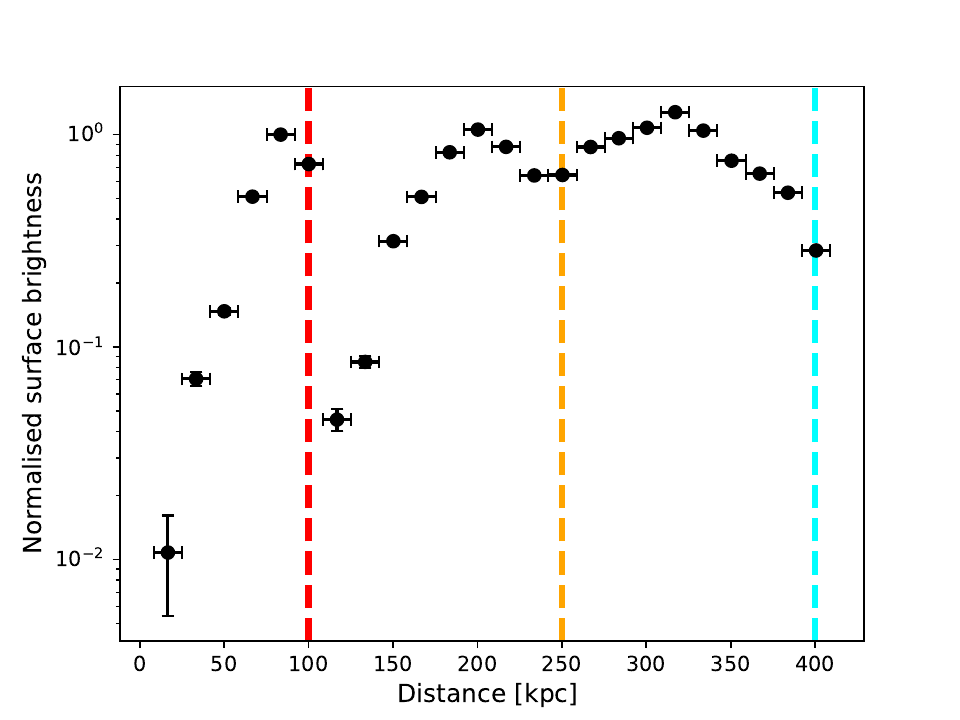}
 	\includegraphics[width=0.49\textwidth]{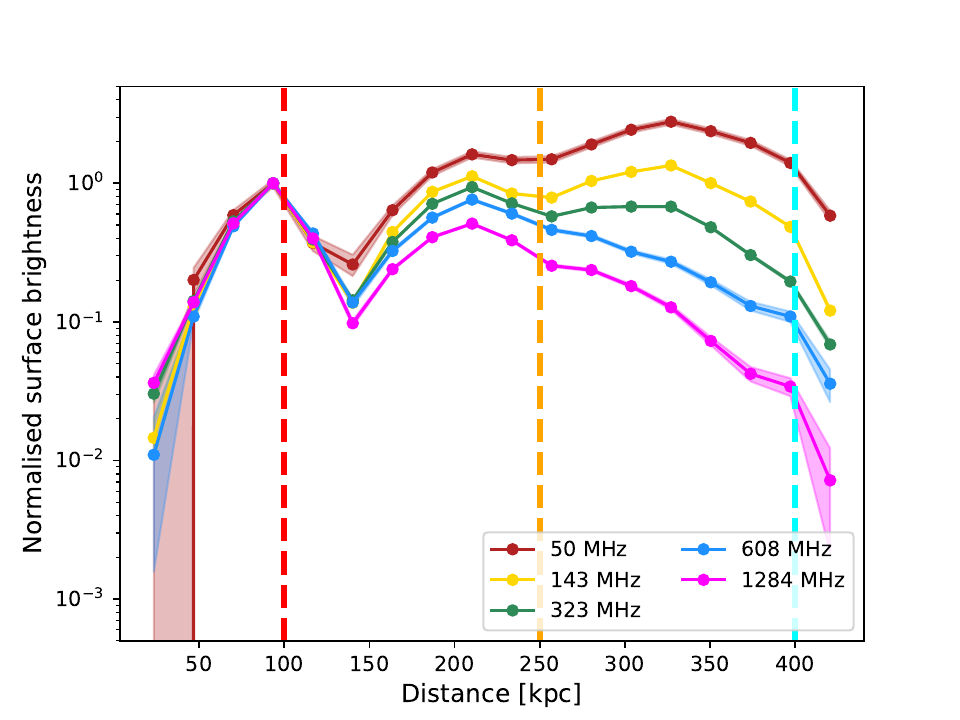}

	\smallskip
	
	\caption{Radio surface brightness profiles of T1 and T2 normalised to the peak value within T1-A and T2-A, respectively. \textit{Left panels}: Profiles of T1 (middle) and T2 (bottom) at 143 MHz (resolution $9''\times 6''$). The sampling boxes are shown in the top panels and have a width of $10''$. \textit{Right panels}: Comparison of profiles at 50 (brown), 143 (yellow), 323 (green), 608 (light blue), and 1284 (magenta) MHz. The resolution of the images and the width of the sampling boxes (not shown) are $14''$. In middle and bottom panels, the vertical lines indicate the boundaries of sub-regions as in Fig. \ref{fig: substructures}.} 
	\label{fig: radio-x profile}
\end{figure*}

To investigate the brightness fluctuations within each sub-region, we computed the surface brightness profiles of T1 and T2. In the left panels of Fig. \ref{fig: radio-x profile} we report the profiles (black data points) measured from the LOFAR HBA image (Figs. \ref{fig: radio images full res}, \ref{fig: radio images full res 2}) in boxes of width $10''$. Each data point of T1 and T2 is normalised by the peak value within T1-A and T2-A, respectively. Analogously, in the right panels of Fig. \ref{fig: radio-x profile} we compare the normalised profiles at 50 (brown), 143 (yellow), 323 (green), 608 (light blue), and 1284 (magenta) MHz, as measured from images convolved at the same resolution of $14''$. The surface brightness profiles are reflective of the sub-regions that we defined in Sect. \ref{sect:Radio morphology} by visual inspection, and reveal interesting features discussed below.

The absolute peak value of T1 is coincident with the radio core (first sampling box). In T1-A, the brightness at 143 MHz rapidly decreases (down to a factor $\sim 10$) with the distance from the core. A discontinuity is visible at $\sim 100$ kpc, where the tail first deviates from its straight path and is slightly compressed. In T1-B (the region of the wiggles), the brightness is enhanced instead of declining, but such growth becomes progressively shallower with the increasing frequency. In T1-C, the declining trend resumes, but we also report the presence of a moderate peak at $\sim 500$ kpc at low-$\nu$. The last peak in T1-D is associated with the arc, which is detected only at 50, 143, and 323 MHz.

The surface brightness profile of T2 is largely unusual for tailed galaxies. The core of T2 is weak (see also Sect. \ref{sect: spectral properties 1}) and is not coincident with the absolute peak of emission. Overall, the normalised profile exhibits three peaks of similar relative amplitude, corresponding to the light bulb (T2-A), the main body of the tail (T2-B), and the plume (T2-C). The first peak is associated with the bright spot at the edge of T2-A. The choke that separates T2-A and T2-B is identified as a sharp discontinuity in our profiles at a distance of $\sim 100-150$ kpc. Within T2-B, the surface brightness increases with the distance up to $\sim 200$ kpc. Finally, the emission of the plume produces the last peak, which becomes progressively shallower with the increasing frequency.

\subsection{Integrated radio spectra}
\label{sect: spectral properties 1}

\begin{figure*}
	\centering
 \includegraphics[width=0.7\textwidth]{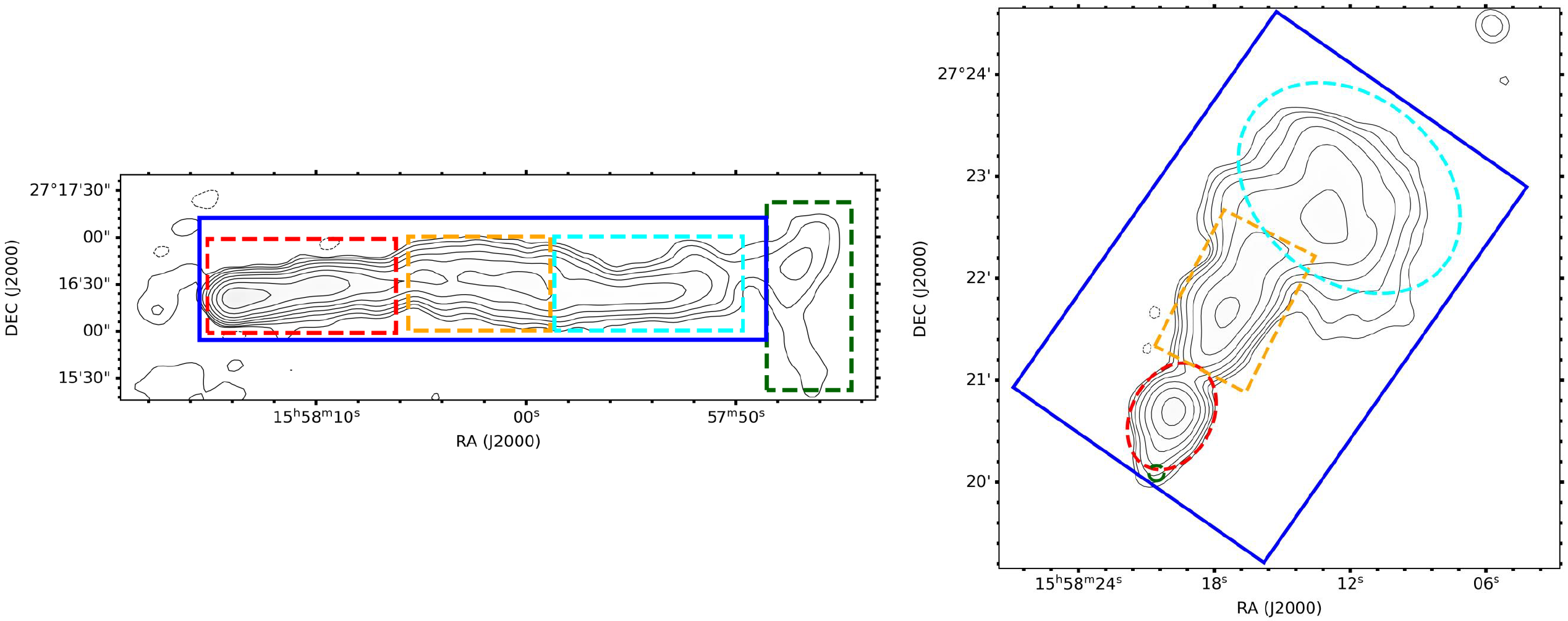} \\
 \includegraphics[width=0.49\textwidth]{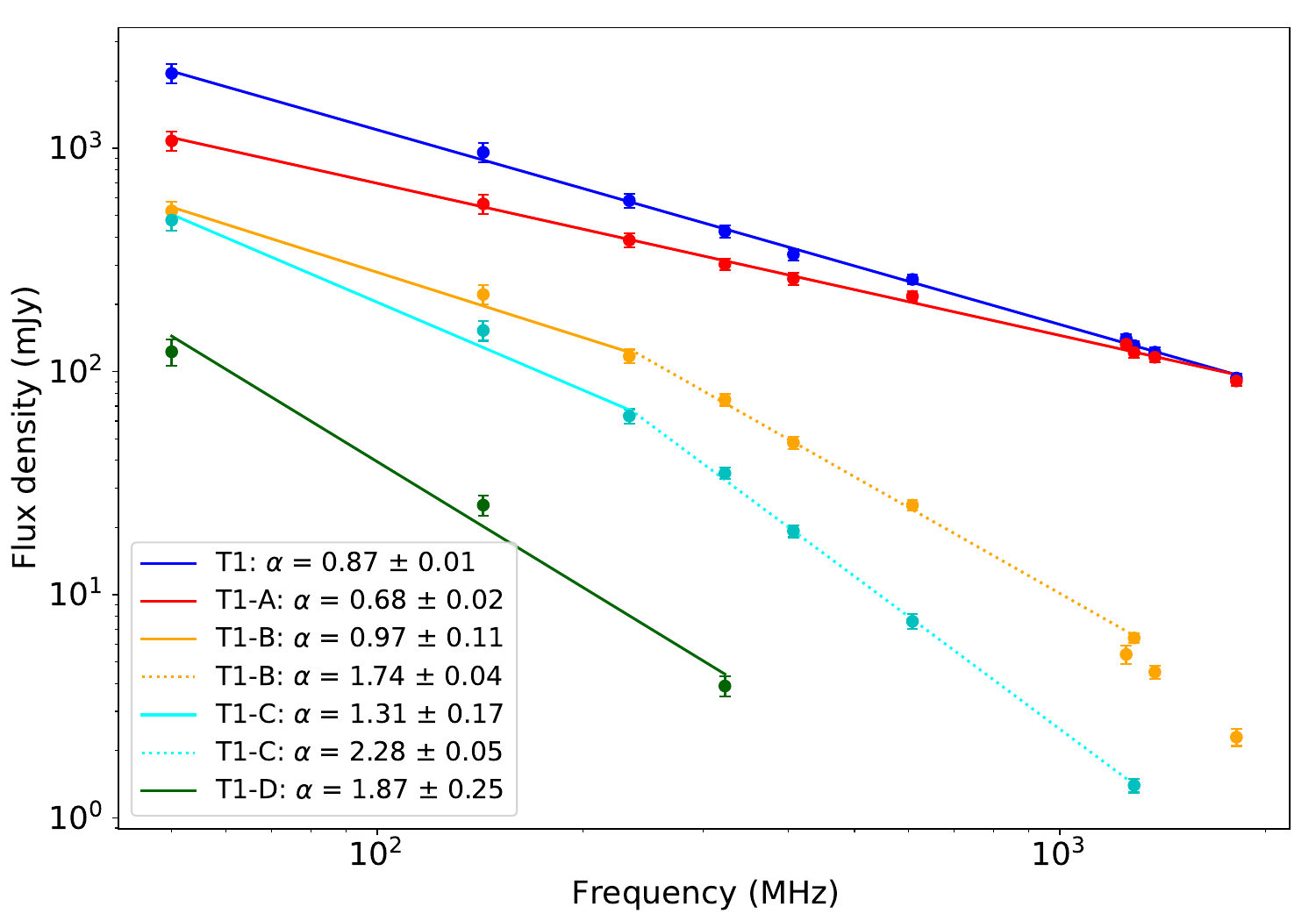}
 \includegraphics[width=0.49\textwidth]{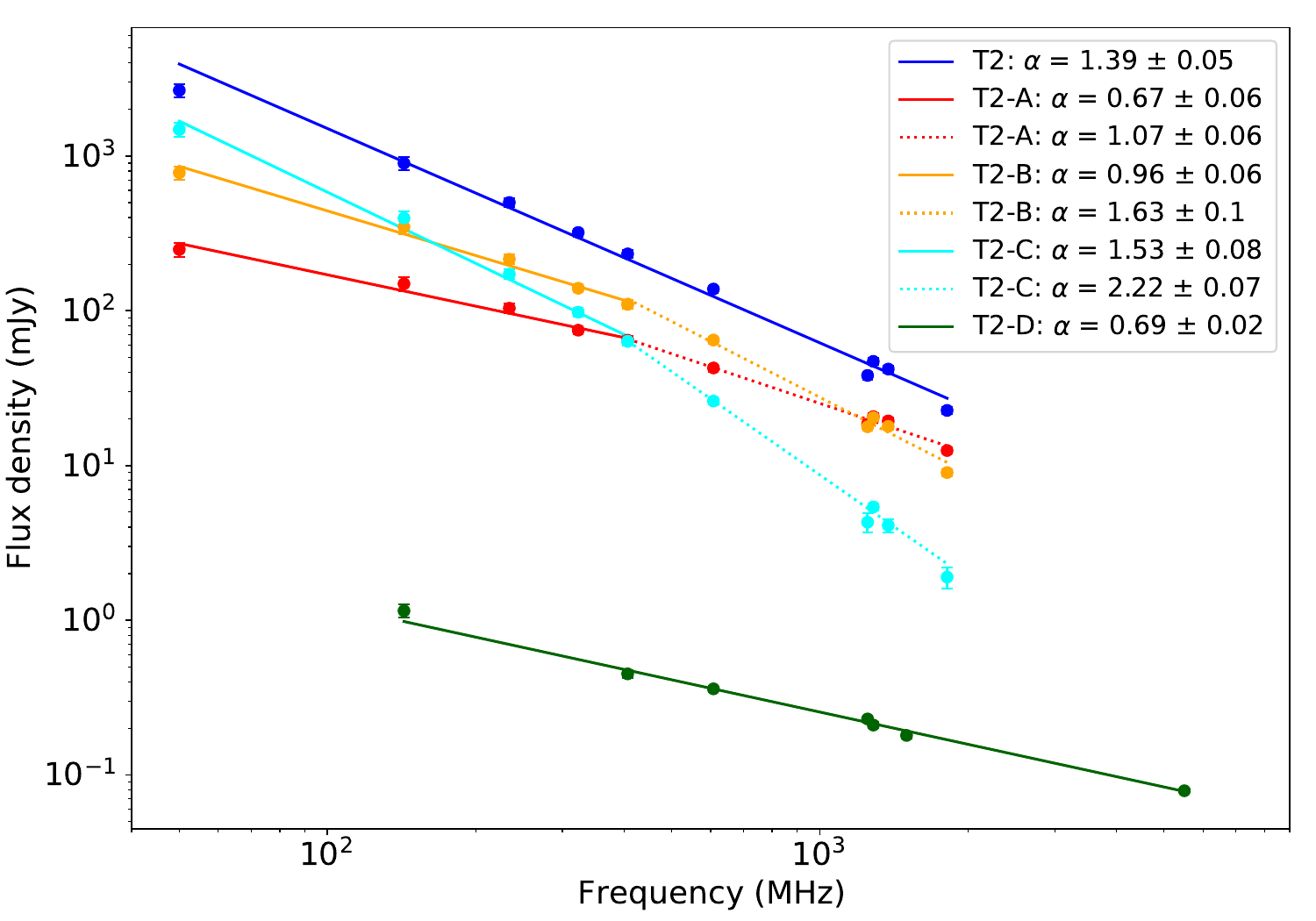}

	\smallskip
	
	\caption{Radio spectra of T1 and T2. Left and right panels refer to T1 and T2, respectively. Data points are the flux densities reported in Table \ref{tab: radio flux}, measured from the regions (including the small green circle used for the core of T2) shown in the top panels following the same colour-code. Flux densities were fitted with single (solid lines) or double (solid plus dotted lines) power-laws, as summarised in Table \ref{tab: fitspix}.} 
	\label{fig: spettro}
\end{figure*}

\begin{table*}
\centering
	\caption[]{Flux densities of T1, T2, and corresponding sub-regions as measured from images at $14''$ resolution (see Fig. \ref{fig: spettro}). The flux density of T2-D was measured from higher-resolution ($<5''$) images. }
	\label{tab: radio flux} 
 \resizebox{\textwidth}{!}{
 \begin{tabular}{ccccccccccc}
	\hline
	\noalign{\smallskip}
	$\nu$ & $S^{\rm T1}$ & $S^{\rm T1-A}$ & $S^{\rm T1-B}$ & $S^{\rm T1-C}$ & $S^{\rm T1-D}$ & $S^{\rm T2}$ & $S^{\rm T2-A}$ & $S^{\rm T2-B}$ & $S^{\rm T2-C}$ &  $S^{\rm T2-D}$ \\  
	(MHz) & (mJy) & (mJy) & (mJy) & (mJy) &  (mJy) & (mJy) & (mJy) & (mJy)  & (mJy) & (mJy) \\  
    \hline
	\noalign{\smallskip}
	50 &  $2168.4 \pm 217.9$ & $1080.5 \pm 108.6$ & $524.7 \pm 53.3$ & $476.8 \pm 48.9$ & $122.6\pm 16.4$ & $2659.0  \pm 267.4$ &  $ 249.1\pm 25.7$ & $ 780.2 \pm 78.6 $ & $ 1484.8\pm 149.1 $ & - \\
    143 &  $959.7 \pm 96.0$ & $563.7 \pm 56.4$ & $221.4 \pm 22.1$ & $152.7 \pm 15.3$ & $25.2\pm 2.6$ & $899.0  \pm 89.9$ &  $ 149.4 \pm 14.9$ & $ 347.9 \pm 34.8 $ &  $397.4 \pm 39.8$ & $  1.15\pm 0.12$\\
    234 &  $583.2 \pm 40.9$ & $387.7 \pm 27.2$ & $117.2 \pm 8.3$ & $63.2 \pm 4.6$ & - & $501.1  \pm 35.2$ &  $ 103.9\pm 7.3$ & $ 215.4 \pm 15.1$ & $ 172.5 \pm 12.2$ & - \\
    323 &  $424.7 \pm 25.5$ & $302.5 \pm 18.2$ & $74.8 \pm 4.5$ & $35.0 \pm 2.1$ & $3.9\pm 0.4$ & $320.6  \pm 19.3$  &  $ 74.9\pm 4.5$ & $ 139.8 \pm 8.4$  & $ 97.9\pm 5.9$ &  - \\ 
    407 &  $334.7 \pm 20.1$ & $260.6 \pm 15.6$ & $48.0 \pm 2.9$ & $19.3 \pm 1.2$ & - & $233.2  \pm 14.0$ &  $ 64.4\pm 3.9$ & $ 110.1 \pm 6.6$ & $ 63.5 \pm 3.8$ & $  0.45 \pm 0.03$ \\
    608 &  $258.7 \pm 13.0$ & $217.3 \pm 10.9$  & $25.2 \pm 1.3$  & $7.6 \pm 0.6$ & - &  $138.2  \pm 7.0$ & $ 42.7\pm 2.2$ & $ 64.7 \pm 3.3$ & $ 26.1 \pm 1.4$  & $  0.36\pm 0.02$\\
    1250 &  $140.4 \pm 7.1$ & $132.5 \pm 6.6$ & $5.4 \pm 0.5$ & - & - & $38.2  \pm 2.2$ &  $ 18.5\pm 1.0$ & $ 17.8 \pm 1.0 $ & $ 4.3 \pm 0.6$ & $ 0.23 \pm 0.01$\\
    1284 &  $130.7 \pm 6.5$ & $121.7 \pm 6.1$ & $6.4 \pm 0.3$ & $1.4 \pm 0.1$ & - &  $47.1  \pm 2.4$ &  $ 20.7\pm 1.0$ & $ 20.4 \pm 1.0 $ & $ 5.4 \pm 0.3$ &  $ 0.21 \pm 0.01$ \\
    1380 &  $122.1 \pm 6.1$ & $115.9 \pm 5.8$ & - & $4.5 \pm 0.3$  & - & $42.0  \pm 2.2$ & $ 19.4\pm 1.0$ & $  17.9 \pm 0.9$ & $ 4.1 \pm 0.4$ & - \\
    1500 &  - & - & - & - &  - & - & - & - & - & $ 0.18\pm 0.01$ \\
    1810  &  $93.5 \pm 4.7$ & $90.7 \pm 4.5$ & - & $2.3 \pm 0.2$ & -  & $22.7  \pm 1.3$  &  $ 12.5\pm 0.6$ & $ 9.0 \pm 0.5 $ & $ 1.9 \pm 0.3$ & - \\
    5500 &  - & - & - & - &  - & - & - & - & - & $ 0.079\pm 0.002$ \\
\noalign{\smallskip}
	\hline
	\end{tabular}
 }
\end{table*}

\begin{table}
\centering
   		\caption{Integrated spectral index of T1 and T2. Col. 1: Considered area. Col. 2: Colour and line style of the fitted power-law as in Fig. \ref{fig: spettro}. Col. 3: Frequency range. Col. 4: Fitted spectral index. Col. 5: Reduced $\chi$-squared.}
    \label{tab: fitspix}
     \resizebox{0.45\textwidth}{!}{
    \begin{tabular}{ccccc}
   	\hline
   	\noalign{\smallskip}
   	Area & Power-law & $\nu$ & $\alpha$ & $\chi^2_{\rm red}$  \\  
	  &  & (MHz) &  &   \\  
   	\noalign{\smallskip}
  	\hline
   	\noalign{\smallskip}
  T1 & Solid blue & 50-1810 & $0.87 \pm 0.01$ & 0.5  \\
  T1-A & Solid red & 50-1810 & $0.68 \pm 0.02$ &  0.7 \\
  T1-B & Solid orange & 50-234 & $0.97 \pm 0.11$ & 1.9  \\
  T1-B & Dotted orange & 234-1284 & $1.74 \pm 0.04$ & 0.9   \\
  T1-C & Solid cyan & 50-234 & $1.31 \pm 0.17$ & 4.2  \\
  T1-C & Dotted cyan & 234-1284 & $2.28 \pm 0.05$ & 0.8  \\
  T1-D & Solid green & 50-323 & $1.87 \pm 0.25$ & 7.5  \\
  T2 & Solid blue & 50-1810 & $1.39 \pm 0.05$ & 5.9  \\
  T2-A & Solid red & 50-407 & $0.67 \pm 0.06$ & 1.2  \\
  T2-A & Dotted red & 407-1810 & $1.07 \pm 0.06$ & 2.1  \\
  T2-B & Solid orange & 50-407 & $0.96 \pm 0.06$ &  1.5 \\
  T2-B & Dotted orange & 407-1810 & $1.63 \pm 0.10$ &  6.5 \\
  T2-C & Solid cyan & 50-407 & $1.53 \pm 0.08$ &  2.3 \\
  T2-C & Dotted cyan & 407-1810 & $2.22 \pm 0.07$ &  1.5 \\
  T2-D & Solid green & 143-5500 & $0.69 \pm 0.02$ & 1.3  \\
   	\noalign{\smallskip}
   	\hline
   	\end{tabular}
    }
   	  	\centering
   \end{table}

To derive the integrated radio spectra of T1 and T2, we imaged all the datasets (except VLA data in A-array) with a common \textit{uv}-range of $350\lambda-16{\rm k\lambda}$. The chosen minimum baseline length provides more uniform \textit{uv}-coverage of our data at short spacings. The obtained images were convolved at the same resolution of $14''$. We report our flux density measurements in Table \ref{tab: radio flux}; the corresponding radio spectra are shown in Fig. \ref{fig: spettro}, and the obtained spectral indices are summarised in Table \ref{tab: fitspix}.

We measured the total flux densities of T1 (excluding T1-D) in a box of size $6.0'\times 1.3'$
(blue box in Fig. \ref{fig: spettro}) and fitted the data points with a single power-law. We found that a single power-law of slope $\alpha=0.87\pm 0.01$ (blue line in Fig. \ref{fig: spettro}) can well reproduce the integrated spectrum of T1 from 50 to 1810 MHz. In addition, we obtained the radio spectrum of each sub-region by measuring the flux densities within boxes of size $2.0'\times 1.0'$, $1.5'\times 1.0'$, $2.0'\times 1.0'$, and $1.0'\times 2.0'$ for T1-A (red), T1-B (orange), T1-C (cyan), and T1-D (green), respectively. Fig. \ref{fig: spettro} clearly shows that the total flux density of T1 is dominated by T1-A at all frequencies. For T1-A, data points can be described by a single power-law of slope $\alpha=0.68\pm 0.02$. However, we find evidence of spectral breaks for T1-B and T1-C, as single power-laws cannot fit our measurements. We thus considered double power-laws, with a fixed break at 234 MHz, as this appears to be roughly the frequency where the spectrum steepens. The details on the fitted spectral indices between 50 and 234 MHz (solid lines) and between 234 and 1284 MHz (dotted lines) are reported in Table \ref{tab: fitspix}. The arc is detected (above $3\sigma$) only by LOFAR and GMRT at 323 MHz, therefore we did not attempt to fit two power-laws, but the poor fit ($\chi^2_{\rm red}=7.5$) suggests the existence of a break for this sub-region as well.

Similarly to T1, we obtained the total flux density of T2 in a box of size $3.0'\times 4.5'$
(blue box in Fig. \ref{fig: spettro}), but a single power-law does not reproduce all our data points ($\chi^2_{\rm red}=5.9$). The flux densities of T2-A (red), T2-B (orange), and T2-C (cyan) were measured in an ellipse of axis $0.8'\times 1.1'$, a box of size $1.0'\times 1.5'$, and an ellipse of axis $1.8'\times 2.4'$, respectively, and then fitted with two power-laws having a fixed break at 407 MHz (see details in Table \ref{tab: fitspix}). Even though we notice that the low-$\nu$ spectrum of T2-B is not accurately fitted ($\chi^2_{\rm red}=6.5$), likely due to contamination from the plume in overlapping areas, our analysis shows that the spectrum of T2 progressively steepens from the inner to the outer sub-regions. Moreover, we produced additional radio images (including VLA data in A-array) with specific combination of \textit{uv}-range and weighting schemes to maximise the resolution ($<5''$) of our images, and derive the spectrum of the core of T2 (T2-D, green line). We used the {\tt imfit} task in {\tt CASA} to derive the peak value of a Gaussian fit to the core at each frequency. These values provide a fitted spectral index of $\alpha=0.69\pm 0.02$ from 143 to 5500 MHz.

\subsection{Resolved spectral properties}
\label{sect: spectral properties 2}

\begin{table}
\centering
   		\caption{Summary of the parameters used to produce images for the spectral index maps in Fig. \ref{fig: spixmap}. Cols. 1-3: considered datasets, \textit{uv}-range, and Gaussian taper. Col. 4: final resolution after convolution of each image with the same beam.}
    \label{tab: spixparam}   
    \begin{tabular}{cccc}
   	\hline
   	\noalign{\smallskip}
   	$\nu$ & \textit{uv}-range & Taper & $\theta$ \\  
	 (MHz) & (${\rm k}\lambda$) & (arcsec) & (arcsec) \\  
   	\noalign{\smallskip}
  	\hline
   	\noalign{\smallskip}
  50, 143, 323 & [0.35, 18] & 20 & 30 \\
   143, 323, 608, 1284 & [0.35, 29] & 8 & 14 \\
   	\noalign{\smallskip}
   	\hline
   	\end{tabular}
   	  	\centering
   \end{table}

\begin{figure*}
	\centering
 
\includegraphics[width=0.45\textwidth]{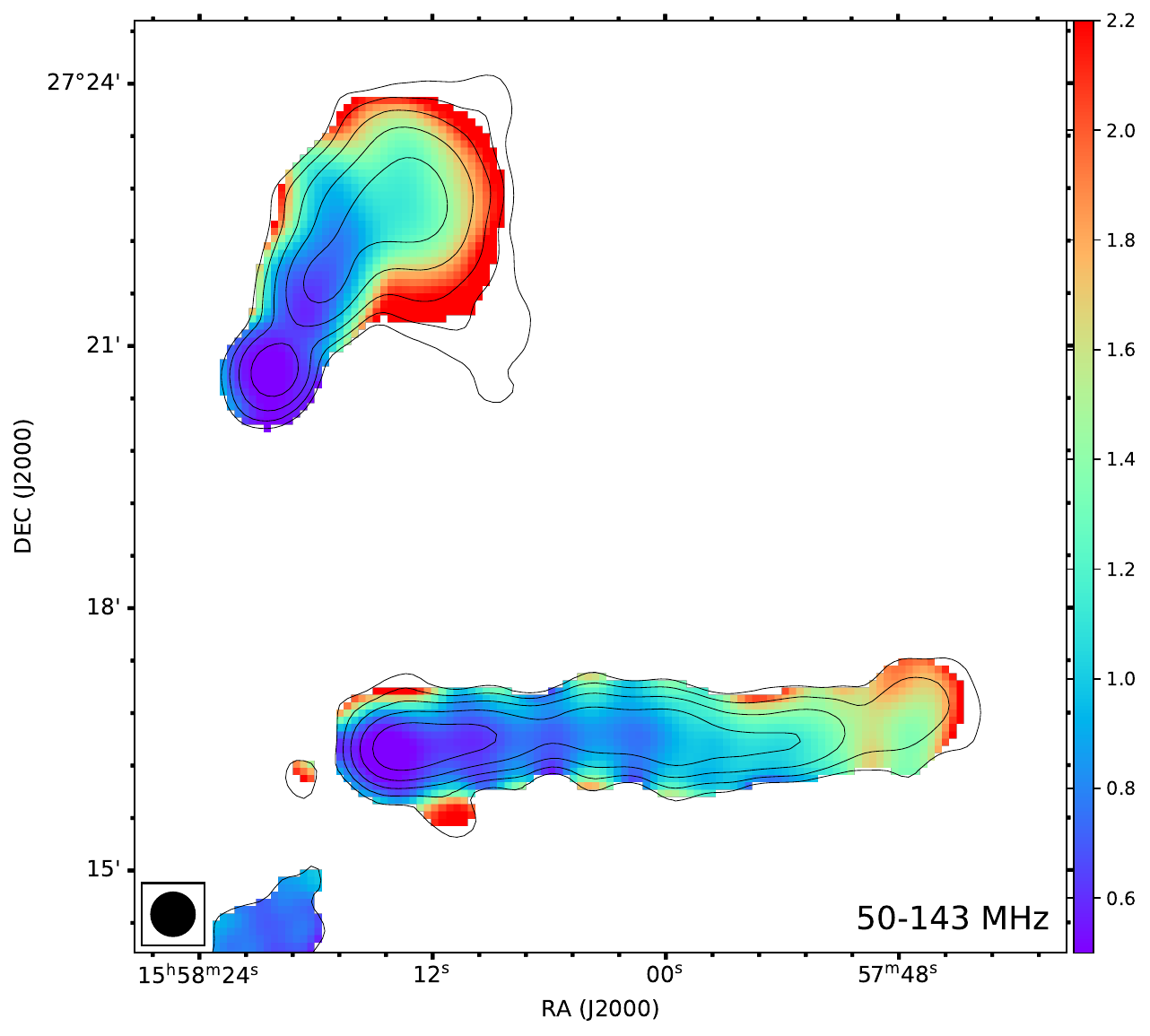}
\includegraphics[width=0.45\textwidth]{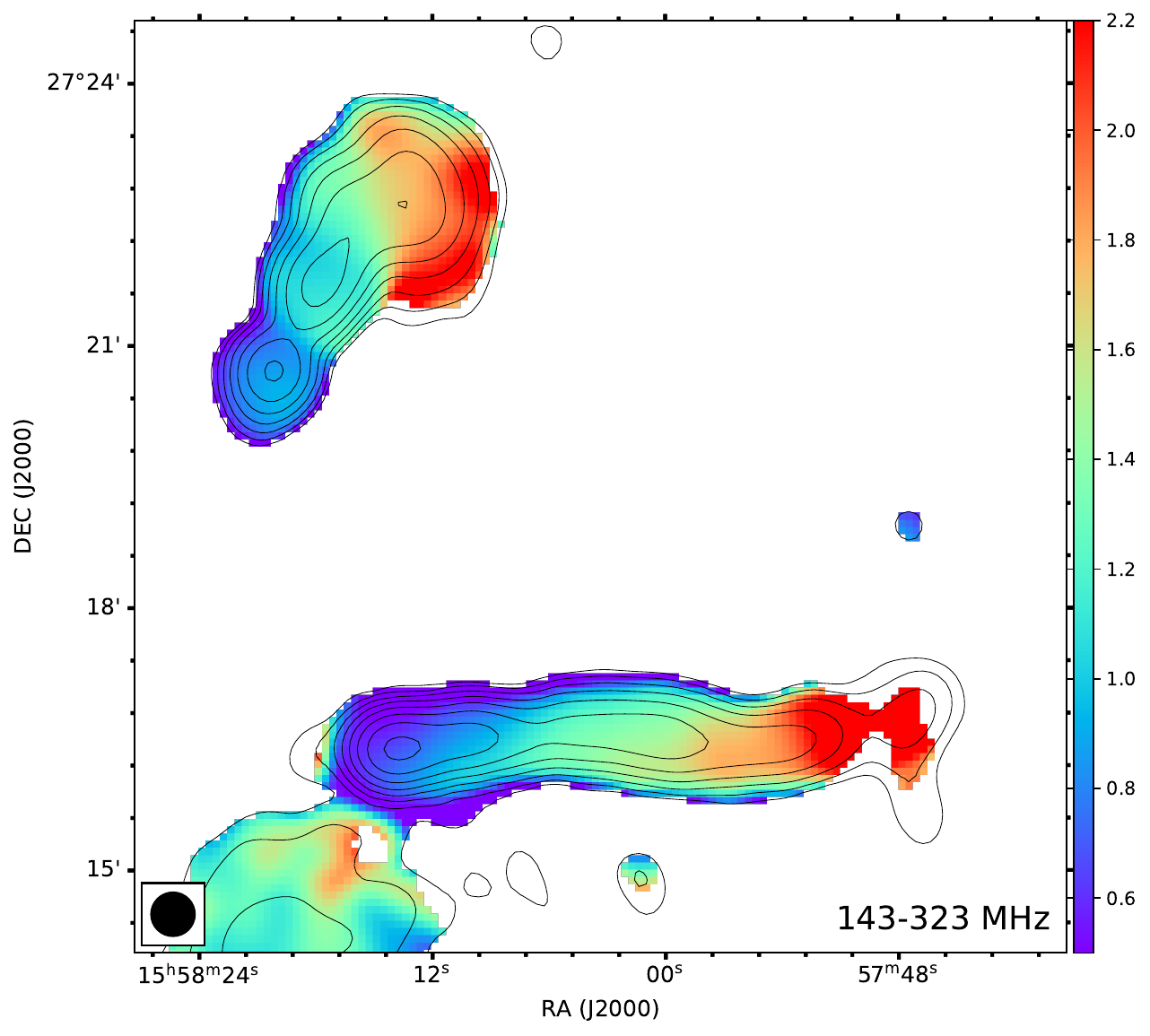}
 \includegraphics[width=0.45\textwidth]{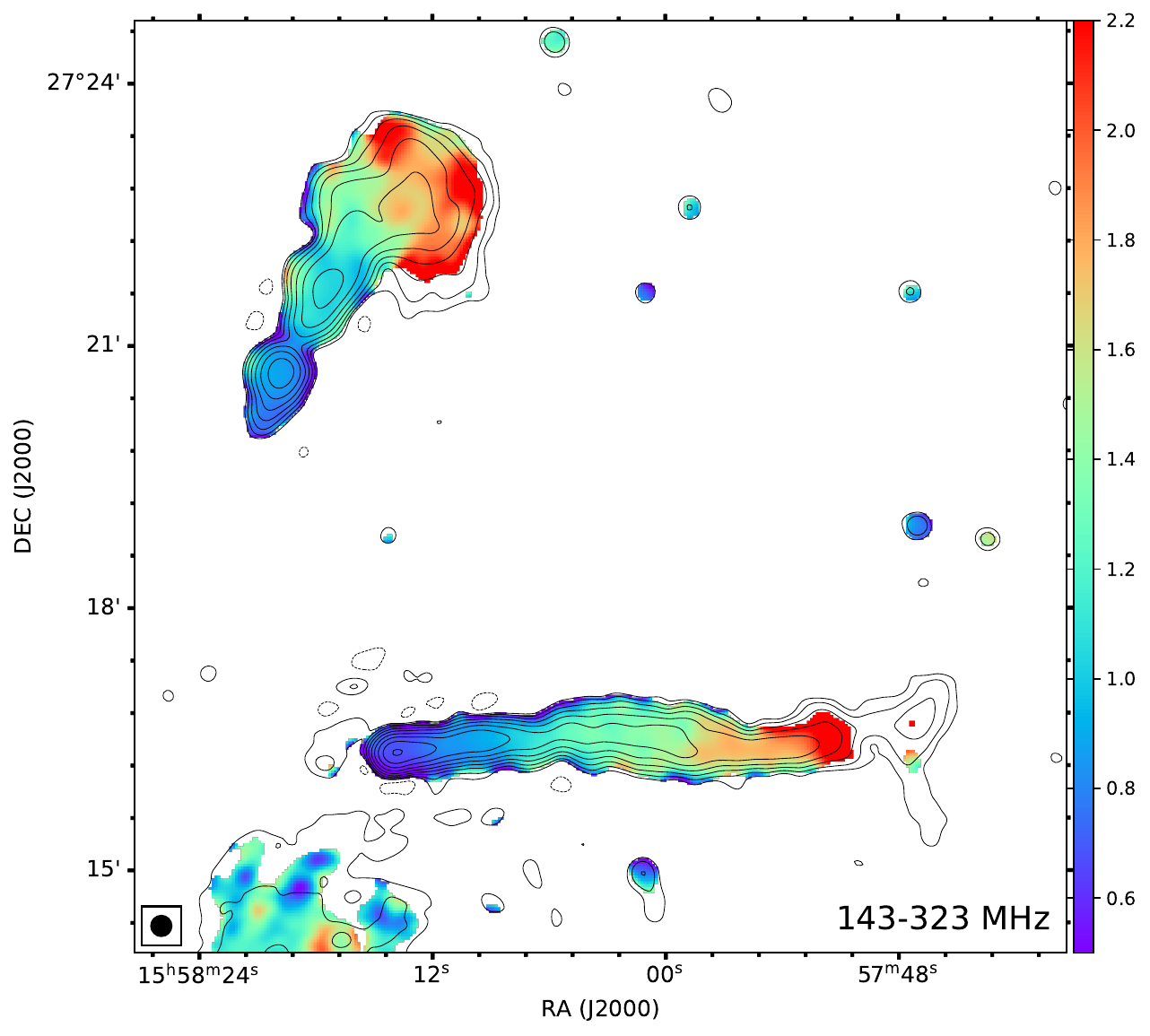}
\includegraphics[width=0.45\textwidth]{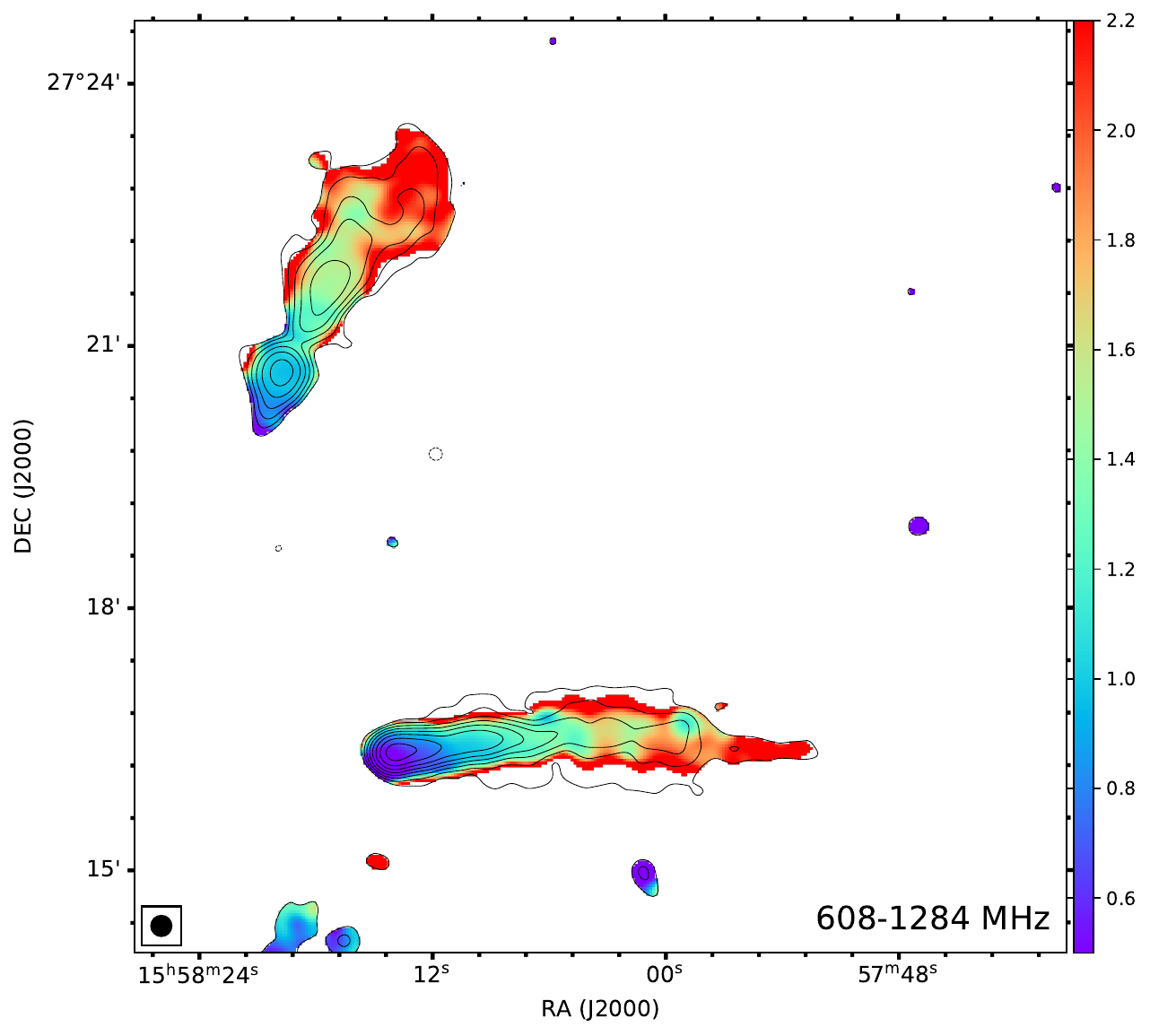}
	\smallskip
	
 \caption{Spectral index maps of T1 and T2 computed with a threshold of $S\geq 5\sigma$. In all the panels, the contour levels are $[\pm5, \;10, \;20, ...]\times \sigma$ of the lowest-frequency image in each combination. \textit{Top}: spectral index maps at $30''$ between 50-143 MHz (left) and 143-323 MHz (right). \textit{Bottom}: spectral index maps at $14''$ between 143-323 MHz (left) and 608-1284 MHz (right). The corresponding error maps are shown in Fig. \ref{fig: errspixmap}.}
	\label{fig: spixmap}
\end{figure*}

\begin{figure*}
	\centering
   \includegraphics[width=0.25\textwidth]{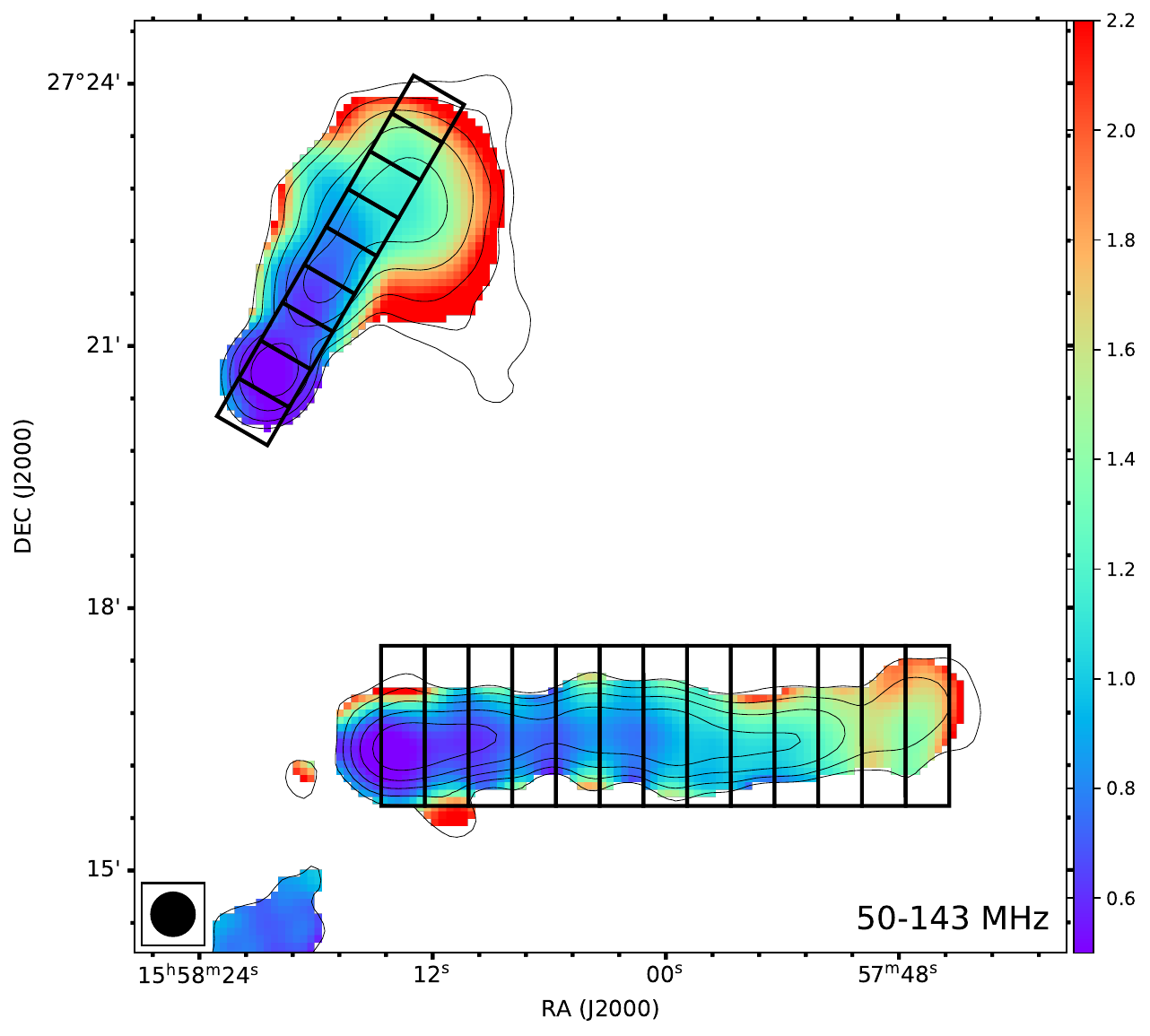}
     \includegraphics[width=0.25\textwidth]{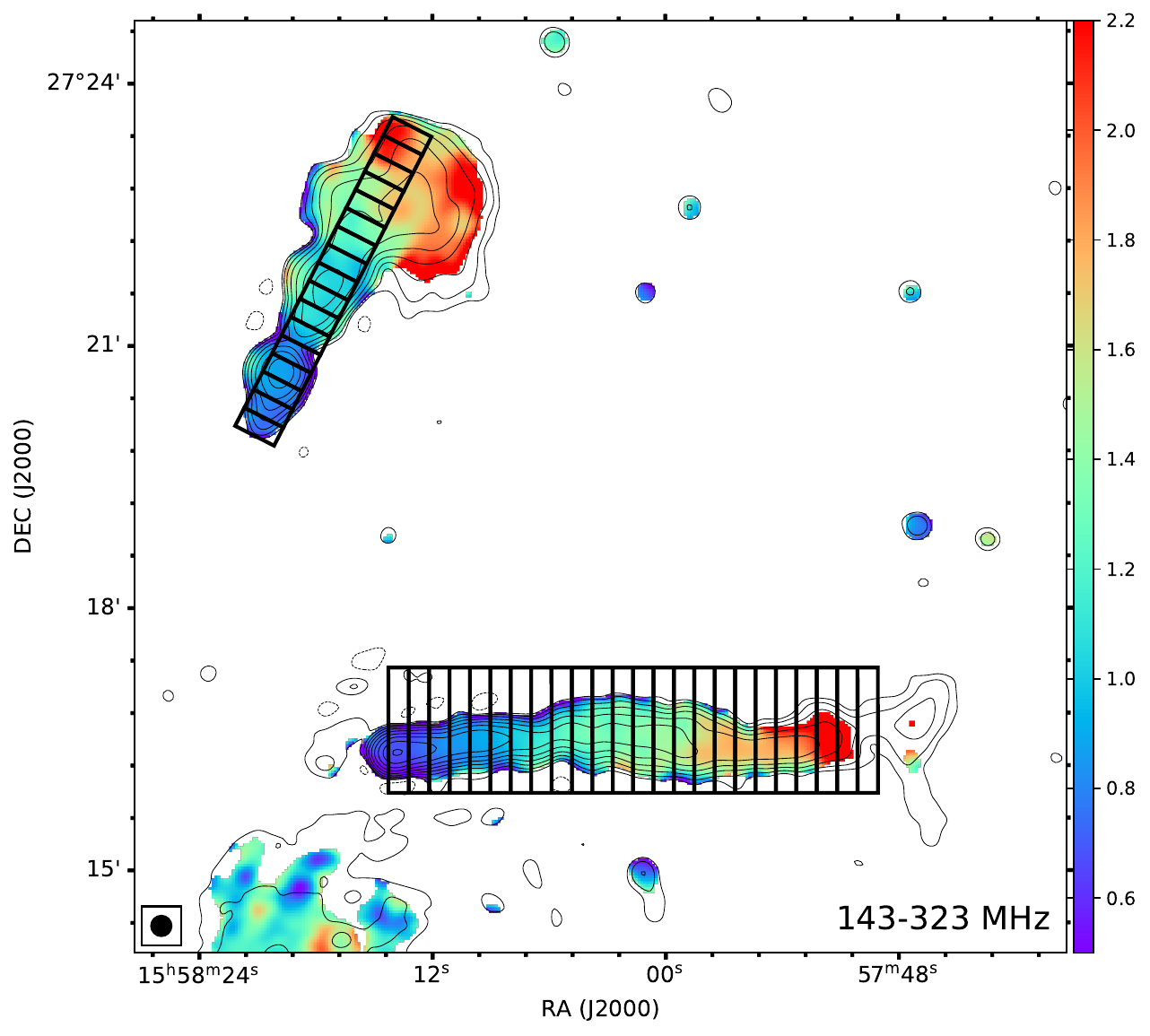}
 
 \includegraphics[width=0.49\textwidth]{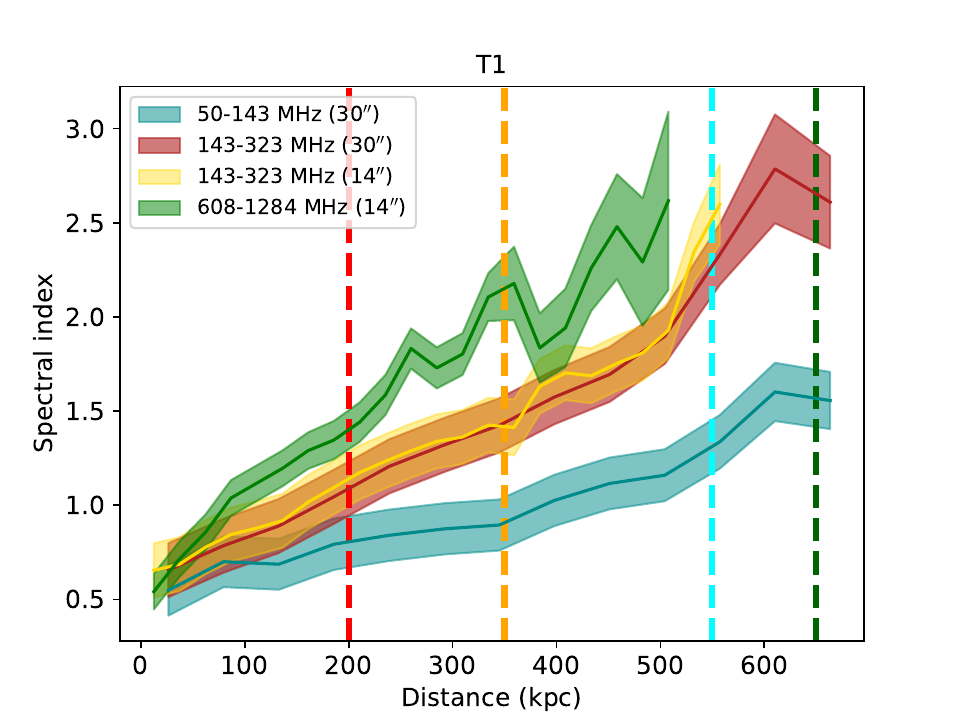}  
\includegraphics[width=0.49\textwidth]{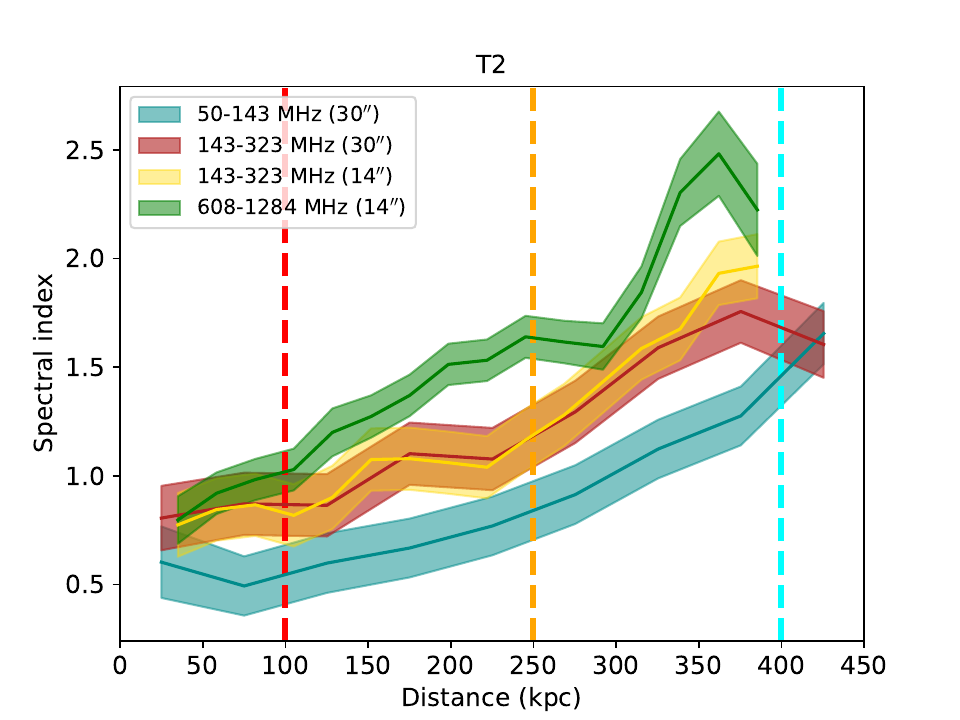}
	\smallskip
	
	\caption{Spectral index profiles as measured from maps in Fig. \ref{fig: spixmap} for T1 (left) and T2 (right). The sampling boxes have beam-size width and are shown in the upper panels. The vertical lines indicate the boundaries of sub-regions as in Fig. \ref{fig: substructures}.}
	\label{fig: spixprofile}
\end{figure*}

The analysis of the integrated radio spectra of T1 and T2 in Sect. \ref{sect: spectral properties 1} has shown that the spectral index is not constant across our targets. Under the hypothesis of pure radiative losses, a gradual steepening of the spectral index is expected along the tail, whereas sudden spectral flattening may suggest re-energising mechanisms. To determine the spectral trend along the tails, we produced spectral index maps by combining sets of radio images (see details in Table \ref{tab: spixparam}) and setting a minimum flux density threshold of $5\sigma$ at each frequency. In Fig. \ref{fig: spixmap} we report the spectral index maps at 50-143 and 143-323 MHz at $30''$ ($\sim 50$ kpc), and at 143-323 MHz and 608-1284 MHz at $14''$ ($\sim 25$ kpc); the associated error maps are shown in Fig. \ref{fig: errspixmap}. By measuring the spectral index within boxes of width equal to the beam size, we derived the corresponding spectral index profiles that are shown in Fig. \ref{fig: spixprofile}. As typically observed along the lobes of FRI galaxies due to ageing of the emitting particles, on average both T1 and T2 exhibit a progressive steepening of the spectral index along the tail. We discuss the spectral trends within each sub-region in the following paragraphs.

Within T1-A, the spectral index ranges from $\alpha \sim 0.5$ (in the core) up to $\alpha \sim 1.5$, with steeper values for higher frequency pairs. The profiles steepen with the distance with approximately constant slopes (even though the trend is flatter at lower frequencies). Along T1-B and T1-C the spectra further steepen, reaching ultra-steep ($\alpha\gtrsim1.5$) values of $\alpha \sim 2$ and $\alpha \sim 2.5$ between 143-323 MHz and 608-1284 MHz, respectively. In the lower-frequency regime ($<323$ MHz), the constant steepening trend with the distance is retained, whereas deviations in the form of flatter and steeper features are clearly visible at 608-1284 MHz both in T1-B and T1-C. Despite the different resolutions ($14''$ and $30''$), we notice that the trends of the 143-323 MHz maps are consistent. Beyond T1-C, emission is detected only below 323 MHz. In T1-D (the arc) the spectral index is $\alpha \sim 1.5$ and $\alpha \sim 2.5$ between 50-143 MHz and 143-323 MHz, respectively. The constant steepening of the spectral index is still preserved (with moderate deviations at 143-323 MHz), thus allowing us to confidently conclude that the arc is the oldest part of T1. As a consequence, the total projected length of T1 is $\sim 700$ kpc.

In T2-A, the inner regions exhibit a spectral index that is flatter ($\alpha \sim 0.5$) at 50-143 MHz than that at higher frequencies ($\alpha \sim 0.7$). For both T2-A and T2-B, the spectral index steepens with the distance with roughly constant and smooth trends at all wavelengths. The spectrum of T2-B reaches values of $\alpha \sim 0.8$, $\alpha \sim 1.1$, and $\alpha \sim 1.5$ at 50-143 MHz, 143-323 MHz, and 608-1284 MHz, respectively. Studying the profile of T2-C is not trivial due to the expansion of the tail into the plume and its western bending at low frequencies. Along our sampling boxes, the measured spectral indices are ultra-steep ($\alpha \sim 1.5-2.5$) for all maps. The smooth trend is retained below 323 MHz, while a rapid steepening is observed at 608-1284 MHz.

\subsection{Testing the radiative ageing}
\label{sect: Radiative ageing}

\begin{figure*}[]
	\centering
 \includegraphics[width=0.40\textwidth]{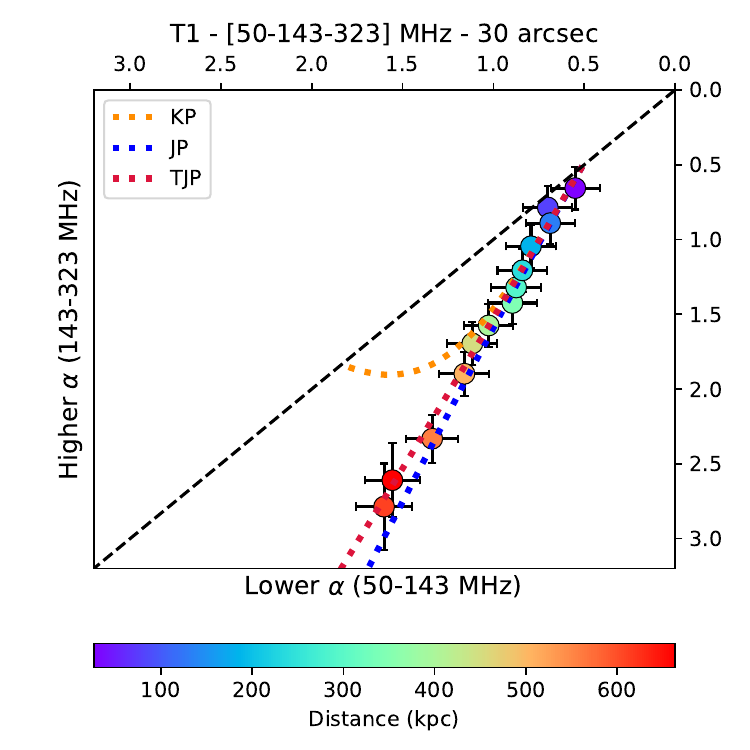}
 \includegraphics[width=0.40\textwidth]{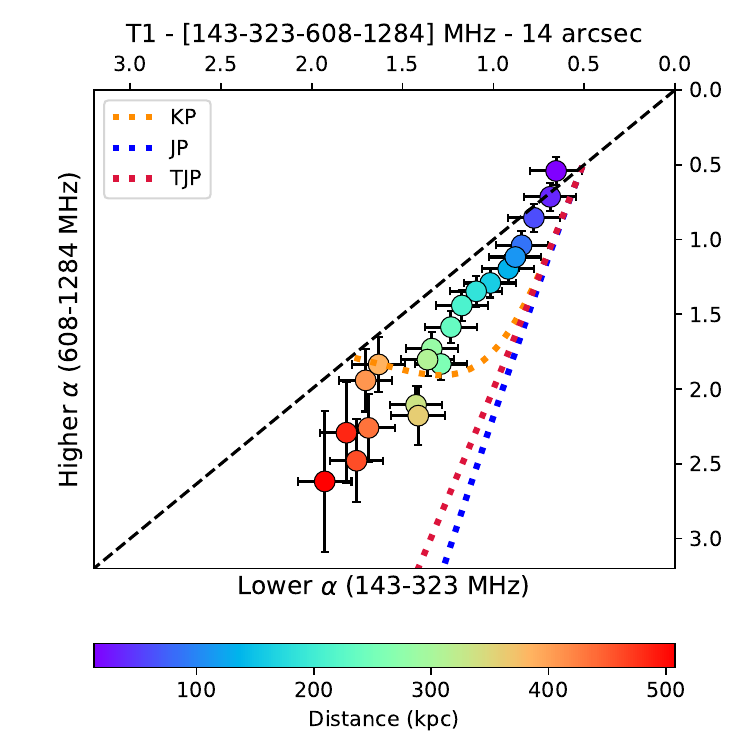}
 \includegraphics[width=0.40\textwidth]{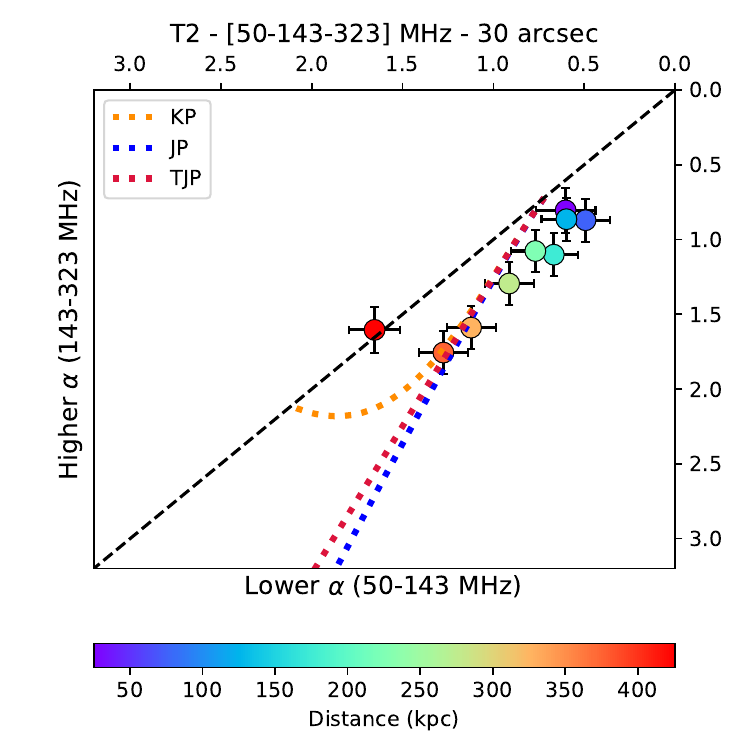}
 \includegraphics[width=0.40\textwidth]{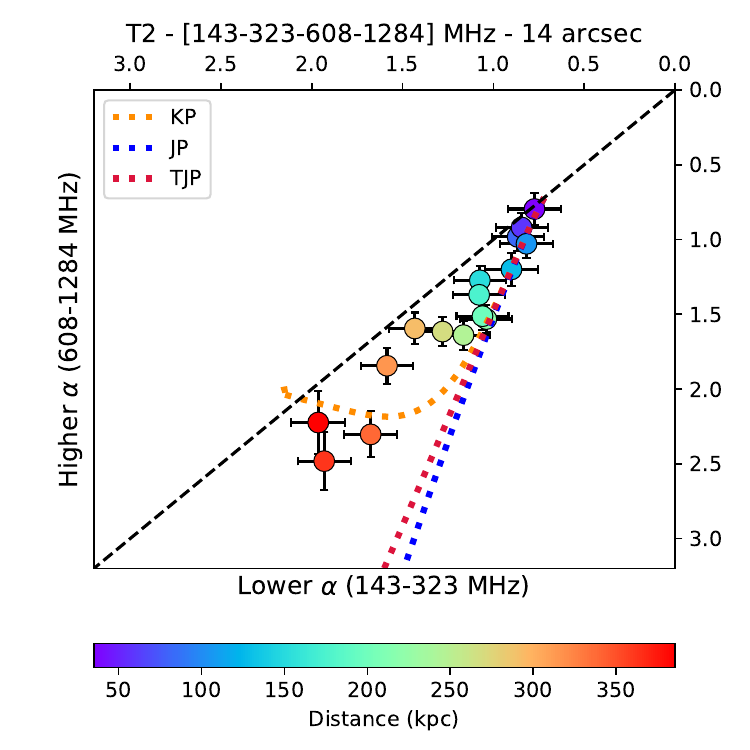}
	\smallskip
	\caption{Radio colour-colour diagrams of T1 (top) and T2 (bottom). Data points are colour-coded based on their increasing distance from the core from blueish to reddish colours. The black dashed line (one-to-one line) indicates a power-law with $\alpha=\alpha_{\rm inj}$. The dotted lines represent the theoretical KP (orange), JP (blue), and TJP (red) ageing curves obtained with $\alpha_{\rm inj}=0.51$ (T1), $\alpha_{\rm inj}=0.72$ (T2), $B_{\rm 0}=2.2 \; {\rm \mu G}$ (T1, T2) as input. The frequency pairs and resolution are the same as in Fig. \ref{fig: spixprofile}. }
	\label{fig: CCP}
\end{figure*}

Relativistic particles are ejected from the core of the radio galaxy and then radiatively age along the tail due to synchrotron and inverse Compton losses. By assuming a constant injection rate, we expect to observe a progressive decline of the flux density and a steepening of the spectral index with the increasing distance. However, in Sects. \ref{sect: Radio surface brightness fluctuations}, \ref{sect: spectral properties 2} we found complex distributions of the surface brightness and spectral index within each sub-region of T1 and T2, which suggest either local deviations from a pure ageing scenario or a varying particle injection rate. In this section we aim to test ageing models in detail.

Ageing models depend on the form of the initial electron energy distribution ($N(E)\propto E^{-\delta_{\rm inj}}$, where $\delta_{\rm inj}=2\alpha_{\rm inj}+1$ is the population injection index, and $\alpha_{\rm inj}$ is the spectral index at age $t=0$, based on the assumption of Fermi I acceleration mechanism) and the spatial distribution of the magnetic field ($B(r)$), which are not known a priori. Models assuming a single injection event are the Kardashev-Pacholczyk \citep[KP;][]{kardashev62,pacholczyk70}, Jaffe-Perola \citep[JP;][]{jaffe73}, and Tribble-Jaffe-Perola \citep[TJP;][]{tribble93}. Both the KP and JP models assume a uniform magnetic field, but differ in terms of the treatment of the pitch angle $\theta_{\rm p}$. In the KP model, a constant and isotropic $\theta_{\rm p}$ is assumed throughout the entire lifetime of the electrons; the JP model considers electron scattering, which leads to isotropic $\theta_{\rm p}$ on short-time scales only, and thus assumes a time-averaged pitch angle. As a consequence of the assumption on $\theta_{\rm p}$, energetic electrons with small pitch angles can live indefinitely for the KP model if emitting synchrotron radiation only, whereas an exponential cut-off arises in the electron distribution at high energy for the JP model \citep[e.g.][]{hardcastle13}. The TJP model is based on the JP model, but introduces Gaussian spatial fluctuations of the magnetic field around a central value ($B_{0}$). The high-energy cut-off is shallower in the TJP model than in the JP model, thus allowing particles to live longer in non-uniform magnetic fields.

In the following, we test the KP, JP, and TJP ageing models. We consider a value for the magnetic field that minimises the radiative losses and maximises the lifetime of the source, which is $B_0=B_{\rm CMB}/\sqrt{3}$, where $B_{\rm CMB}=3.25(1+z)^2 \; \mu{\rm G}$ is the equivalent magnetic field of the cosmic microwave background (CMB); this yields to $B_{\rm 0}\sim 2.2 \; \mu{\rm G}$ for both T1 and T2 at the cluster redshift. The injection spectral index was derived by means of the {\tt findinject} task of the Broadband Radio Astronomy ToolS ({\tt BRATS\footnote{\url{https://www.askanastronomer.co.uk/brats/}};} \citealt{harwood13,harwood15}) software, which fits the radio spectrum of the target from multi-frequency radio images and outputs the value of $\alpha_{\rm inj}$ that minimises the distribution of $\chi^2$ for the fitted model; we obtained $\alpha_{\rm inj}^{\rm T1}=0.51\pm 0.01$ for T1 and $\alpha_{\rm inj}^{\rm T2}= 0.72\pm0.01$ for T2, in agreement with the measured spectral index of the radio cores (see Sects. \ref{sect: spectral properties 1}, \ref{sect: spectral properties 2}), thus suggesting that their spectra are representative of the injection distributions.

Radio colour-colour diagrams (RCCDs; \citealt{katz-stone93}) are diagnostic plots to probe the local shape of radio spectra computed from two pairs of frequencies, which is independent of the magnetic field and possible adiabatic expansion and compression (these can only shift the spectrum in frequency and affect the age), and thus test theoretical ageing models. In RCCDs, the one-to-one line represents a power-law spectrum with $\alpha=\alpha_{\rm inj}$, whereas data points lying below the one-to-one line ($\alpha>\alpha_{\rm inj}$) indicate particle ageing. By considering the spectral index maps and sampling boxes as in Sect. \ref{sect: spectral properties 2}, we obtained the RCCDs that are shown in Fig. \ref{fig: CCP}. We overlaid the theoretical ageing curves (dotted lines) as obtained from the {\tt kpdata}, {\tt jpdata}, and {\tt tribbledata} tasks in {\tt BRATS} under the assumptions discussed above. 

For the low-resolution ($30''$) and low-$\nu$ (50-143-323 MHz) set of images of T1 (upper left panel in Fig. \ref{fig: CCP}), both the JP and TJP models reproduce the emission of the source in each sub-region, whereas the KP model cannot describe data points beyond $\sim 400$ kpc (T1-C and T1-D). With $14''$-resolution (upper right panel), the observed 143-323-608-1284 MHz spectral distribution can be barely described by the models in T1-A (inner $ \sim 100$ kpc), and progressively increasing deviations are found in T1-B and T1-C. Although a steeper injection index ($\alpha_{\rm inj}\sim 0.6-0.8$) would shift the ageing curves towards our measurements, models still fail to reproduce data points at large distance from the core. We will further discuss the results and implications of the RCCDs for T1 in Sects. \ref{sect: On the discrepancy of RCCDs for T1}, \ref{sect: Velocity of T1}.

The bottom panels in Fig. \ref{fig: CCP} report the complex distribution of the data points in the RCCD for T2. In the low-$\nu$ (50-143-323 MHz) set at $30''$ (left), data points align roughly parallel to the one-to-one line (except for an outlier associated with the plume). In the high-$\nu$ (143-323-608-1284 MHz) set at $14''$ (right), the KP, JP, and TJP models can approximately reproduce the data points associated with T2-A, but prominent deviations are found for T2-B and T2-C. In summary, none of the considered models can entirely describe the spectral behaviour of T2 (see further discussion in Sect. \ref{sect: On the origin of T2}).

\subsection{Local ICM conditions}
\label{sect: Local ICM conditions}

\begin{figure*}
	\centering
 
 \includegraphics[width=0.4\textwidth]{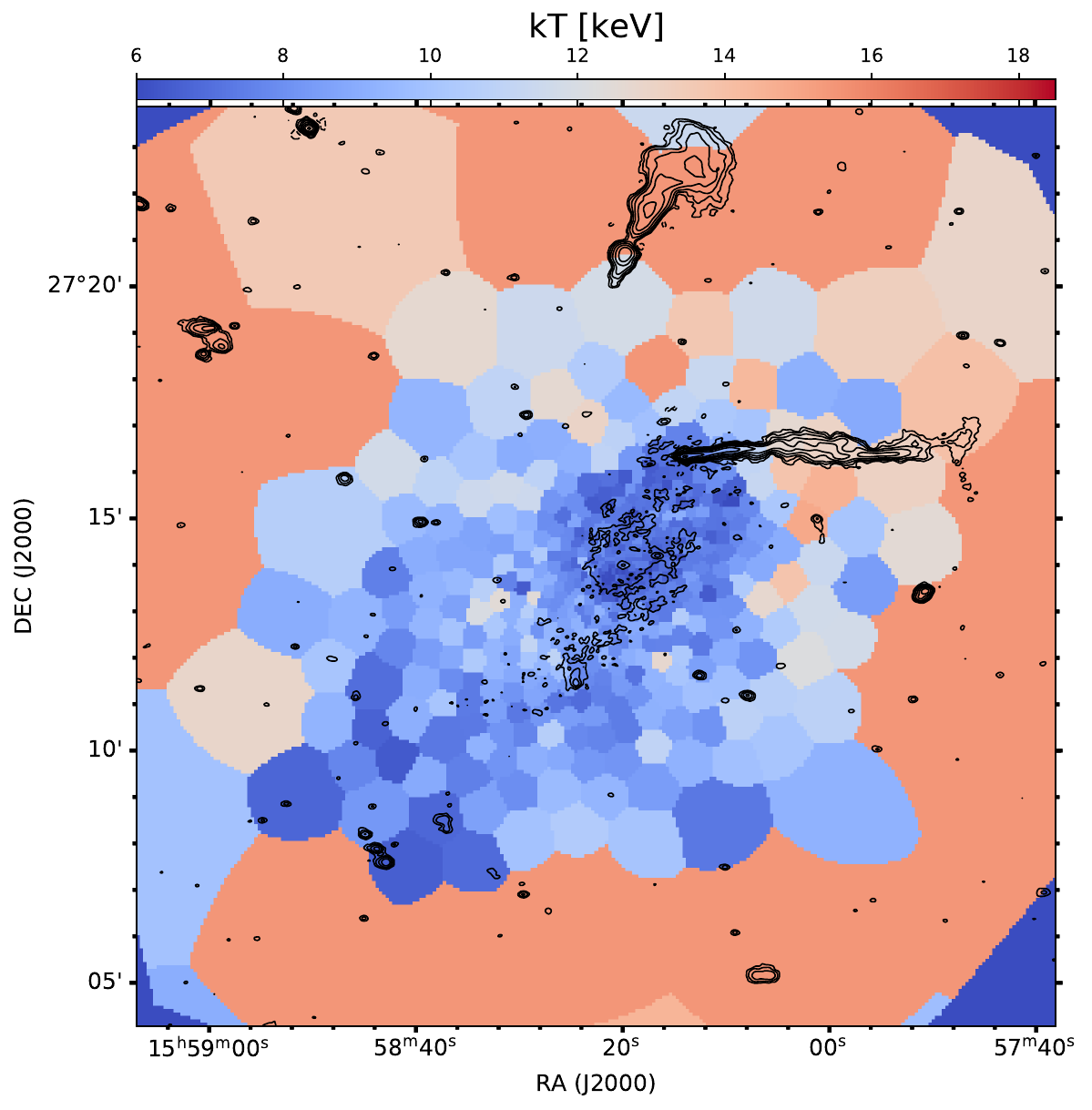}
\includegraphics[width=0.48\textwidth]{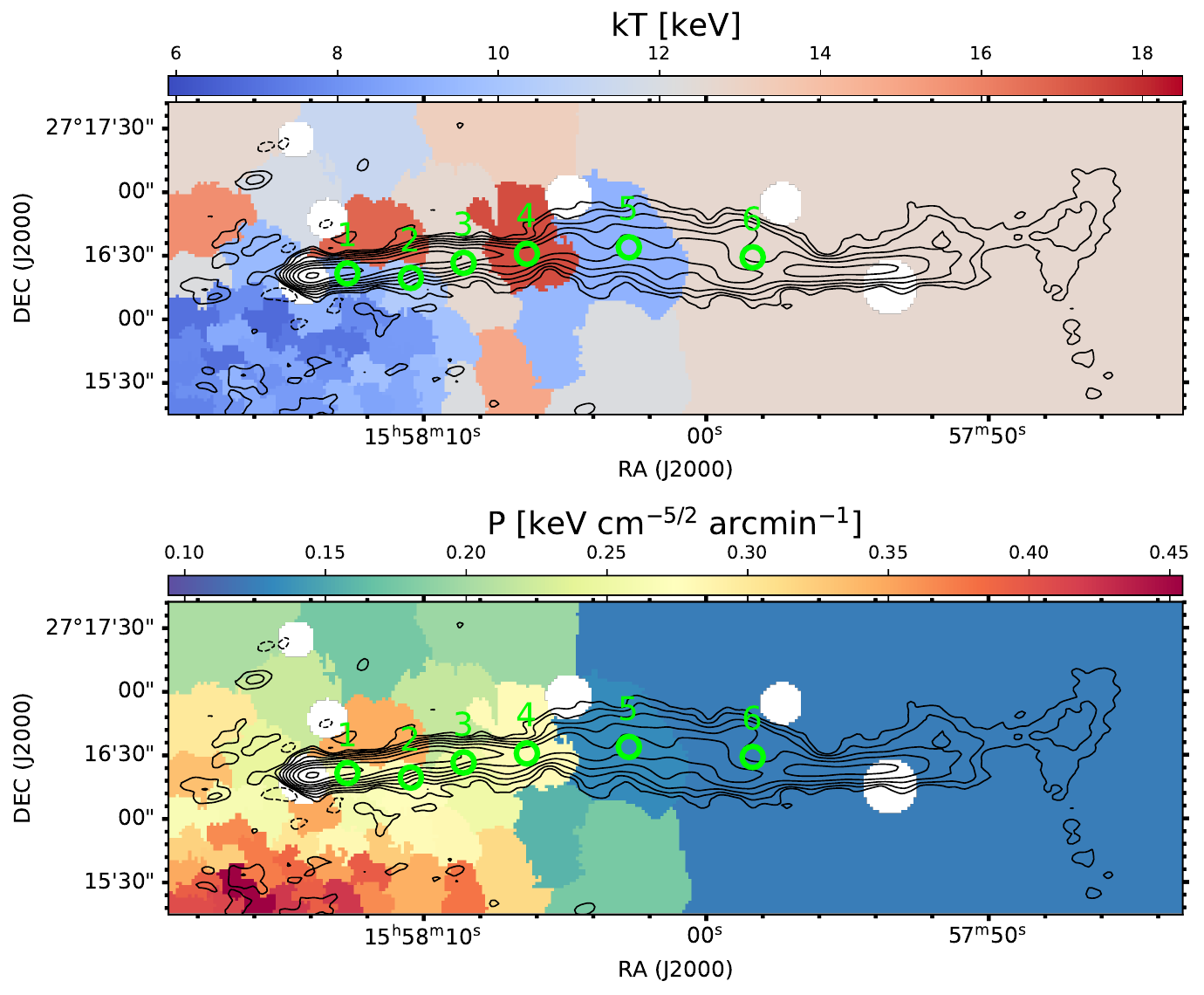}
  \includegraphics[width=0.4\textwidth]{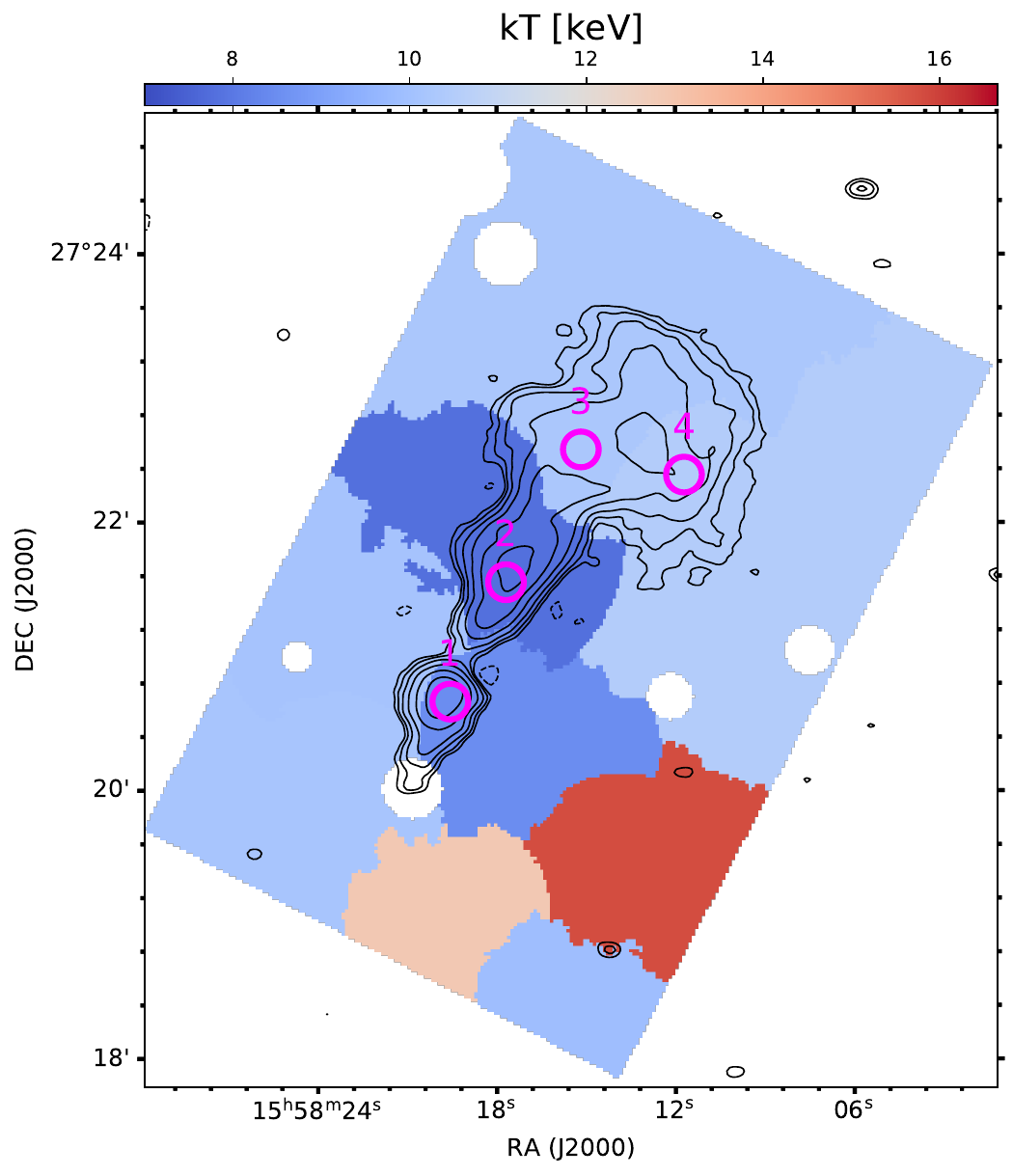}
 \includegraphics[width=0.40\textwidth]{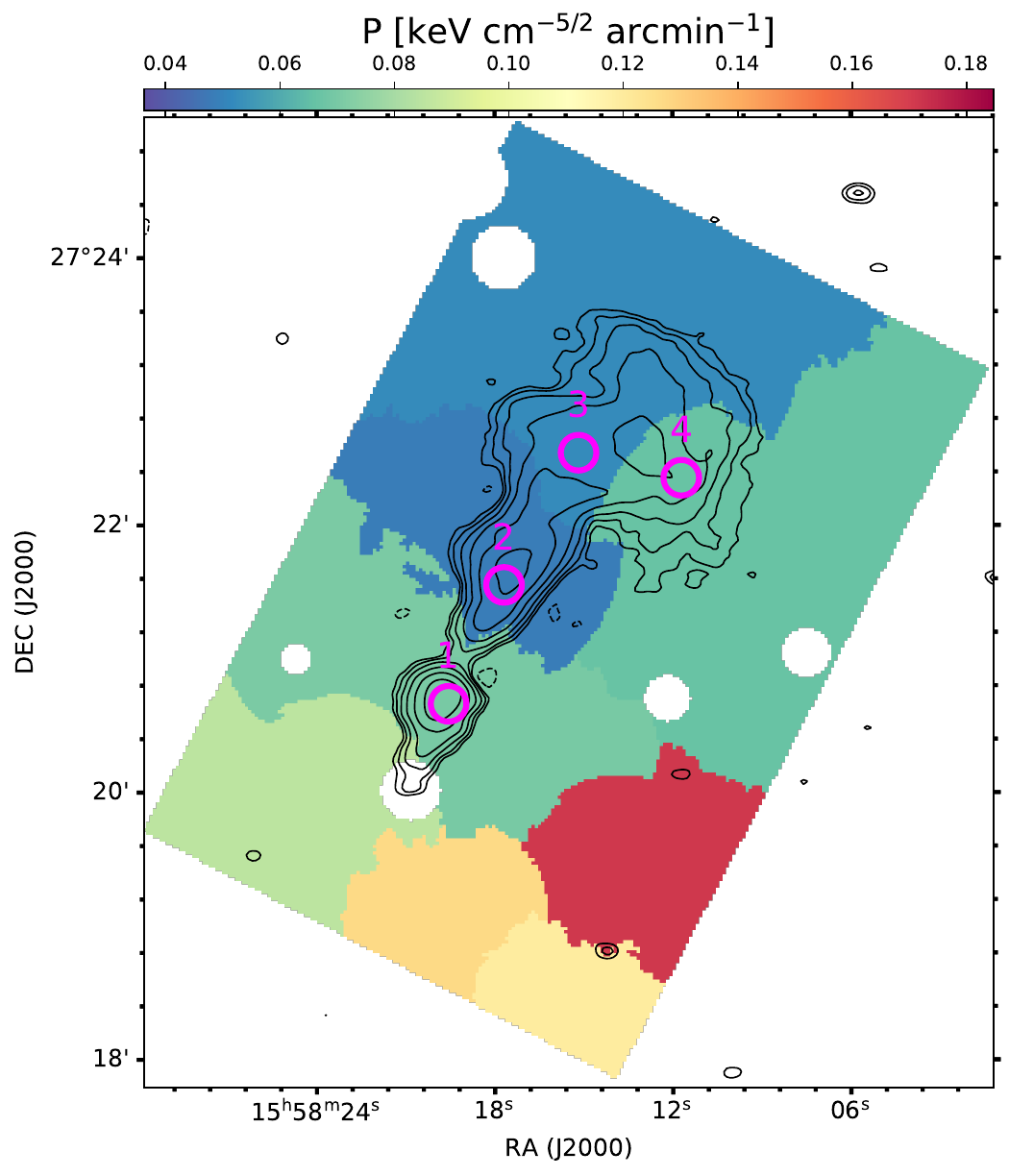}
\includegraphics[width=0.245\textwidth]{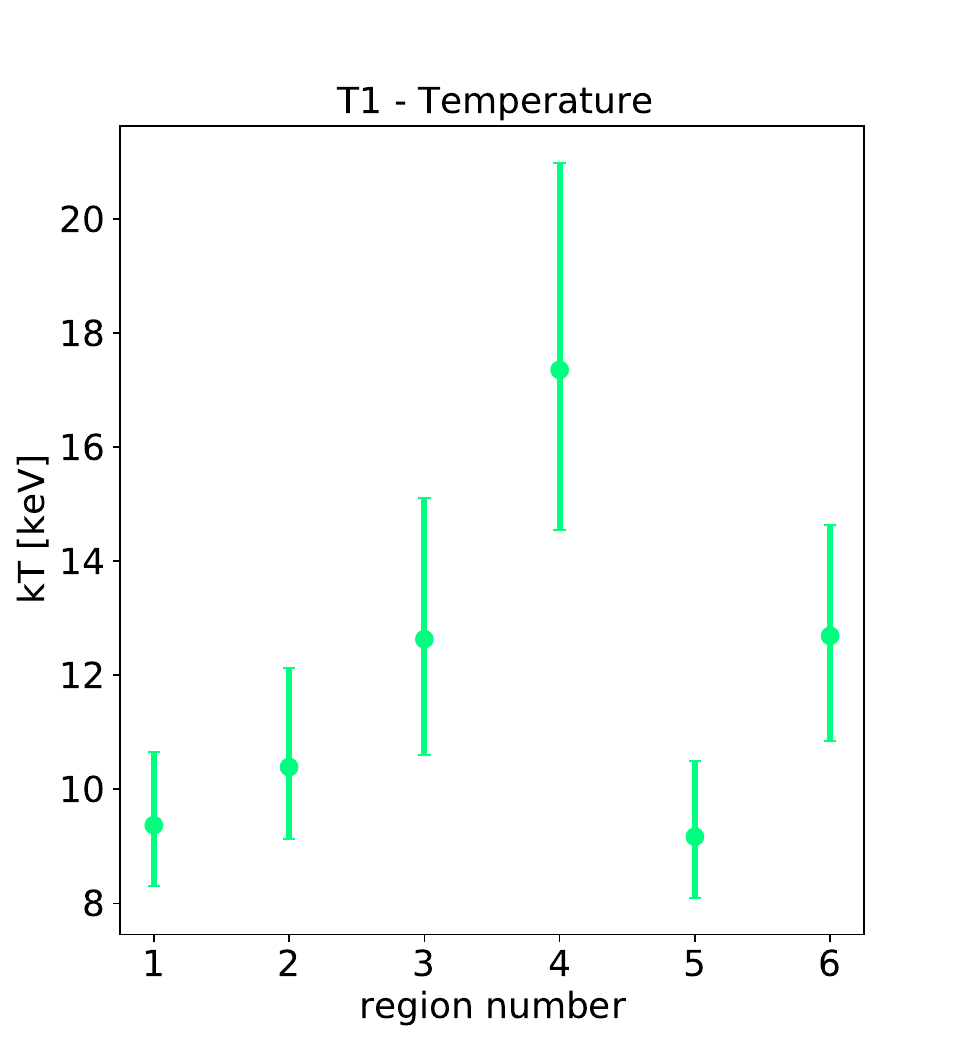}
 \includegraphics[width=0.245\textwidth]{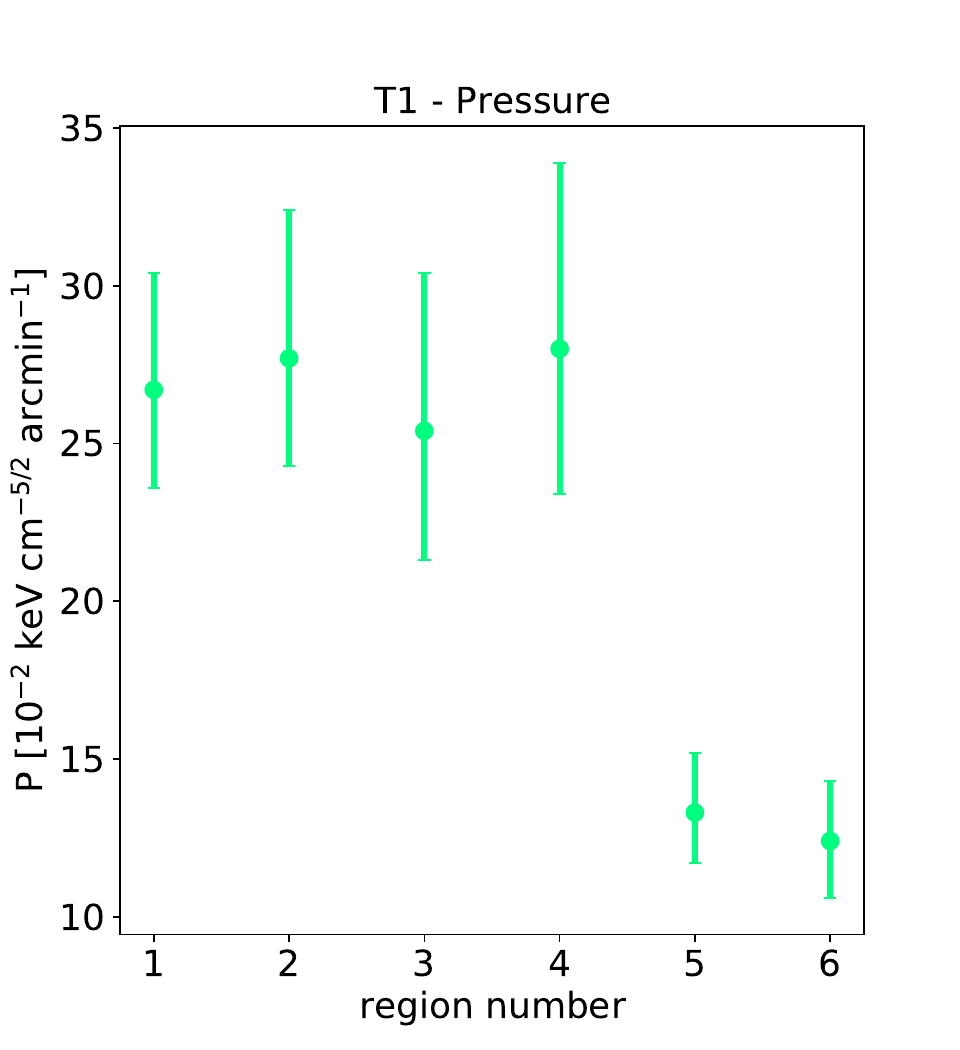}
 \includegraphics[width=0.245\textwidth]{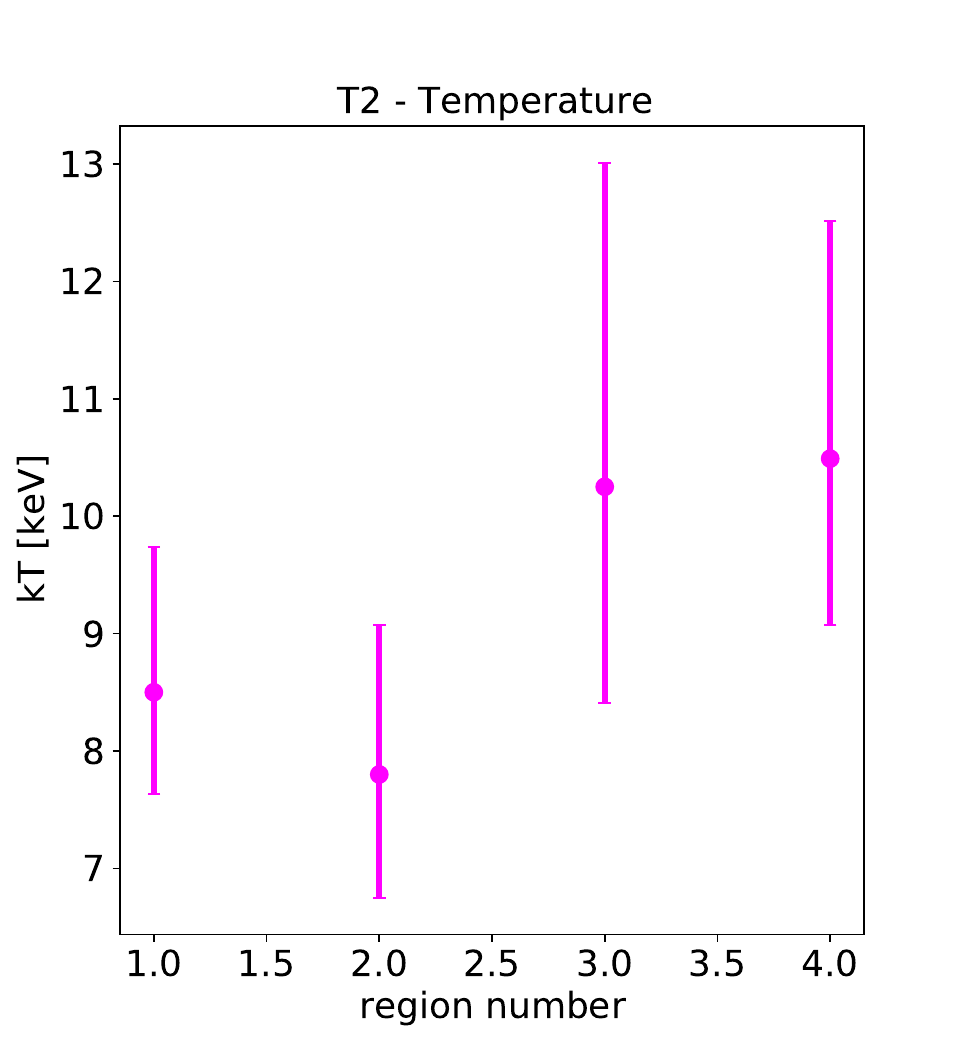}
 \includegraphics[width=0.245\textwidth]{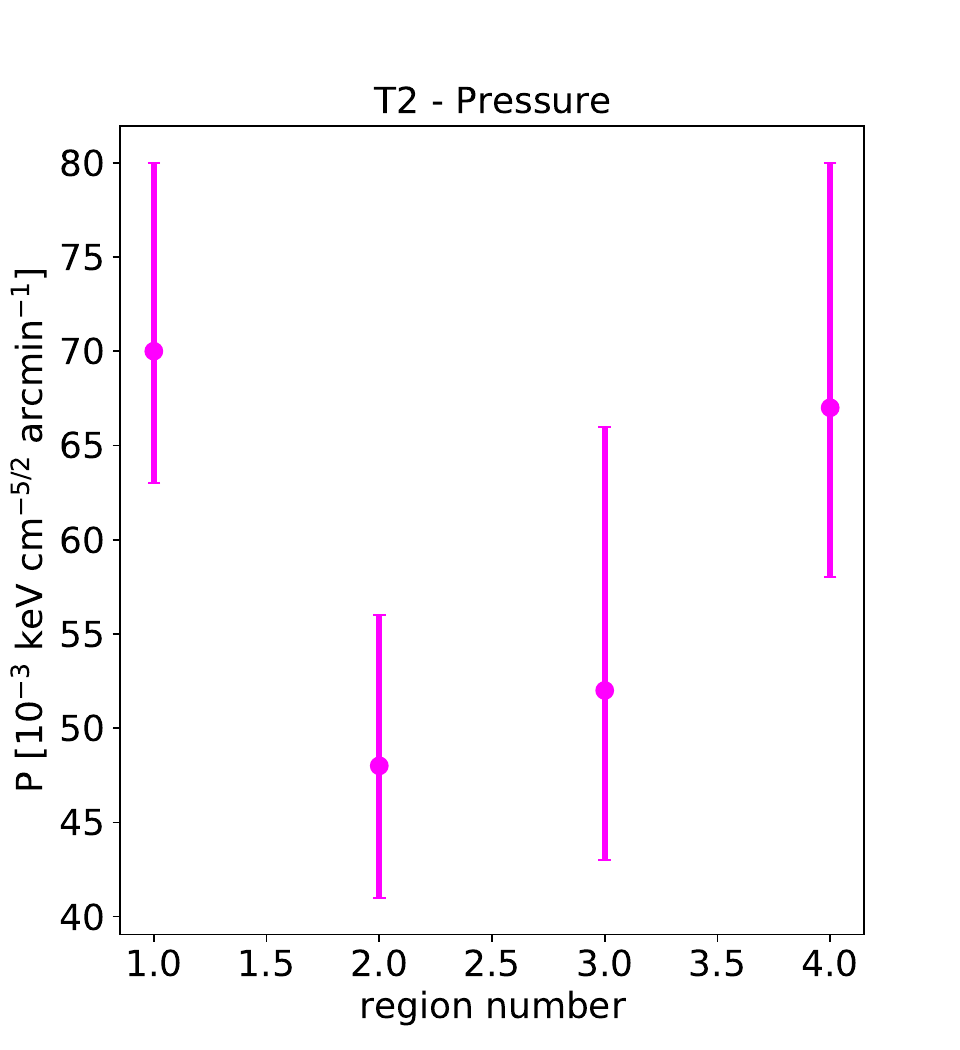}
	\smallskip
	
	\caption{Projected thermodynamic maps. \textit{Top left}: Temperature map from XMM-Newton photometric analysis covering the whole cluster \citep[adapted from][]{rossetti13}. \textit{Top right}: Temperature and pressure maps from \textit{Chandra} spectral analysis towards T1. Bin `4' roughly marks the transition T1-A/T1-B. \textit{Middle}: Temperature and pressure maps from \textit{Chandra} spectral analysis towards T2. Bin `1' includes T2-A and the choke, bin `2' includes T2-B. \textit{Bottom}: Temperature and pressure profiles for T1 (green) and T2 (magenta), as measured from the bins that are labelled in the corresponding maps with circles. The white dots in the \textit{Chandra} maps indicate point sources excluded from the analysis. The LOFAR HBA radio contours are overlaid in black.}
	\label{fig: thermomaps}
\end{figure*}

\begin{figure*}
	\centering
  \includegraphics[width=0.45\textwidth]{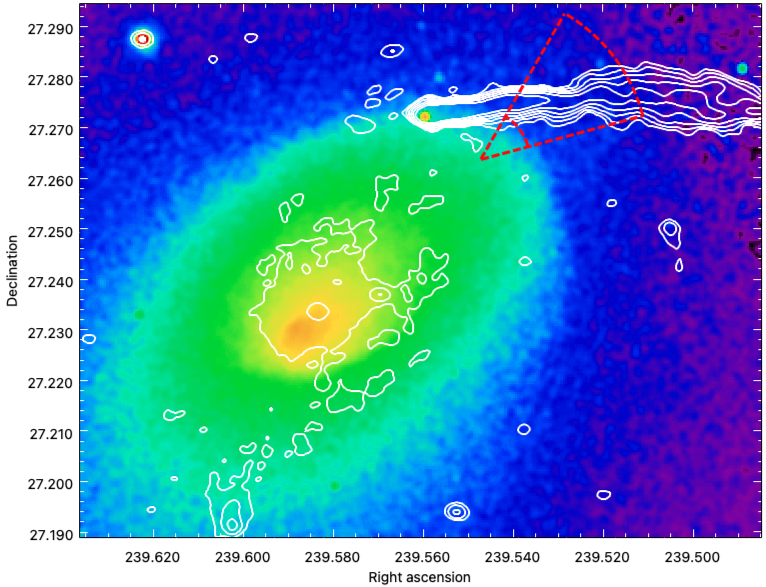}
 \includegraphics[width=0.45\textwidth]{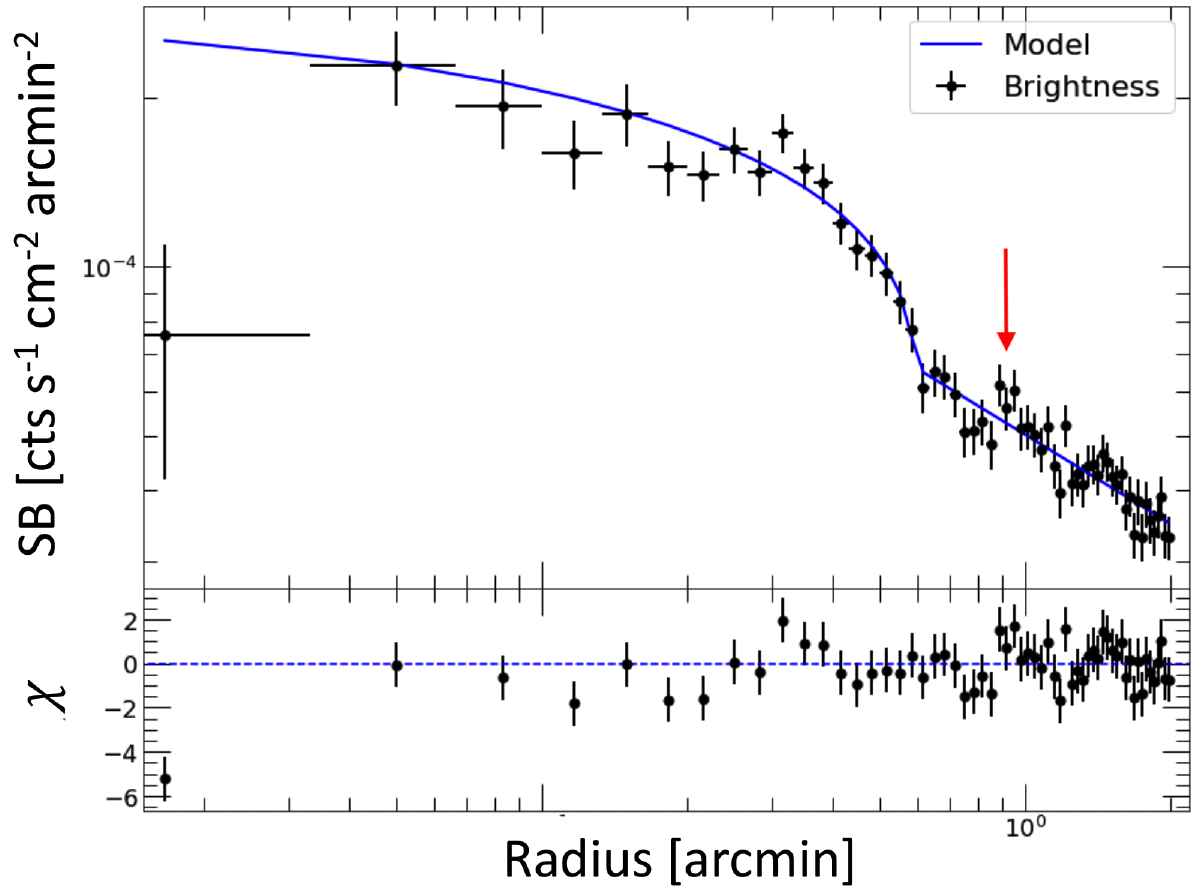}

	\smallskip
	
	\caption{X-ray surface brightness analysis in the direction of T1. \textit{Left}: \textit{Chandra} X-ray image in the 0.5-2 keV band with overlaid radio contours. The surface brightness was measured from the red sector in annuli of width $2''$. The fitted position of the discontinuity is shown within the sector, and is coincident with the NW cold front. \textit{Right}: Data points fitted with a broken power-law (blue line). The red arrow indicates the peak discussed in Sect. \ref{sect: Local ICM conditions}.  }
	\label{fig: CF}
\end{figure*}

Throughout this section we aim to search for direct evidence of interplay between radio emission and local ICM conditions. By means of XMM-Newton images in different energy bands, \cite{rossetti13} produced the projected temperature map that we show in the upper left panel of Fig. \ref{fig: thermomaps} as it provides a useful overview of the ICM temperature over the whole cluster. Nevertheless, this map does not allow us to probe the peripheral regions with sufficient resolution. In this respect, following the same procedure described in \cite{bruno23b}, we produced maps of projected thermodynamic quantities towards T1 and T2 by performing a spectral analysis with our \textit{Chandra} data. {\tt CONTBIN}\footnote{\url{ https://github.com/jeremysanders/contbin}} \citep{sanders06} was used to bin the 0.5-2 keV exposure-corrected \textit{Chandra} image in regions with a minimum ${\rm S/N}=40$. We extracted the spectra in each region of the event and blanksky files, subtracted the background from the ICM emission, and jointly fitted the resulting spectra in {\tt XSPEC} \citep{arnaud96xspec} with an absorbed thermal plasma component ({\tt phabs} $\times$ {\tt apec}) by fixing the Galactic hydrogen column density in the direction of A2142 and the ICM metal abundance to values of $N_{\rm H}=3.8\times10^{20} \; {\rm cm^{-2}}$ and $Z=0.28 \; Z_{\odot}$ \citep{markevitch00}. The Cash statistics \citep[Cstat;][]{cash79} was adopted for fitting. 

Each spectrum provides values of temperature $kT$ (in units of keV) and normalisation $\mathcal{N}$ (in units of ${\rm cm^{-5}}$), which is proportional to the squared numerical density integrated over the volume. We derived the (projected) pressure as
\begin{equation}
p = kT \times \left( \frac{\mathcal{N}}{A} \right)^{\frac{1}{2}}  \; \; \; \; \;  {\rm [keV \; cm^{-5/2} \; arcmin^{-1}]} \; \; \; ,
\label{p}
\end{equation}
where $A$ is the area of each region (in units of arcmin$^2$). By propagating errors, uncertainties on $p$ are computed as
\begin{equation}
\Delta p= p \sqrt{\left(\frac{\Delta kT}{k T}\right)^2  + \frac{1}{4} \left(\frac{\Delta \mathcal{N}}{\mathcal{N}}\right)^2}  \; \; \; .
\label{p_err}
\end{equation}    
In Fig. \ref{fig: thermomaps} we report the temperature and pressure maps in the direction of T1 (top right panels) and T2 (middle panels), respectively, and the corresponding profiles (bottom panels) as measured from regions that are labelled with green and magenta circles. While the chosen high ${\rm S/N}$ ensures accurate spectral fitting, the extraction regions are relatively large and limit our analysis to minimum spatial scales of $\sim 30$ kpc for T1 and $\sim 100$ kpc for T2.

For T1, we found drops in temperature and pressure from regions 4 to 5, which correspond to the transition from T1-A to T1-B. We further probed these trends by extracting and fitting the X-ray surface brightness with a broken power-law by means of {\tt pyproffit}\footnote{\url{https://github.com/domeckert/pyproffit}} \citep{eckert20}. As shown in Fig. \ref{fig: CF}, the profile is dominated by the presence of the prominent NW cold front, which we detect as a density jump of $C=1.80\pm 0.14$ (${\rm Cstat/dof}=45/49$), consistent with the value reported by \cite{rossetti13}. Furthermore, we observe a peak at a distance of $\sim 1'$ (indicated by the red arrow in the profile), which is co-spatial with the transition from T1-A to T1-B. This feature is likely responsible for the observed $kT$ and $p$ drops and is suggestive of local ICM compression, but we were not able to constrain its nature through fitting procedures. While our analysis is inconclusive, it is unlikely that such compression is driven by a shock, as its passage along the tail would have left signatures in the spectral index distribution with a discontinuity between T1-A and T1-B that we do not observe (Fig. \ref{fig: spixprofile}).

For T2, both $kT$ and $p$ profiles are roughly continuous within errors. Even though we notice a tentative (significance $\gtrsim 1\sigma$) drop in pressure between T2-A and T2-B, we cannot draw any solid conclusion on possible discontinuities from these plots. A complementary analysis (not shown) of the X-ray surface brightness profile in the direction of T2 with high spatial resolution (ranging from $2''$ to $12''$) is consistent with continuous trends of the thermodynamic properties. We notice that modelling of the background \citep[e.g.][]{gastaldello07,bartalucci14,botteon18a,bruno21} may provide more accurate spectral results rather than subtraction, especially for the cluster outskirts (which is the case of T2), but this is beyond the aim of the present work.

\section{Discussion}
\label{sect: Discussion}

In the previous sections we analysed the morphological and spectral properties of T1 and T2. In the following, we discuss possible scenarios explaining the observed features.

\subsection{On the discrepancy of RCCD analysis for T1}
\label{sect: On the discrepancy of RCCDs for T1}

Through the RCCD analysis in Sect. \ref{sect: Radiative ageing} we showed that the observed spectral distribution of T1 at low frequency ($\lesssim 300$ MHz) can be reproduced by the JP and TJP models under simple assumptions on the magnetic field (Fig. \ref{fig: CCP}). Nevertheless, an inconsistent spectral trend is measured at higher frequencies. In RCCDs, data points are expected to follow the same spectral shape in a pure radiative ageing scenario, regardless of the considered frequency pairs. Therefore, the observed inconsistency requires a deeper understanding. 

Offsets in the flux density scale and calibration artefacts can systematically shift the data points. The spectra of compact sources in the field do not reveal clear offsets in our datasets, and although images at 608 MHz show spurious emission around the tail ($\sim 4\sigma$ significance, likely generated by self-calibration errors), this is not driving the observed spectral distribution in the RCCD. A combination of projection effects and mixing of particles with different energies within each sampling region can broaden the intrinsic spectral distribution, mimicking  our observed trend. While this hypothesis is plausible, the discrepancy among RCCDs is preserved when using images produced with different weighting schemes and resolutions, changing the considered frequency pairs, and sampling with beam-size circular regions to reduce possible transverse mixing across the boxes. Physical phenomena, such as compression/expansion and re-energising, can also alter the spectral distribution in RCCDs. However, compression and expansion are frequency-independent processes, while standard re-energising mechanisms via shock and turbulence predominantly affect the low-energy particles emitting at low-frequency. In this respect, unknown exotic processes favouring the re-acceleration of high-energy against low-energy particles would be necessary to solve the RCCD discrepancy. 

In summary, we did not identify any obvious systematic effects or physical conditions explaining the observed spectral discrepancy. The possible role of subtle calibration effects should be investigated through independent reprocessing and methods. Mixed (either intrinsic or projected) energy distributions are plausible, but difficult to be further disentangled.

\subsection{Dynamics of T1}
\label{sect: Velocity of T1}

\begin{figure}
	\centering
 \includegraphics[width=0.49\textwidth]{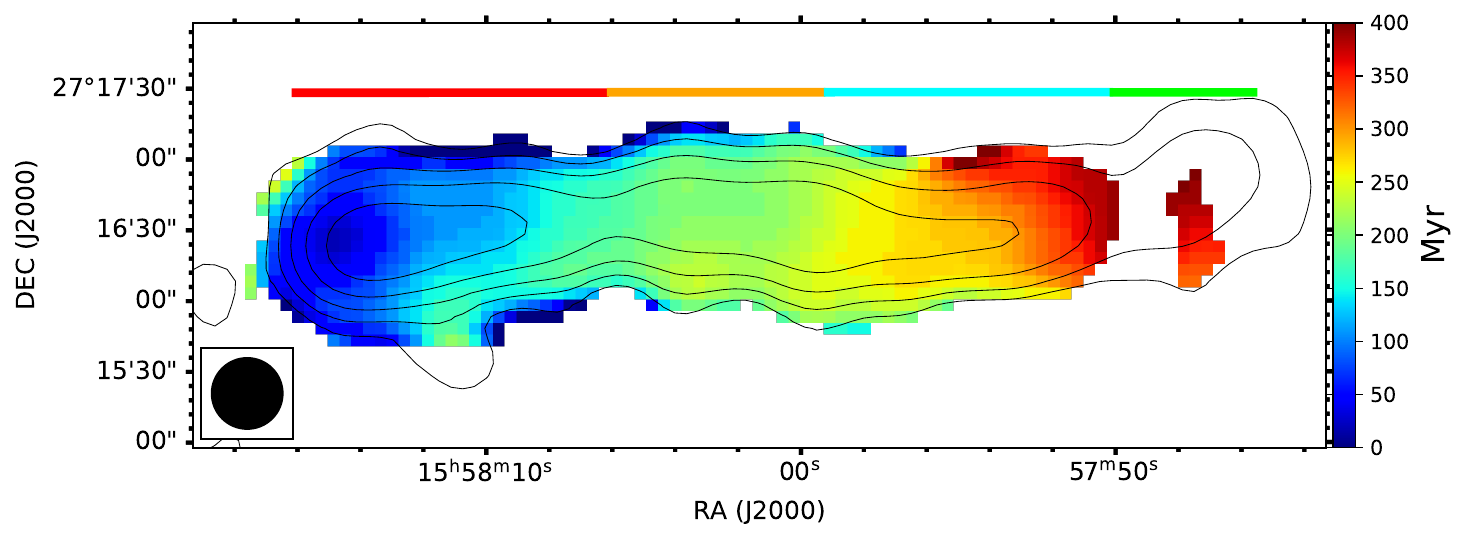}
  \includegraphics[width=0.49\textwidth]{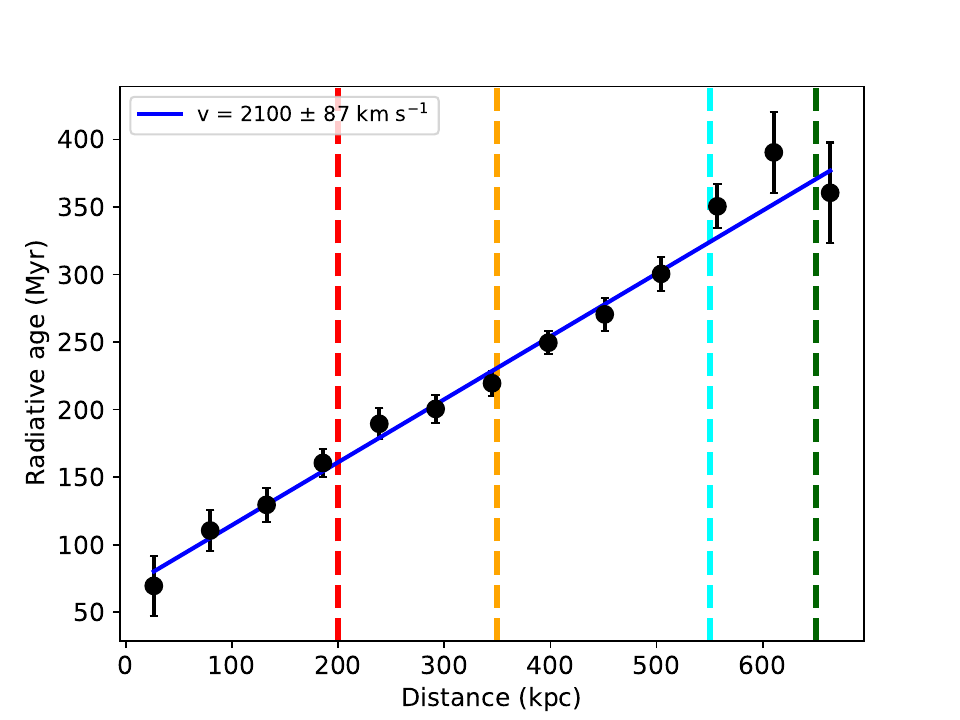}

	\smallskip
	
	\caption{Radiative age of T1. \textit{Top}: Radiative age map at $30''$ obtained by fitting radio images at 50, 143, and 323 MHz with the TJP model ($\alpha_{\rm inj}=0.51$, $B_{\rm 0}=2.2 \; {\rm \mu G}$). Typical errors are $\sim 20$ Myr. The sub-regions are indicated by horizontal bars. \textit{Bottom}: Age profile measured in boxes of beam-size width. The boundaries of the sub-regions are indicated with dashed vertical lines. The blue solid line is the result of a linear fit to the data points.  }
	\label{fig: vel-age}
\end{figure}

In Sect. \ref{sect: On the discrepancy of RCCDs for T1} we discussed possible solutions to the RCCD discrepancy of T1. Throughout this section, we only consider the low frequency RCCD in Sect. \ref{sect: Radiative ageing}, as we showed that the measured spectral trend is predicted by the JP and TJP ageing curves.

Following up on the results of the RCCD, we performed a pixel-by-pixel fitting of a TJP spectrum to our 50, 143, and 323 MHz images at $30''$-resolution by means of {\tt BRATS}, and obtained the radiative age map of T1 that is shown in the top panel of Fig. \ref{fig: vel-age}. The lower panel of Fig. \ref{fig: vel-age} reports the corresponding age profile. Fitted ages are in the range $t\sim 50-150$ Myr in T1-A, $t\sim 150-200$ Myr in T1-B, $t\sim 200-300$ Myr in T1-C, and $t\sim 350$ Myr in T1-D, with typical errors of $\sim 20$ Myr. 

We fitted the age profile data points with a linear relation, which provides the tangential velocity ${\rm v}_{\rm sky}$ (in the plane of the sky) under the assumptions that T1 is moving at a constant velocity, the radiative age coincides with the dynamical age (implying ${\rm v}_{\rm sky}\sim L/t$), and there are not compression/expansion, re-energising, or bulk motions. We obtained a fitted velocity of ${\rm v}_{\rm sky}=2100 \pm 87 \; {\rm km \; s^{-1}}$, which has to be considered as a lower limit because the value of $B_{\rm 0}$ that we used provides an upper limit to the radiative age and we ignored projection effects. The peculiar radial velocity (along the line of sight) of T1 is computed from the spectroscopic redshift of its host galaxy ($z_{\rm T1}=0.0954$) and that of A2142 ($z_{\rm A2142}=0.0894$) in units of the light speed $c$ as \citep[e.g.][]{davis&scrimgeour14}:  
\begin{equation}
    {\rm v}_{\rm los} = c \left(\frac{z_{\rm T1}-z_{\rm A2142}}{1+ z_{\rm A2142}}\right) \; \; \sim  1650\; {\rm km \; s^{-1}}  \;.
\end{equation}
The inferred velocity components provide constraints on the deprojected dynamics of T1, as we derive a 3D velocity of ${\rm v}=\sqrt{{\rm v}_{\rm los}^2 + {\rm v}_{\rm sky}^2}=2670 \; {\rm km \; s^{-1}}$ and a viewing angle of $i = \arctan{\left({\rm v}_{\rm sky}/{\rm v}_{\rm los}\right)} =52^{\rm o}$. For a comparison, the radial velocity dispersion of A2142 is $\sigma_{\rm A2142} = 1193 \; {\rm km \; s^{-1}}$ \citep{munari14}, meaning that ${\rm v}\sim  2.2 \sigma_{\rm A2142}$, which is a reasonable value for head-tail galaxies.

\subsection{Transitions along T1}
\label{sect: On the transitions along T1}

\begin{figure*}
	\centering
 \includegraphics[width=0.49\textwidth]{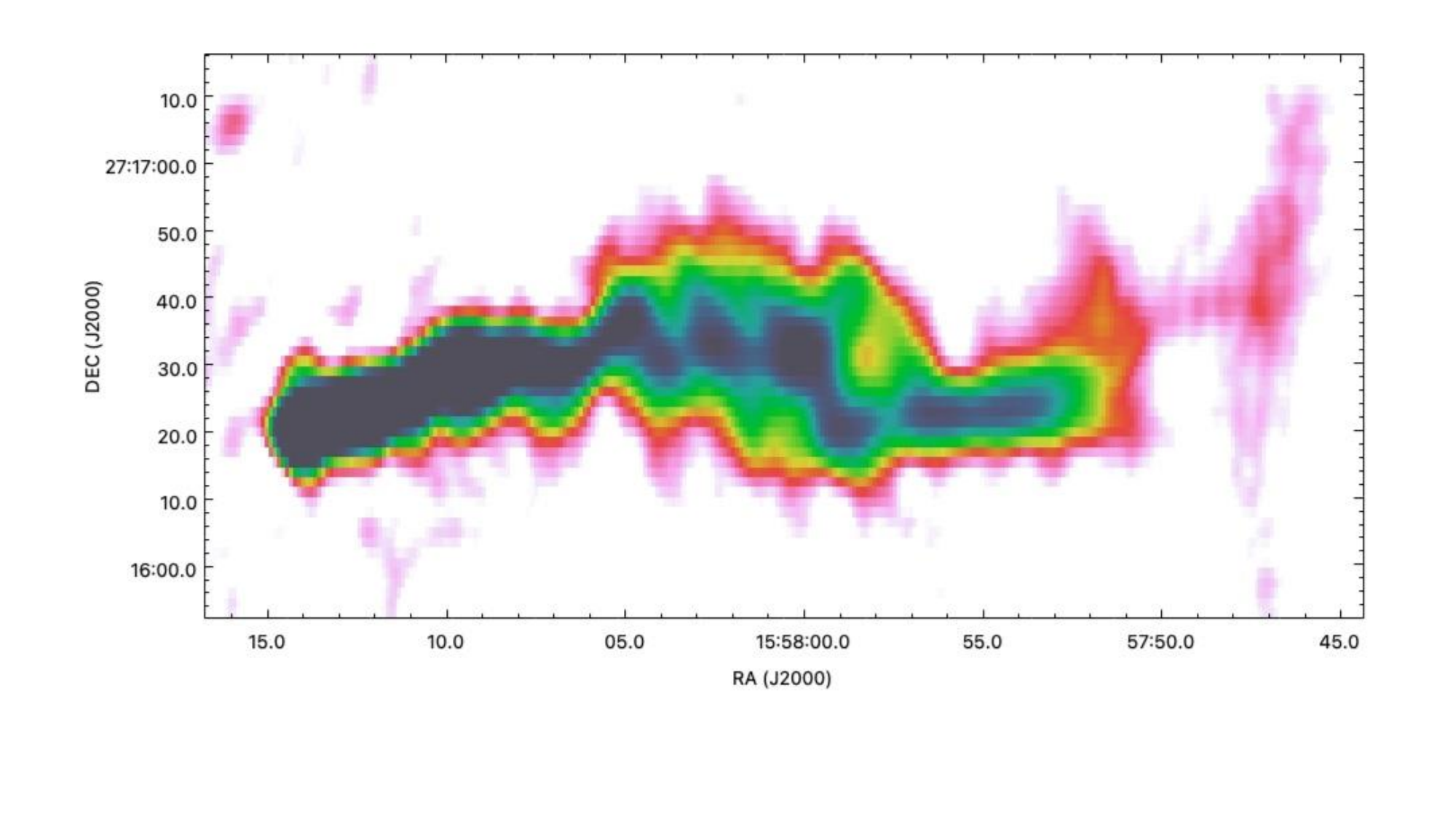}
  \includegraphics[width=0.45\textwidth]{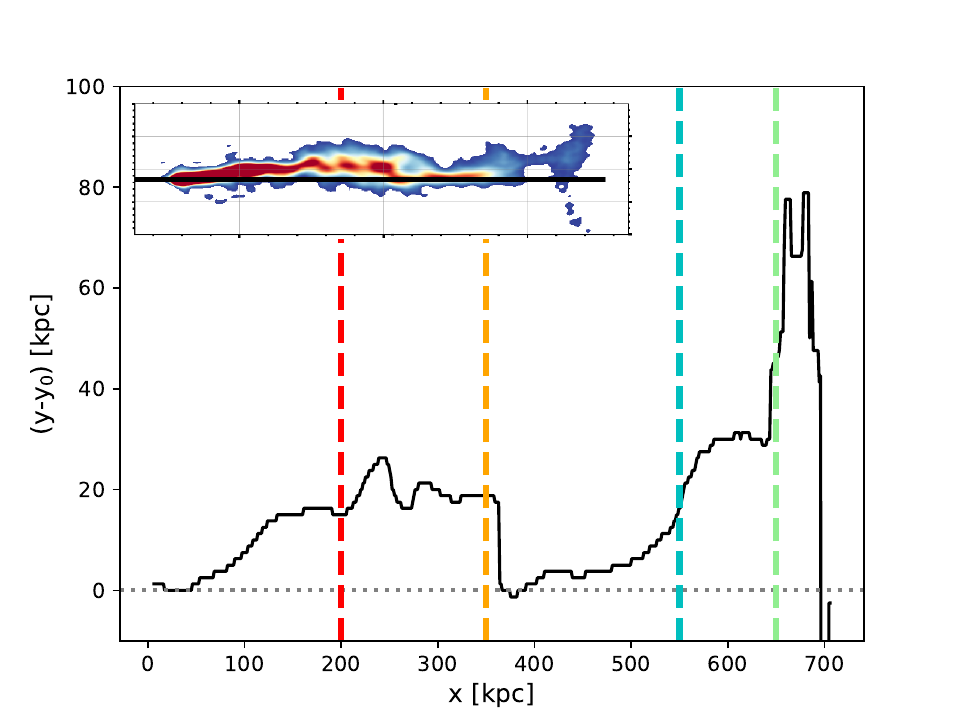}

	\smallskip
	
	\caption{Transitions along T1. \textit{Left}: 143 MHz image (Fig. \ref{fig: radio images full res}) stretched by a DEC to RA axis ratio of 2.5 to visually emphasise the transitions of each sub-region. \textit{Right}: Position (on the $y$-axis) of the emission peak along the tail ($x$-axis) with respect to the zero-level horizontal line shown in the inset (in black) and passing through the radio core.}
	\label{fig: transitionT1}
\end{figure*}

In Sect. \ref{sect: Velocity of T1} we demonstrated that a simple TJP ageing scenario with a constant velocity is a good and physically reasonable representation of the global emission of T1. In line with this scenario, the absence of spectral flattening (Sect. \ref{sect: spectral properties 2}) rules out ongoing large-scale re-acceleration processes, and our analysis in Sect. \ref{sect: Local ICM conditions} does not conclusively support evidence of interaction with the ICM (Fig. \ref{fig: thermomaps}). However, it is evident that local phenomena of unclear nature are taking place within its sub-regions. Indeed, we observed several major discontinuities in the surface brightness profile of T1 (Fig. \ref{fig: radio-x profile}), which we further emphasise in Fig. \ref{fig: transitionT1}. In this section we briefly speculate on the possible origin of such discontinuities.

In T1-B, the tail shows a distinct set of wiggles, but they disappear as soon as the tail fades abruptly at the beginning of T1-C. A possible explanation to the wiggles is the precession of the jets before their downstream bending, but this scenario is unlikely. Indeed, we do not detect similar wiggles in T1-A, and such scenario would thus require the  unlikely condition of sudden stabilising of the jets in a more recent epoch than the age of the plasma in T1-B.

Plausibly, the wiggles result from the development of Kelvin-Helmholtz (KH) instabilities, which are caused by the velocity difference across the interface between the jets and surrounding medium. These may be triggered either in T1-A, which is unresolved in our images, or at the transition in jet properties at the beginning of T1-B. With reference to the latter possibility, we recall the tantalising presence of a galaxy visible in the transition from T1-A to T1-B (Fig. \ref{fig: substructures}). We stress that the growth rate of KH instabilities depends on the magnitude of the velocity gradient and on the effective viscosity and magnetic fields in the interface. Therefore, for the development of such instabilities, the jets have to move fast enough through the ICM, and the displacement of the jets downstream from the nucleus should result from the combination of the motion of the host galaxy through the ICM and the jet flow \citep{oneill19a}. As the jet decelerates and comes to rest in the ICM, the instabilities are not efficiently driven anymore, resulting in a turbulent tail which starts mixing with the ICM; this is a promising scenario for the plasma in T1-C and T1-D. Possibly, some local re-energising processes are occurring in T1-D, where the brightness increases and slight departures from the constant velocity scenario are found (Fig. \ref{fig: vel-age}). Higher resolution images of T1-A could further shed light on the onset of the instabilities.

\subsection{Nuclear emission of T2}
\label{sect: On the core region of T2}

In this section we discuss the properties of the host and core of T2 at different wavelengths. These are important to probe the overall origin of the source, as outlined in Sect. \ref{sect: On the origin of T2}. 

The host of T2 is an elliptical galaxy with no optical emission lines (Sect. \ref{sect: The galaxy cluster Abell 2142}). It emits in the soft X-ray band, but the analysis of its spectrum (see details in Appendix \ref{app: X-ray spectrum of T2 host}) indicates that the observed emission comes from thermal gas. Therefore, the optical and X-ray spectra indicate that the accretion of the black hole is either currently off-state or radiatively inefficient.

Useful information is provided by the comparison of the core prominence at radio frequencies, which is ${\rm CP}=S^{\rm T2-D}_{1400}/S^{\rm T2}_{150}= 2\times 10^{-4}$, when considering the ratio of 1.4 GHz flux density of the core and the 150 MHz total flux density, and the total 1.4 GHz radio power in logarithmic scale $\log_{10}{P_{1400}}=23.9 \; {\rm W \; Hz^{-1}} $. These values are particularly low \citep[see e.g.][for works using these quantities or their combinations]{giovannini88,deruiter90,jurlin21} and suggest that T2 is a remnant radio galaxy. Similar values were reported by \cite{riseley22b} for a remnant radio galaxy with a tailed morphology, whereas, for a comparison, we obtained ${\rm CP}= 4\times 10^{-2}$ and $\log_{10}{P_{1400}}=24.4 \; {\rm W \; Hz^{-1}}$ for T1, in line with active FRI galaxies.

\subsection{Origin of T2}
\label{sect: On the origin of T2}

As highlighted in the previous sections, the global morphology of T2 is consistent with that of a HT galaxy, but it also exhibits unusual features. Moreover, its core prominence (Sect. \ref{sect: On the core region of T2}) suggests that T2 is a remnant tailed galaxy. During its lifetime, the AGN may have experienced either a single or multiple outbursts, which we invoke to discuss possible scenarios explaining the radiatively old components of T2 (Sects. \ref{sect: spectral properties 1}, \ref{sect: spectral properties 2}) and some peculiar features.

In the context of a single AGN outburst, the core launched two radio jets in opposite directions, which were then bent by the ram pressure, and originated a tail as in classical HT galaxies. Afterwards, regions of the tail at different distances from the core passed through diverse ICM phases during the infalling towards the cluster centre. Such gas phases reshaped the sub-regions of T2 into the present T2-A and T2-B due to density and pressure gradients. This scenario is supported by the observed radial steepening of the spectral index (Fig. \ref{fig: spixprofile}), in line with standard HT galaxies, but it relies on specific conditions of the ICM. Indeed, the relativistic plasma should be tremendously compressed at the location of the choke to explain the abrupt separation between T2-A and T2-B. Even though our projected thermodynamic maps (Fig. \ref{fig: thermomaps}) do not provide solid conclusions owing to a combination of poor resolution and low ICM counts in the cluster outskirts, the existence of a thin layer where the thermal pressure is dramatically enhanced is disfavoured and appears unlikely.

A multiple AGN outburst scenario is more plausible and can be reconciled with the choke without invoking thermal pressure exerted by particular ICM layers. We can assume that a first AGN outburst was triggered when the host of T2 was far from its present position (beyond T2-B in projection). During the infall, the radio galaxy developed a tail, which we observe today as T2-B, and the core switched off at the location of the present choke. In proximity of its present location, the galaxy experienced a second AGN outburst, which would be responsible for the formation of T2-A. This scenario naturally explains the double-peaked surface brightness profile of T2 (Fig. \ref{fig: radio-x profile}) and the choke, and is in line with the complex spectral distribution in the RCCDs that deviates from a single injection event\footnote{The complexity of the observed spectral distribution would require ad-hoc ageing modelling that we did not attempt in this work. A useful starting point might be the KGJP \citep{komissarov&gubanov94} model, as it assumes a continuous injection of fresh electrons for a certain period, followed by a passive ageing.}. As a first approximation, the integrated spectra of T2-A and T2-B (Fig. \ref{fig: spettro}) suggest similar break frequencies that yield consistent radiative ages for a uniform magnetic field. This might be indicative of a short period between the two phases, possibly triggered by ram pressure itself \citep[e.g.][]{poggianti21,peluso22}, but firm conclusions cannot be drawn (see e.g. \citealt{rudnick02} for a discussion on complex spectra). 

In the context of the two scenarios described above, T2-A may represent the superposition of the two radio lobes in projection. The remnant scenario indicated by the core prominence analysis is supported by the non-detection of radio jets, which suggests that they are switched off on large scales (at least down to $\sim 4$ kpc). However, a completely different interpretation is also viable. Indeed, the conical morphology of the light bulb and the sharp transition at its edge (Fig. \ref{fig: substructures}) are reminiscent of the structure of FRI-type jets reported in \cite{laing&bridle14} (see e.g. the case of M84 in their Fig. 3), which results from their deceleration on kiloparsec-scales. In other words, T2-A itself could potentially be a radio jet caught in its initial deceleration phase throughout the ambient medium. We notice that the sharp edge of T2-A suggests that the source approximately lies in the plane of the sky, and implying that it has a one-sided jet. This is unlikely, therefore a possibility is that a combination of deceleration of the two jets and their downstream bending due to ram pressure during the infall is ongoing. In summary, the nature of T2-A remains unconfirmed, and higher-resolution radio data towards the nuclear regions could be helpful for further investigation.

Regardless of a single or double AGN outburst, the plume is interpreted as the oldest part of the tail. The aged relativistic plasma progressively expanded, diffused, mixed with the thermal ICM, and originated T2-C. Our high-resolution spectral index maps (Fig. \ref{fig: spixmap}) show flatter (but still ultra-steep) patches that deviate from the expected radial steepening of the tail. The morphology and spectral behaviour of T2-C may be indicative of some kind of interplay between thermal and non-thermal components and/or trace substructures of the magnetic field.

\section{Summary and conclusions}
\label{sect: Summary and conclusions}

In this work we reported on the study of two tailed radio galaxies, T1 and T2, in the galaxy cluster A2142. These targets show interesting morphological features that are suggestive of a complex dynamics and interaction with the thermal ICM. By means of LOFAR, uGMRT, VLA, and MeerKAT radio data, we provided a detailed spectral analysis of T1 and T2. Auxiliary  \textit{Chandra} X-ray observations were used to investigate the local conditions of the ICM in the direction of the targets. In this section we summarise our results and discuss future prospects.

T1 (Fig. \ref{fig: radio images full res}) is a long HT galaxy extending for $\sim 700$ kpc and exhibiting clear surface brightness fluctuations and discontinuities that define four sub-regions (Figs. \ref{fig: substructures}, \ref{fig: transitionT1}). A single power-law of slope $\alpha=0.87\pm 0.01$ can fit the flux density measurements of T1 from 50 to 1810 MHz, but spectral breaks are found within its sub-regions (Fig. \ref{fig: spettro}). The overall spectral index profile steepens with the increasing distance from the core (Fig. \ref{fig: spixprofile}). In the low-frequency (50-143-323 MHz) regime that we considered, standard ageing models (JP, TJP) can well reproduce the observed spectral behaviour of T1 (Fig. \ref{fig: CCP}). Under simple assumptions on the magnetic field, we produced a radiative age map (Fig. \ref{fig: vel-age}), which we used to constrain the tangential velocity of the target (Sect. \ref{sect: Velocity of T1}). We computed a lower limit on the 3D velocity of the galaxy of ${\rm v}>2670 \; {\rm km \; s^{-1}}$, which is a factor of $\sim2$ higher than the radial velocity dispersion of the cluster. Although we showed that a pure radiative ageing scenario and a constant velocity are reasonable approximations, each sub-region shows signs of a complex phenomenology. Indeed, we detected spots, labelled as wiggles due to their oscillating pattern, where the surface brightness is locally enhanced (Fig. \ref{fig: radio-x profile}), sharp transitions along the tail, and fossil emission visible only at the lowest frequencies. All these features might be connected with the development and evolution of KH instabilities along the tail at different distances from the core. Moreover, we found that the spectral shape at high frequencies is surprisingly inconsistent with that at low frequencies (Fig. \ref{sect: spectral properties 2}), hinting at subtle calibration effects, mixing of older and younger emitting electrons, or peculiar physical conditions (Sect. \ref{sect: On the discrepancy of RCCDs for T1}).

T2 (Fig. \ref{fig: radio images full res 2}) is a HT galaxy extending for $\sim 400$ kpc and featuring morphological properties that are unusual for tailed sources. The whole structure of T2 can be decomposed into three sub-regions (Fig. \ref{fig: substructures}) identified by distinct peaks in the surface brightness profile (Fig. \ref{fig: radio-x profile}). The radio core is not coincident with one of these peaks, and our spectral analysis suggests that T2 is a remnant radio galaxy. While we observe a spectral steepening along the tail that is in line with classical HT galaxies (Fig. \ref{fig: spixprofile}), none of the standard (single injection) ageing models can reproduce the spectral measurements (Fig. \ref{fig: CCP}), and non-trivial ad-hoc modelling would be required. The most intriguing feature of T2 is the choke, which is a sharp depletion of radio emission that separates the sub-regions T2-A and T2-B. We discussed possible scenarios to explain such feature, which involve either a single or double AGN outburst event during the infall towards the cluster centre (Sect. \ref{sect: On the origin of T2}). The simplest scenario that we proposed assumes two AGN outbursts that produced the tailed morphology of T2-B and the light bulb structure of T2-A, respectively. The X-ray analysis of the ICM disfavours exceptional thermal compression at the location of the choke (Sect. \ref{sect: Local ICM conditions}), but deeper data in this direction are necessary to definitely confirm our findings.

In conclusion, our work further highlights the increasing complexity in the phenomenology of tailed galaxies that has been shown by recent studies with the advent of deep and high-fidelity images from low ($\sim 100$ MHz) to high ($\sim 1$ GHz) frequencies. These works are providing valuable constraints on physical parameters, but theoretical modelling and numerical simulations are also necessary to shed light on the complex mechanisms that shape the properties of radio galaxies in clusters, such as re-energising processes, hydrodynamical instabilities, plasma mixing, magnetic field structure, and multiple AGN outbursts. As a follow-up work, we suggest the exploitation of LOFAR HBA data acquired with the international stations, which are available in the archive as part of the observations (with the core and remote stations) analysed here. These sensitive data could provide an insightful view of the sub-regions of T1 and T2 down to spatial scales of $\sim 500$ pc,  possibly providing information on the onset of instabilities along T1 and the presence of jets and the choke in T2. Additionally, polarisation studies at gigahertz frequencies with MeerKAT are ongoing and will be used to constrain the structure of the magnetic field across T1 and T2.

\begin{acknowledgements}
We thank the referee for their comments and suggestions. M.B. acknowledges support from the agreement ASI-INAF n. 2017-14-H.O and from the PRIN MIUR 2017PH3WAT 'Blackout'. A.I. acknowledges the European Research Council (ERC) programme (grant agreement No. 833824, PI B. Poggianti), and the INAF founding program 'Ricerca Fondamentale 2022' (project 'Exploring the physics of ram pressure stripping in galaxy clusters with Chandra and LOFAR', PI A. Ignesti). C.J.R. acknowledges financial support from the ERC Starting Grant ‘DRANOEL’, number 714245. M.R., F.G., and G.B. acknowledge support from INAF mainstream project 'Galaxy Clusters Science with LOFAR'. A.B. acknowledges financial support from the European Union - Next Generation EU. R.J.vW. acknowledges support from the ERC Starting Grant ClusterWeb 804208. D.V.L acknowledges support of the Department of Atomic Energy, Government of India, under project No. 12-R\&D-TFR-5.02-0700. The research leading to these results has received funding from the European Unions Horizon 2020 Programme under the AHEAD project (grant agreement No. 654215).    

LOFAR \citep{vanhaarlem13} is the Low Frequency Array designed and constructed by ASTRON. It has observing, data processing, and data storage facilities in several countries, which are owned by various parties (each with their own funding sources), and that are collectively operated by the ILT foundation under a joint scientific policy. The ILT resources have benefited from the following recent major funding sources: CNRS-INSU, Observatoire de Paris and Universit\'e d’Orl\'eans, France; BMBF, MIWF- NRW, MPG, Germany; Science Foundation Ireland (SFI), Department of Business, Enterprise and Innovation (DBEI), Ireland; NWO, The Netherlands; The Science and Technology Facilities Council, UK; Ministry of Science and Higher Education, Poland; The Istituto Nazionale di Astrofisica (INAF), Italy. This research made use of the Dutch national e-infrastructure with support of the SURF Cooperative (e-infra 180169) and the LOFAR e-infra group. The J\"ulich LOFAR Long Term Archive and the German LOFAR network are both coordinated and operated by the J\"ulich Supercomputing Centre (JSC), and computing resources on the supercomputer JUWELS at JSC were provided by the Gauss Centre for Supercomputing e.V. (grant CHTB00) through the John von Neumann Institute for Computing (NIC). This research made use of the University of Hertfordshire high-performance computing facility and the LOFAR-UK computing facility located at the University of Hertfordshire and supported by STFC [ST/P000096/1], and of the Italian LOFAR IT computing infrastructure supported and operated by INAF, and by the Physics Department of Turin University (under an agreement with Consorzio Interuniversitario per la Fisica Spaziale) at the C3S Supercomputing Centre, Italy. This research made use of the HOTCAT cluster \citep{bertocco20,taffoni20} at Osservatorio Astronomico di Trieste. The National Radio Astronomy Observatory is a facility of the National Science Foundation operated under cooperative agreement by Associated Universities, Inc. We thank the staff of the GMRT that made these observations possible. GMRT is run by the National Centre for Radio Astrophysics of the Tata Institute of Fundamental Research. The MeerKAT telescope is operated by the South African Radio Astronomy Observatory, which is a facility of the National Research Foundation, an agency of the Department of Science and Innovation. The scientific results reported in this article are based on observations made by the Chandra X-ray Observatory data obtained from the Chandra Data Archive. This research has made use of SAOImageDS9, developed by Smithsonian Astrophysical Observatory \citep{ds9}. This research has made use of the VizieR catalogue access tool, CDS, Strasbourg Astronomical Observatory, France (DOI: 10.26093/cds/vizier). This research made use of APLpy, an open-source plotting package for Python \citep{robitaille&bressert12APLPY}, Astropy, a community-developed core Python package for Astronomy \citep{astropycollaboration13,astropycollaboration18}, Matplotlib \citep{hunter07MATPLOTLIB}, Numpy \citep{harris20NUMPY}. 

\end{acknowledgements}

\bibliographystyle{aa}
\bibliography{bibliografia}

\begin{appendix}

\section{Spectral index error maps}
\label{app: Spectral index error maps}

\FloatBarrier

\begin{figure*}[!htbp]
	\centering
\includegraphics[width=0.45\textwidth]{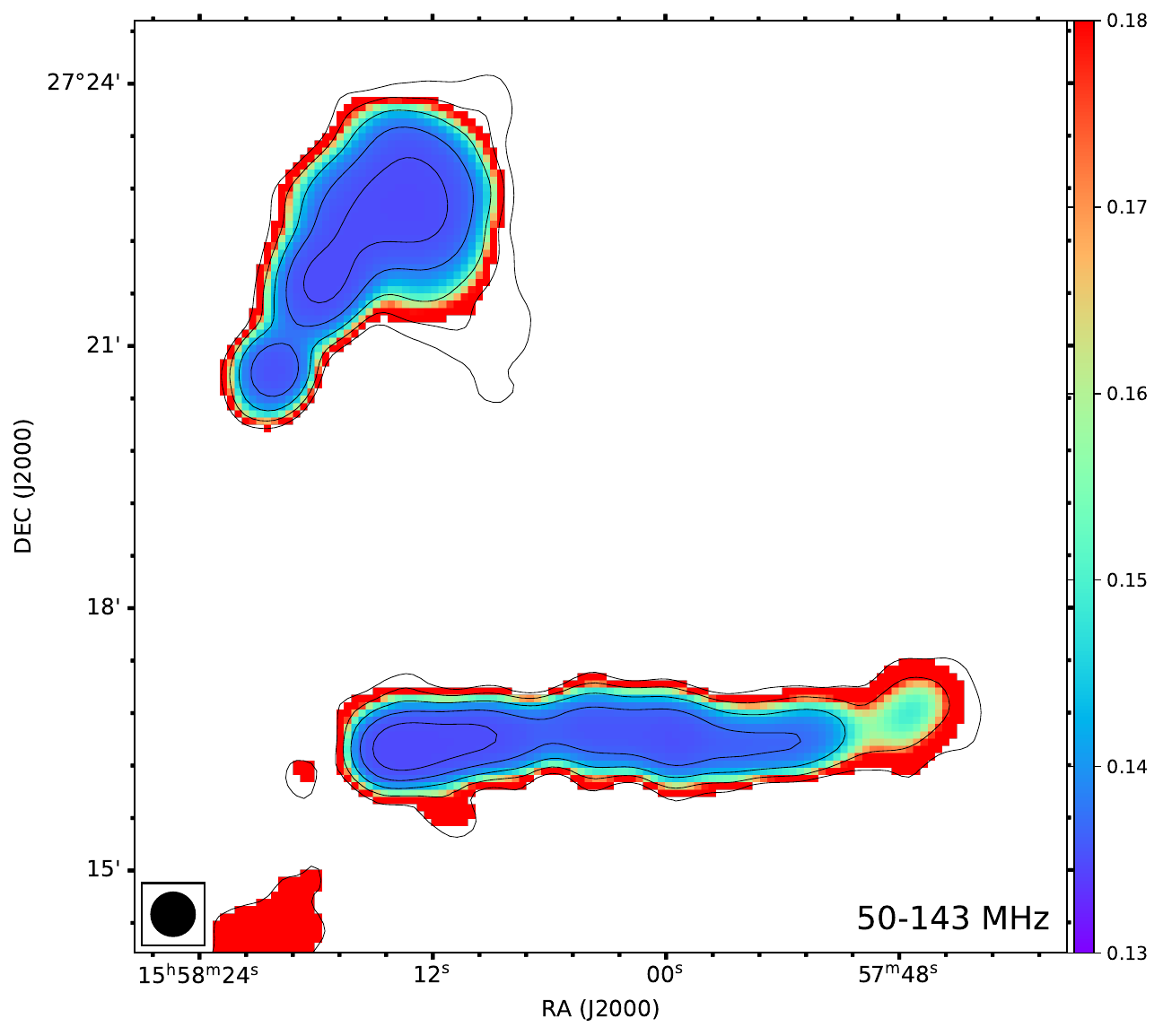}
\includegraphics[width=0.45\textwidth]{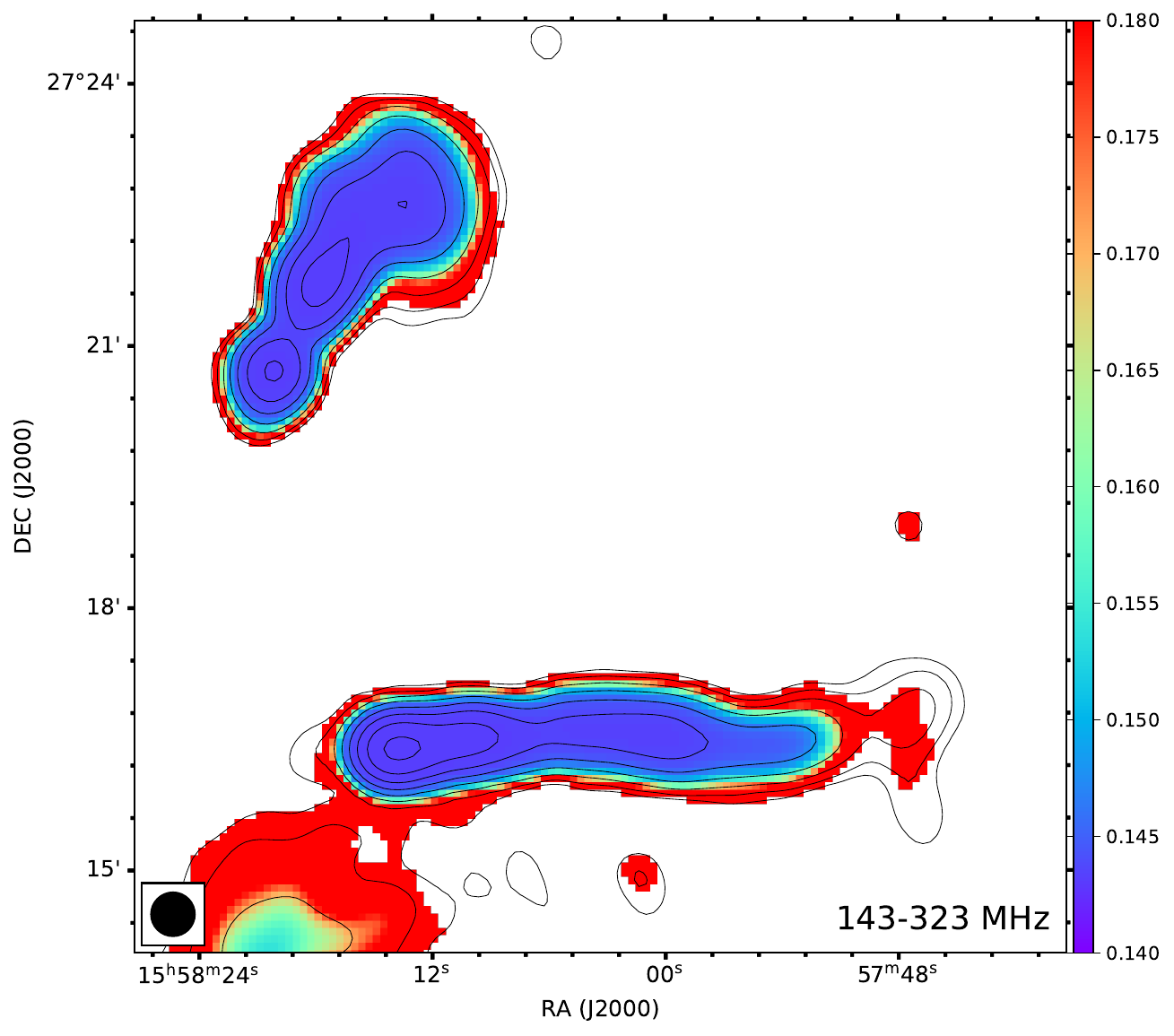}
 \includegraphics[width=0.45\textwidth]{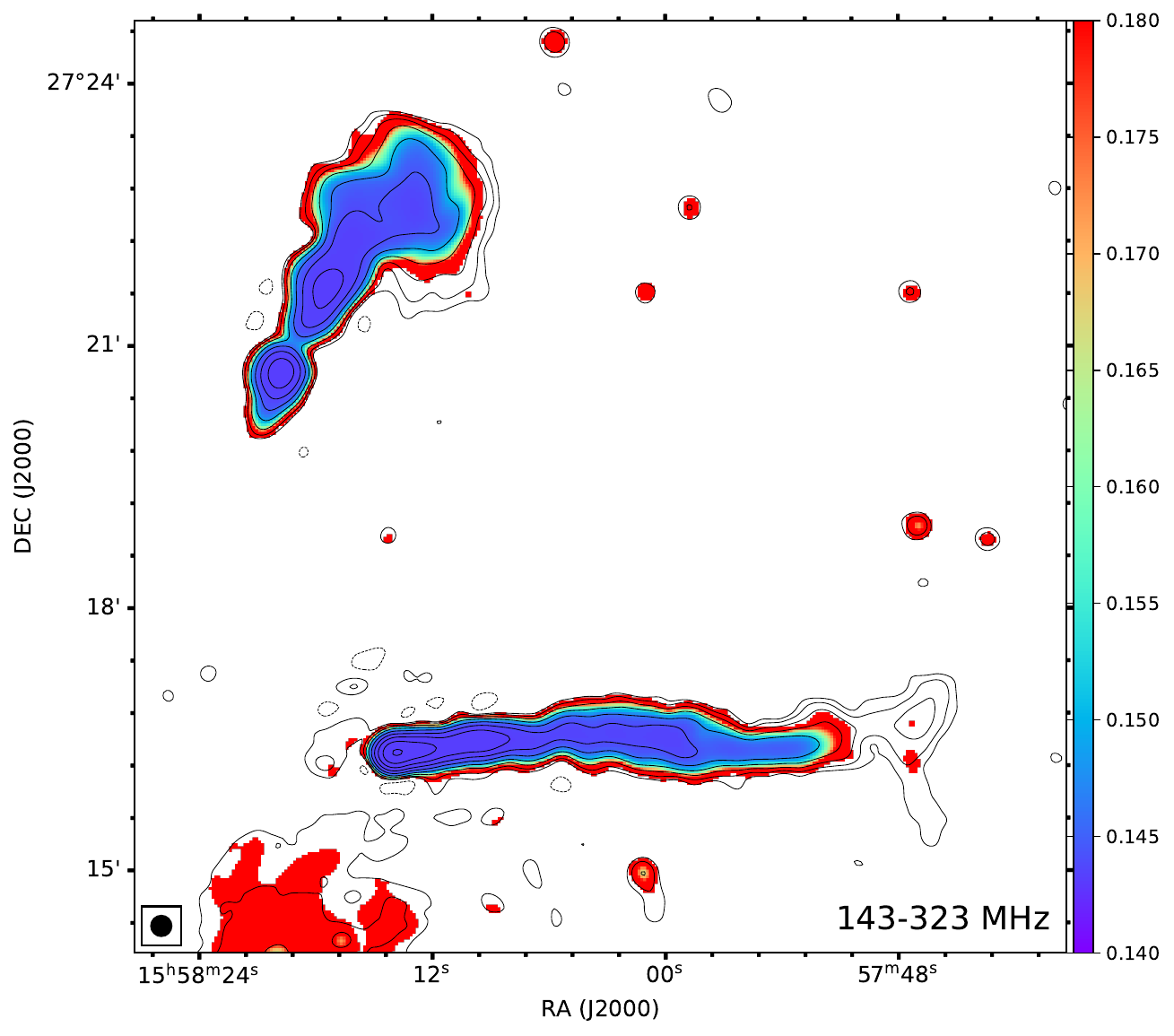}
 \includegraphics[width=0.45\textwidth]{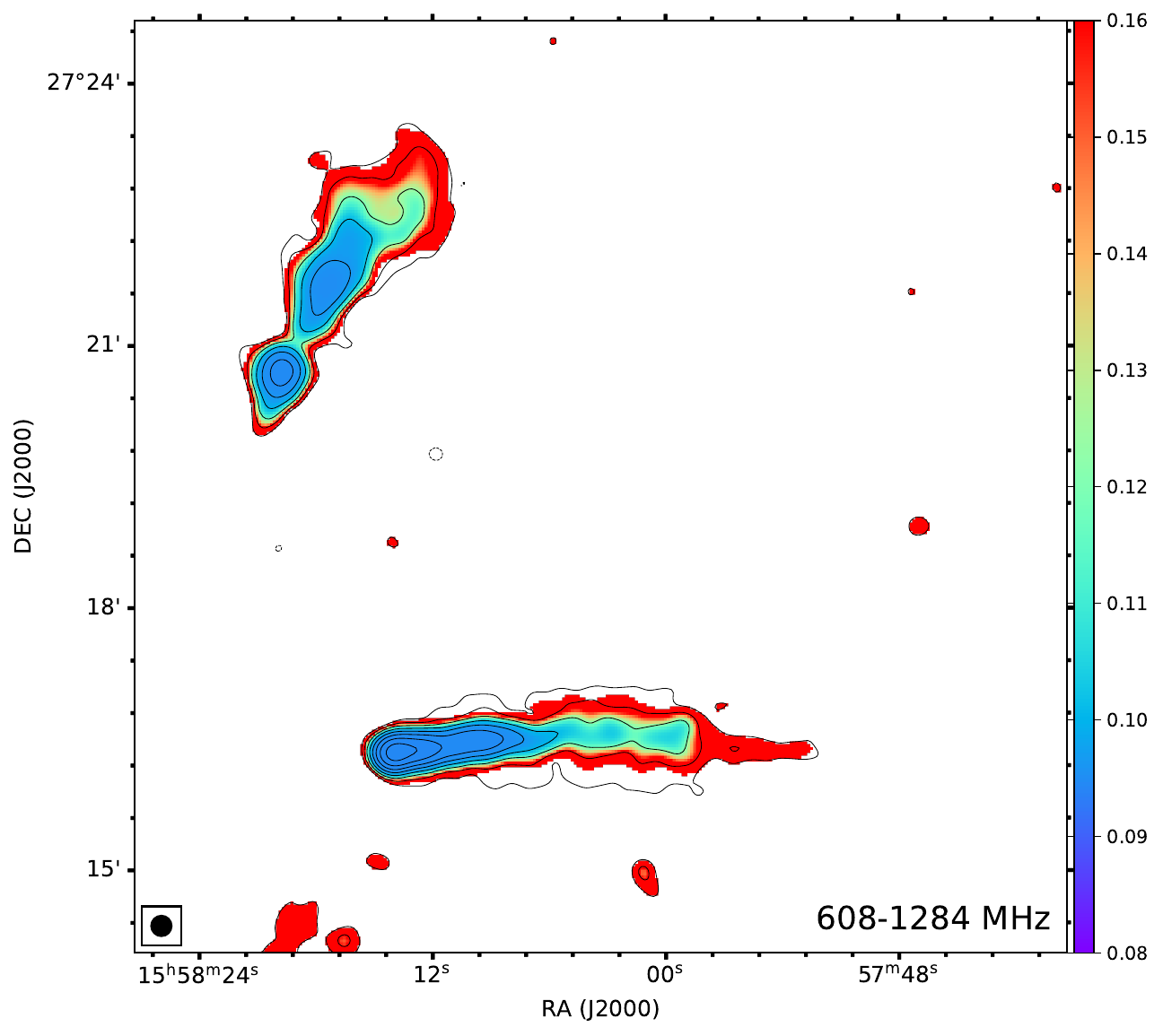}

	\smallskip
	
	\caption{Spectral index error maps of T1 and T2, corresponding to the maps shown in Fig \ref{fig: spixmap}.}
	\label{fig: errspixmap}
\end{figure*}

In Fig. \ref{fig: errspixmap} we report the error maps associated with the spectral index maps shown in Fig. \ref{fig: spixmap}. Errors are computed from the standard error propagation formula as 
\begin{equation}
\Delta \alpha =  \left\lvert \frac{1}{\ln{ \left( \frac{\nu_{\rm 1}}{\nu_{\rm 2} } \right) }}\right\lvert \sqrt{ \left( \frac{\Delta S_{\rm 1}}{S_{\rm 1}}\right)^2 + \left( \frac{\Delta S_{\rm 2}}{S_{\rm 2}}\right)^2 } \; \; \; ,
\label{eq:spectralindexerrorformula}
\end{equation}
where $\Delta S$ is obtained as in Eq. \ref{eq: erroronflux}.

\FloatBarrier

\section{X-ray spectrum of T2 host}
\label{app: X-ray spectrum of T2 host}

\FloatBarrier

\begin{figure}[!htb]
	\centering
	\includegraphics[width=0.4\textwidth]{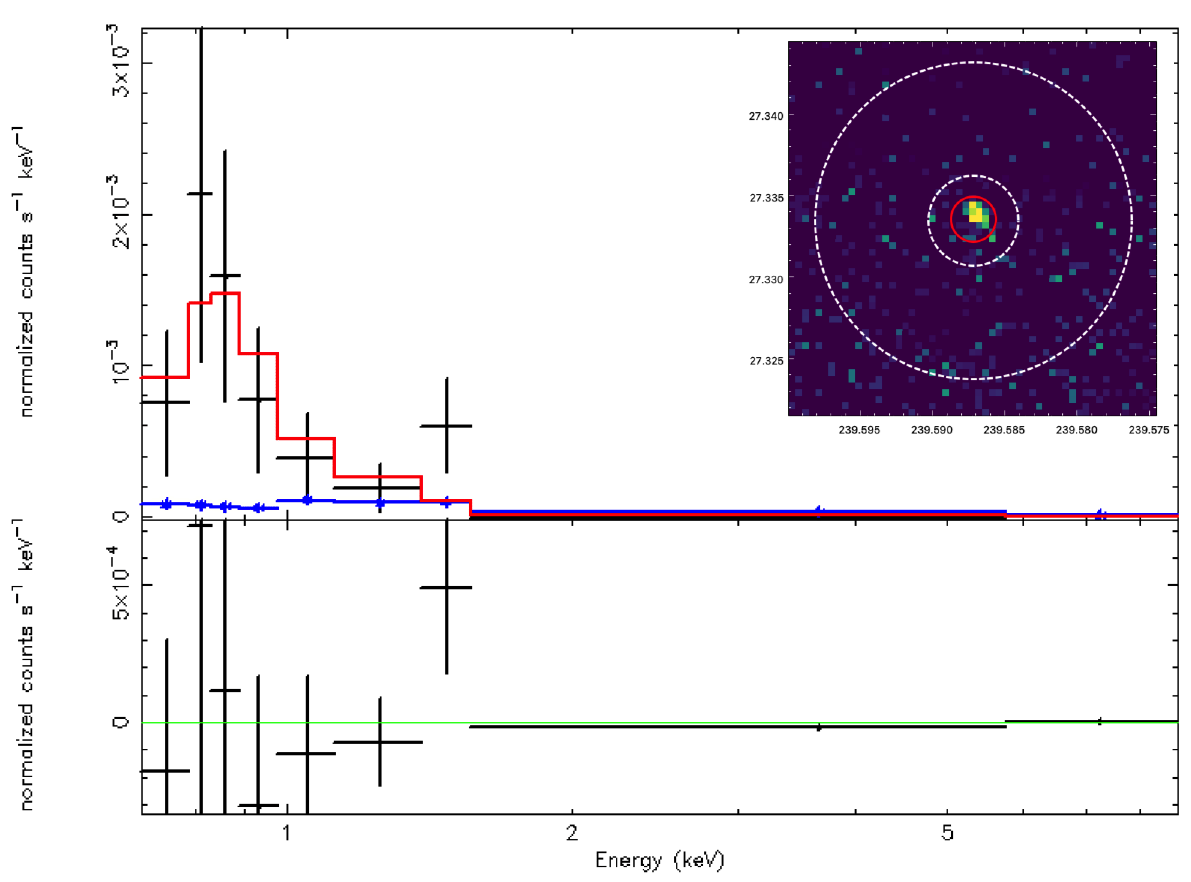}
	\smallskip
	
	\caption{\textit{Chandra} X-ray spectrum of the host of T2. The upper inset shows the extraction regions used for the source (red circle) and background (with annulus). Spectra were extracted for all available observations and jointly fitted, but a single spectrum is shown for clearer inspection. Data points were fitted with an absorbed thermal component (red curve). The subtracted background is shown in blue. The ratio of the data to the model is reported in the lower inset.}
	\label{fig: T2coreX}
\end{figure}

The host of T2 emits in the X-ray band (see Fig. \ref{ROX}). In this section we analyse its \textit{Chandra} X-ray spectrum, which is shown in Fig. \ref{fig: T2coreX}.    

We extracted the spectrum of the background from a circular annulus of width $25''$ centred on the target. The spectrum of the source was extracted from a circular region of radius $5''$ ($8.5$ kpc), chosen as the one that maximises the ${\rm S/N}$, being:
\begin{equation}
{\rm S/N} = \frac{N_{\rm cnt}-f_{\rm area}N_{\rm bkg}}{\sqrt{N_{\rm cnt}+f_{\rm area}N_{\rm bkg}}} \; \; \; ,
\label{eq: snrX}
\end{equation}
where $N_{\rm cnt}$ is the total (target plus background) count number within the circle, $N_{\rm bkg}$ is the background count number within the annulus, and $f_{\rm area}$ is the ratio of target to background extraction region areas. We extracted the corresponding background and source spectra from each pointing covering the region of T2, and we jointly fitted the background-subtracted spectra with an absorbed thermal component ({\tt phabs} $\times$ {\tt apec}). The Galactic hydrogen column density, redshift, and gas metallicity were kept fixed to the values considered for A2142 as in Sect. \ref{sect: Local ICM conditions}. The thermal component is sufficient to reproduce the spectrum of the target, as no hard X-ray ($>2$ keV) emission, which is distinctive of AGN, is detected. An additional power-law component ({\tt phabs} $\times$ ({\tt apec}$+${\tt po})) modelling possible soft X-ray emission from AGN does not improve the fit (significance below $1\sigma$) and is thus rejected. 

The best-fit temperature, density (computed from the {\tt apec} normalisation) are $kT=1.0\pm 0.1$ keV and $n_{\rm e}=(1.7\pm 0.4)\times 10^{-3} \; {\rm cm^{-3}}$. These values are consistent with those of the hot ionised medium in elliptical galaxies \citep[e.g.][]{mathews&brighenti03,sun07} and suggest that the observed X-ray emission comes from thermal gas rather than AGN activity. 

\FloatBarrier

\end{appendix}

\end{document}